\documentclass[preprint,10pt]{elsarticle}

\usepackage{amsmath}
\usepackage{amssymb}
\usepackage{anysize}
\usepackage{color}
\usepackage{multirow}
\usepackage{aas_macros}

\journal{Journal of Computational Physics}

\begin{document}
\newcommand{\HALF}{\frac{1}{2}}
\newcommand{\pd}[2]{\frac{\partial #1}{\partial #2}}
\newcommand{\DS}{\displaystyle}
\newcommand{\tens}[1]{\mathsf{#1}}
\renewcommand{\vec}[1]{\mathbf{#1}}
\newcommand{\hvec}[1]{\hat{\mathbf{#1}}}
\newcommand{\vol}{{\cal V}}
\newcommand{\av}[1]{\left<{#1}\right>}

\begin{frontmatter}



\title{High-order conservative reconstruction schemes for finite volume methods in cylindrical and spherical coordinates}


\author{A. Mignone}

\address{Dipartimento di Fisica, Universit\'a di Torino, via Pietro Giuria 1, 10125 Torino, Italy}

\begin{abstract}
High-order reconstruction schemes for the solution of hyperbolic conservation laws in orthogonal curvilinear coordinates are revised in the finite volume approach.
The formulation employs a piecewise polynomial approximation to the zone-average values to reconstruct left and right interface states from within a computational zone to arbitrary order of accuracy by inverting a Vandermonde-like linear system of equations with spatially varying coefficients.
The approach is general and can be used on uniform and non-uniform meshes although explicit expressions are derived for polynomials from second to fifth degree in cylindrical and spherical geometries with uniform grid spacing.
It is shown that, in regions of large curvature, the resulting expressions differ considerably from their Cartesian counterparts and that the lack of such corrections can severely degrade the accuracy of the solution close to the coordinate origin.
Limiting techniques and monotonicity constraints are revised for conventional reconstruction schemes, namely, the piecewise linear method (PLM), third-order weighted essentially non-oscillatory (WENO) scheme and the piecewise parabolic method (PPM).

The performance of the improved reconstruction schemes is investigated in a number of selected numerical benchmarks involving the solution of both scalar and systems of nonlinear equations (such as the equations of gas dynamics and magnetohydrodynamics) in cylindrical and spherical geometries in one and two dimensions.
Results confirm that the proposed approach yields considerably smaller errors, higher convergence rates and it avoid spurious numerical effects at a symmetry axis.
\end{abstract}


\end{frontmatter}


\section{Introduction}
%
%
%

Unsteady, time-dependent compressible flows often involve complex flow interactions featuring both continuous and discontinuous waves.
Numerical computations based on finite-volume (FV) discretizations have now established as a reliable tool to model such flows and delivering oscillation-free stable solutions while preserving conservation of relevant physical quantities such as mass, momentum and energy.
FV methods (for a review see the books from \cite{Toro.1999, LeVeque.2002}) rely on a conservative discretization based on the integral formulation of the underlying system of partial differential equations (PDEs) where volume averages (rather than point values) are evolved in time.
Average quantities can thus vary only when an unbalance exists between the fluxes entering and leaving the region boundary.
The computation of the interface flux is the heart of these methods and it is usually achieved by employing proper upwinding techniques that rely on the solution of a Riemann problem between discontinuous left and right states at cell interfaces.
These states are reconstructed from the volume averages of the solution and several techniques are available in literature, e.g., second-order TVD methods  (\cite{vanLeer.1977_IV, Harten.1983, Sweby.1984, Toro.1999, LeVeque.2002}), third-order piecewise parabolic method (\cite{ColWoo.1984}), essentially non-oscillatory (ENO, \cite{Harten_etal.1987}) and weighted essentially non-oscillatory (WENO \cite{Liu_Osher_Chan.1994, JiaShu.1996}, see also \cite{Balsara_Shu.2000}, \cite{Titarev_Toro.2004}, \cite{Balsara_etal.2009}, \cite{YamCar.2009} and references therein), monotonicity preserving (MP, \cite{SurHuy.1997}) schemes.
Generally speaking, the reconstruction is a two-step process where a high-order accurate estimate of the interface values is first provided and later modified (or limited) to fulfill monotonicity constraints. 

Traditionally, most reconstruction techniques have been devised for Cartesian geometry and a vast literature exists on this subject.
Curvilinear systems, nevertheless, are often preferred and employed in modeling many scientific applications such as, for instance, geophysical or atmospheric flows, flows in turbomachinery, astrophysical accretion disks orbiting around a central object or, more simply, flows with rotational symmetry around a vertical axis.

In this respect, it should be stressed that little attention has been devoted to the development of high-order finite volume methods in curvilinear coordinate systems \cite{MonchMull.1989, Falle.1991, BloLuf.1993, SkiOst.2010, Ziegler.2011} and that straightforward application of Cartesian-based reconstruction schemes to a curvilinear grid may suffer from a number of drawbacks and inconsistencies that have often been overlooked.
For second-order accurate schemes, this has already been demonstrated by a number of authors (e.g., \cite{MonchMull.1989, Falle.1991} and, more recently, \cite{Ziegler.2011}) who recognized that volume averages should be assigned to the centroid of volume rather than the geometrical cell center.
Higher than second-order schemes, on the other hand, still deserve a more careful treatment since simple-minded extensions of plane-parallel reconstruction methods may easily lead to incorrect results and severely compromise the accuracy of the solution in proximity of a symmetry axis.
In the original formulation of the Piecewise Parabolic Method (PPM, \cite{ColWoo.1984}), for instance, the authors suggested to perform the reconstruction in the volume coordinate (rather than the linear one) so that the same algorithm used for a Cartesian mesh could be employed on a cylindrical or spherical radial grid.
In doing so, however, the resulting interface states become formally first-order accurate even for smooth flows.
This shortcoming was addressed in \cite{BloLuf.1993} (see also \cite{SkiOst.2010}) and corrected by first interpolating the indefinite integral of a conserved fluid quantity and then differentiating the resulting polynomial with respect to the linear coordinate to obtain the desired point values at a cell interface.
Although formally correct, this approach has the disadvantage of being potentially singular at the coordinate origin and that the resulting interface states may loose one or even two orders of accuracy.


The intent of the present work is to re-formulate and improve, in the context of orthogonal curvilinear coordinates, some of the most widely used reconstruction techniques employed by FV methods.
The proposed formulation is based on a piecewise polynomial reconstruction from the volume averages of conserved quantities lying in adjacent zones yielding, in one dimension, interface states that are formally correct to arbitrary order of accuracy.
This approach is presented in Section \ref{sec:problem_formulation} for a scalar conservation law and it is general enough to be employed on regularly- as well as irregularly- spaced grids.
Closed form solutions are derived on uniform radial grids in cylindrical and spherical geometries for polynomials of second up to fifth degree.
For a more complex coordinate system and/or non-uniform grids the reconstruction process can still be carried out by using numerical quadrature and/or the solution of a linear system of equations at the beginning of the computation.
In order to suppress spurious oscillations, conventional limiting techniques for second-order TVD, WENO and PPM schemes are revised in Section \ref{sec:limiting} for the case of cylindrical and spherical geometries.
Extension to nonlinear systems of equations is treated in Section \ref{sec:nonlinear_systems} where reconstruction from primitive variables and integration of geometrical source terms are discussed.
Finally, in Section \ref{sec:numerical_benchmarks}, the proposed schemes are tested and compared on one- and two-dimensional selected test problems in cylindrical and spherical geometries.
Both scalar hyperbolic conservation laws and nonlinear systems of equation are considered.

\section{Problem formulation}
\label{sec:problem_formulation}
%
%
%
 
\subsection{Finite volume discretization in curvilinear coordinates}
\label{sec:fv_scalar}
%
%
%

Given an orthogonal system of coordinates $(x_1,\, x_2,\, x_3)$ with unit vectors $\hvec{e}_1,\, \hvec{e}_2,\, \hvec{e}_3$ and scale factors $(h_1,\, h_2,\, h_3)$, we now wish to solve the scalar conservation law
\begin{equation}\label{eq:prototype}
  \pd{Q}{t} + \nabla\cdot\vec{F} = S\,,
\end{equation}
where $Q$ is a conserved fluid quantity, $\vec{F}=(F_1,\, F_2,\, F_3)$ is the corresponding flux vector, $S$ is a source term while the divergence operator takes the form
\begin{equation}
  \nabla\cdot\vec{F} = \frac{1}{h_1h_2h_3}\left[
     \pd{}{x_1}(h_2h_3F_1) + \pd{}{x_2}(h_1h_3F_2)  + \pd{}{x_3}(h_1h_2F_3) 
     \right]\,.
\end{equation}

Eq. (\ref{eq:prototype}) is discretized on a computational domain divided into $N_1\times N_2\times N_3$ cells (or zones) with lower and upper coordinate bounds respectively given by $(x_{1,i-\HALF},\, x_{2,j-\HALF},\, x_{3,k-\HALF})$ and $(x_{1,i+\HALF},\, x_{2,j+\HALF},\, x_{3,k+\HALF})$ so that the mesh spacings are denoted with 
\begin{equation}
  \Delta x_{1,i} = x_{1,i+\HALF} - x_{1,i-\HALF} \,;\qquad
  \Delta x_{2,j} = x_{2,j+\HALF} - x_{2,j-\HALF} \,;\qquad 
  \Delta x_{3,k} = x_{3,k+\HALF} - x_{3,k-\HALF} \,,\qquad
\end{equation}
while the cell volume is defined by
\begin{equation}
  \Delta\vol_{i,j,k} = \int_{x_{3,k-\HALF}}^{x_{3,k+\HALF}} 
                       \int_{x_{2,j-\HALF}}^{x_{2,j+\HALF}}
                       \int_{x_{1,i-\HALF}}^{x_{1,i+\HALF}}  h_1h_2h_3\,dx_1dx_2dx_3\,.
\end{equation}
For convenience, let $\vec{i}=(i,\, j,\, k) \in \mathbb{Z}^3$ be a vector of integer numbers giving the position of a computational zone in a three-dimensional lattice with $1\le i \le N_1$, $1\le j \le N_2$, $1\le k \le N_3$.
The position of the cell interfaces orthogonal to the direction given by $\hvec{e}_d$ will therefore be denoted with $\vec{i}\pm\HALF\hvec{e}_d$.
Integrating Eq. (\ref{eq:prototype}) over the cell volume and applying Gauss's theorem yields the conservative discretization
\begin{equation}\label{eq:prototype_integrated}
  \frac{d}{dt}\av{Q}_{\vec{i}} + \frac{1}{\Delta\vol_{\vec{i}}}
   \sum_d \left[ 
    \left(A_d\tilde{F}_d\right)_{\vec{i}+\HALF\hvec{e}_d} 
  - \left(A_d\tilde{F}_d\right)_{\vec{i}-\HALF\hvec{e}_d}
   \right] = \av{S}_{\vec{i}} \,,
\end{equation}
where $\av{Q}_{\vec{i}}$ denotes the volume average of $Q$ over the control volume, $d=1,2,3$ spans across multiple directions, $\tilde{F}_d$ is the flux averaged over the surface $A_d$ with outward normal pointing in the direction $\hvec{e}_d$.
Thus, when $d=1$ (for instance), one has $\vec{i}+\HALF\hvec{e}_d = (i+\HALF,\, j,\, k)$ and
\begin{equation}\label{eq:flux_average}
 \tilde{F}_{1,\vec{i}+\HALF\hvec{e}_1} = \frac{1}{A_{1,\vec{i}+\HALF\hvec{e}_1}}
   \int_{x_{2,j-\HALF}}^{x_{2,j+\HALF}}  \int_{x_{3,k-\HALF}}^{x_{3,k+\HALF}}
   F_1h_2 h_3 dx_2dx_3
  \,;\qquad
  A_{1,\vec{i}+\HALF\hvec{e}_1} = \int_{x_{2,j-\HALF}}^{x_{2,j+\HALF}}  
     \int_{x_{3,k-\HALF}}^{x_{3,k+\HALF}} h_2h_3dx_2dx_3 \,.
\end{equation}
where both $F_1$ and the scale factors $h_1,\, h_2,\, h_3$ are functions of the position vector at the interface $(x_{1,i+\HALF},\, x_2,\, x_3)$.
Similar expressions hold for $d=2$ and $d=3$ by cyclic index permutation.

In cylindrical coordinates $(x_1,\, x_2,\, x_3) \equiv (R,\phi,z)$, $(h_1,\,h_2,\,h_3) \equiv (1,\, R,\, 1)$ and Eq. (\ref{eq:prototype_integrated}) takes the form
\begin{equation}\label{eq:cons_law_cylindrical}
\begin{split}
 \frac{d\av{Q}_\vec{i}}{dt} = &
  - \frac{ \left(\tilde{F}_RR\right)_{\vec{i}+\HALF\hvec{e}_r} 
         - \left(\tilde{F}_RR\right)_{\vec{i}-\HALF\hvec{e}_r}} 
         {\Delta\vol_{R,i}} 
  - \frac{ \left(\tilde{F}_\phi\right)_{\vec{i}+\HALF\hvec{e}_\phi} 
           - \left(\tilde{F}_\phi\right)_{\vec{i}-\HALF\hvec{e}_\phi}}{R_i\Delta\phi_j}
  \\ & 
  - \frac{ \left(\tilde{F}_z\right)_{\vec{i}+\HALF\hvec{e}_z} 
         - \left(\tilde{F}_z\right)_{\vec{i}-\HALF\hvec{e}_z}}{\Delta z_k}
  + \av{S}_{\vec{i}}
\end{split}
\end{equation}
where $(\tilde{F}_R,\, \tilde{F}_\phi,\, \tilde{F}_z)$  are the surface-averaged flux components in the three directions and $\Delta\vol_{R,i} = (R_{i+\HALF}^2 - R_{i-\HALF}^2)/2$ is the cell radial volume.

Similarly, in spherical coordinates $(x_1,\,x_2,\,x_3)\equiv(r,\,\theta,\,\phi)$, $(h_1,\, h_2,\, h_3) = (1,\, r,\, r\sin\theta)$ and one obtains
\begin{equation}\label{eq:cons_law_spherical}
\begin{split}
   \frac{d\av{Q}_{\vec{i}}}{dt} = &
  -\frac{  \left(\tilde{F}_rr^2\right)_{\vec{i}+\HALF\hvec{e}_r} 
         - \left(\tilde{F}_rr^2\right)_{\vec{i}-\HALF\hvec{e}_r}}{\Delta\vol_{r,i}}
  -\frac{  \left(\tilde{F}_\theta\sin\theta\right)_{\vec{i}+\HALF\hvec{e}_\theta} 
         - \left(\tilde{F}_\theta\sin\theta\right)_{\vec{i}-\HALF\hvec{e}_\theta}}  
        {\tilde{r}_i\Delta\mu_j} \\  &
  -\frac{\Delta\theta_j}{\Delta\mu_j}
   \frac{  \left(\tilde{F}_\phi\right)_{\vec{i}+\HALF\hvec{e}_\phi} 
         - \left(\tilde{F}_\phi\right)_{\vec{i}-\HALF\hvec{e}_\phi}}
           {\tilde{r}_i\Delta\phi_k}
  + \av{S}_{\vec{i}}
\end{split}
\end{equation}
where $(\tilde{F}_r, \tilde{F}_\theta, \tilde{F}_\phi)$ are the surface-averaged vector components of the flux $\vec{F}$ in the three coordinate directions and
\begin{equation}
\Delta\vol_{r,i} = \frac{r_{i+\HALF}^3 - r_{i-\HALF}^3}{3}
  \,;\qquad
  \tilde{r}_i = \frac{2}{3}\frac{r_{i+\HALF}^3 - r_{i-\HALF}^3}
                               {r_{i+\HALF}^2 - r_{i-\HALF}^2}
  \,;\qquad
\Delta\mu_j = \cos\theta_{j-\HALF} - \cos\theta_{j+\HALF} \,
\end{equation}
are geometrical factors.

The interface fluxes are normally computed by solving a Riemann problem between adjacent discontinuous states.
Using, for example, a simple midpoint quadrature rule one has
\begin{equation}\label{eq:interface_flux}
  \tilde{F}_{d,\vec{i}+\HALF\hvec{e}_d} \approx
  {\cal R}\left(Q^{L}_{\vec{i}+\HALF\hvec{e}_d},\, 
                Q^{R}_{\vec{i}+\HALF\hvec{e}_d}\right)  \,,
\end{equation}
where ${\cal R}(\cdot,\cdot)$ is an approximate Riemann solver flux whereas $Q^L_{\vec{i}+\HALF\hvec{e}_d}$ and $Q^R_{\vec{i}+\HALF\hvec{e}_d}$ are the one-sided limit values (from below and from above, respectively) of the piecewise polynomial reconstructions from within the two adjacent zones $\vec{i}$ and $\vec{i}+\hvec{e}_d$.
In compact notations:
\begin{equation}\label{eq:interface_limit}
  Q^{L}_{\vec{i}+\HALF\vec{e}_d} = 
  \lim_{\vec{x}\to \vec{x}^{-}_{\vec{i}+\HALF\hvec{e}_d}}
      Q_{\vec{i}}\left(\vec{x}\right)   
  \,;\qquad
  Q^{R}_{\vec{i}+\HALF\vec{e}_d} = 
   \lim_{\vec{x}\to \vec{x}^{+}_{\vec{i}+\HALF\hvec{e}_d}}
  Q_{\vec{i}+\hvec{e}_d}\left(\vec{x}\right)   \,,
\end{equation}
where $\vec{x}_{\vec{i}}$ is a short-hand notation for $(x_{1,i},\, x_{2,j},\, x_{3,k})$.

In the rest of this work we restrict our attention to one-dimensional reconstruction methods and leave multi-dimensional high-order approximations to forthcoming papers.

\subsection{Conservative reconstruction from volume averages}
\label{sec:conservative_reconstruction}
%
%
%

Consider a non-uniform grid spacing with zone width $\Delta \xi_i= \xi_{i+\HALF} - \xi_{i-\HALF}$, where $\xi \in (x_1,\, x_2,\, x_3)$ is the coordinate along the reconstruction direction and $\xi_{i+\HALF}$ denotes the location of the cell interface between zones $i$ and $i+1$.
Let $\av{Q}_i$ be the cell average of $Q$ inside zone $i$ at some given time, that is,
\begin{equation}
 \av{Q}_i = \frac{1}{\Delta\vol_i}
 \int_{\xi_{i-\HALF}}^{\xi_{i+\HALF}} Q(\xi) \pd{\vol}{\xi} d\xi 
 \qquad\mathrm{where}\qquad
 \Delta\vol_i = \int_{\xi_{i-\HALF}}^{\xi_{i+\HALF}} \pd{\vol}{\xi} d\xi \,,
\end{equation}
and $\partial\vol/\partial\xi$ is a one-dimensional Jacobian and $\Delta\vol_i$ is the local cell volume.

We now wish to find a $p$-th accurate approximation to the actual solution by constructing, within a given zone $i$, a polynomial distribution 
\begin{equation}\label{eq:polynomial}
  Q_i(\xi) = a_{i,0} + a_{i,1}(\xi-\xi^c_i) + a_{i,2}(\xi-\xi^c_i)^2 + \cdots 
                 + a_{i,p-1}(\xi-\xi^c_i)^{p-1}
\end{equation}
where $\{a_{i,n}\}$ are coefficients to be determined and $\xi^c_i$ may be taken to be the cell center although the final interface values do not depend on the particular choice of $\xi^c_i$ and one may as well set $\xi^c_i = 0$.
The method has to be locally conservative meaning that the polynomial $Q_i(\xi)$ must fit the neighboring cell-averages:
\begin{equation}\label{eq:averages}
 \int_{\xi_{i+s-\HALF}}^{\xi_{i+s+\HALF}} Q_i(\xi) \pd{\vol}{\xi} d\xi 
 = \Delta\vol_{i+s}\av{Q}_{i+s}
  \qquad\mathrm{for}\qquad
  -i_L\le s\le i_R
\end{equation}
where the stencil includes $i_L$ cells to the left and $i_R$ cells to the right of the $i$-th zone so that $i_L+i_R + 1= p$.
Straightforward manipulation of Eq. (\ref{eq:averages}) leads to the following $p\times p$ linear system in the coefficients $\{a_{i,n}\}$:
\begin{equation}\label{eq:linear_system}
\left(\begin{array}{ccc}
  \beta_{i-i_L,0}  & \cdots  & \beta_{i-i_L,p-1}  \\ \noalign{\medskip} 
  \vdots           & \ddots  & \vdots                \\ \noalign{\medskip}
  \beta_{i+i_R,0}  & \cdots  & \beta_{i+i_R,p-1} 
\end{array}\right)
 \left(\begin{array}{c}
   a_{i,0} \\ \noalign{\medskip}
   \vdots \\ \noalign{\medskip}
   a_{i,p-1}  \end{array}\right)
  = 
 \left(\begin{array}{c}
   \av{Q}_{i-i_L} \\ \noalign{\medskip}
   \vdots \\ \noalign{\medskip}
   \av{Q}_{i+i_R}  \end{array}\right)
\end{equation}
where 
\begin{equation}\label{eq:beta_coeff}
 \beta_{i+s,n} = \frac{1}{\Delta\vol_{i+s}}
               \int_{\xi_{i+s-\HALF}}^{\xi_{i+s+\HALF}} 
                    (\xi-\xi^c_i)^n \pd{\vol}{\xi}\,d\xi
\end{equation}
are geometry-dependent coefficients.

When $\xi^c_i = 0$ one has, by definition, that $\beta_{i+s,0} = 1$ while $\beta_{i,1}$ defines the centroid of volume $\bar{\xi}_i$ (also called the center of gravity or barycenter)  which in cylindrical and spherical coordinates is given by 
\begin{equation}\label{eq:centroid_of_volume}
  \beta_{i,1} \equiv \bar{\xi}_i = \left\{\begin{array}{ll}
  \DS  R_i + \frac{\Delta R^2}{12 R_i} 
       \qquad & \mathrm{cylindrical},\,\xi=R  \\ \noalign{\bigskip}
  \DS  r_i + \frac{2r_i\Delta r^2}{12r_i^2 + \Delta r^2}
       \qquad & \mathrm{spherical},\,\xi=r  \\ \noalign{\bigskip}
  \DS  \frac{\theta_{i-\HALF}\cos\theta_{i-\HALF} - \sin\theta_{i-\HALF} 
            -\theta_{i+\HALF}\cos\theta_{i+\HALF} + \sin\theta_{i+\HALF}}
            {\cos\theta_{i-\HALF} - \cos\theta_{i+\HALF}}
       \qquad & \mathrm{spherical},\,\xi=\theta
  \end{array}\right.
\end{equation}
Unlike the actual cell center, the centroid of volume is no longer equidistant from the cell interfaces.
However, its importance stems from the fact that, up to second-order accuracy, the volume average of any function $Q$ can be interchanged with the function point-value evaluated at the centroid: 
\begin{equation}\label{eq:volume_point_err}
  \av{Q}_i - Q_i(\bar{\xi}_i) \approx \frac{\Delta\xi^2}{24}
                                       \left.\frac{d^2Q}{d\xi^2}\right|_{\xi=\bar{\xi}_i}\,.
\end{equation}
This result follows from Eqs. (\ref{eq:polynomial}) and (\ref{eq:averages}) with $\xi^c_i=\bar{\xi}_i$ and $s=0$.
Equation (\ref{eq:volume_point_err}) demonstrates that zone-averaged quantities should be assigned to the centroid of volume rather than the cell center as it is ordinarily done in Cartesian coordinates, see also \cite{MonchMull.1989, Falle.1991, SkiOst.2010}.

Once the coefficients $\{a_{i,n}\}$ are found by inverting Eq. (\ref{eq:linear_system}), the leftmost and rightmost interface values $Q^{+}_{i}$ and $Q^{-}_{i}$ may be obtained through:
\begin{equation}\label{eq:interface_limit_1D}
  Q^+_i = Q^L_{i+\HALF} = \lim_{\xi\to\xi^{(-)}_{i+\HALF}} Q_i(\xi)
  \,;\qquad
  Q^-_i = Q^R_{i-\HALF} = \lim_{\xi\to\xi^{(+)}_{i-\HALF}} Q_i(\xi)\,.
\end{equation}
Although formally correct, the main disadvantage of the previous approach is that Eq. (\ref{eq:linear_system}) has to be solved for each grid cell and at each time step during the numerical computation.
This can be a rather time-consuming task.

A more efficient formulation consists in rewriting the leftmost and rightmost interface values directly as a linear combination of the adjacent cell averages,
\begin{equation}\label{eq:pm_states}
 Q^{\pm}_{i} = \sum_{s=-i_L}^{i_R} w^{\pm}_{i,s} \av{Q}_{i+s} \,,
\end{equation}
where, after combining Eqs. (\ref{eq:polynomial}) and (\ref{eq:linear_system}), the weights $w^{\pm}_{i,s}$ can be shown to satisfy (see Appendix \ref{app:linear_system_w})
\begin{equation}\label{eq:linear_system_w}
\left(\begin{array}{ccc}
  \beta_{i-i_L,0}  & \cdots  & \beta_{i-i_L,p-1}  \\ \noalign{\medskip} 
  \vdots           & \ddots  & \vdots                \\ \noalign{\medskip}
  \beta_{i+i_R,0}  & \cdots  & \beta_{i+i_R,p-1} 
\end{array}\right)^T
 \left(\begin{array}{c}
   w^{\pm}_{i,-i_L} \\ \noalign{\medskip}
   \vdots \\ \noalign{\medskip}
   w^{\pm}_{i,i_R}  \end{array}\right)
  = 
 \left(\begin{array}{c}
   1 \\ \noalign{\medskip}
   \vdots \\ \noalign{\medskip}
   (\xi_{i\pm\HALF}-\xi^c_i)^{p-1} \end{array}\right) \,,
\end{equation}
where $T$ denotes transpose and the coefficients automatically satisfy the normalization condition
\begin{equation}\label{eq:w_normalization}
   \sum_{s=-i_L}^{s=i_R} w^\pm_{i,s} = 1\,.
\end{equation}

In this way, the computation of left and right states does not depend on the solution values but only on the geometry and grid structure. 
For this reason, Eq. (\ref{eq:linear_system_w}) may be solved just once after the grid has been constructed and the weights $w_{i,k}^\pm$ can be stored into memory (at a very modest cost) for later re-use.
A number of remarks are worth making.

\begin{enumerate}
\item The reconstruction stencil can be chosen to be symmetric with respect to the cell center when $i_L = i_R$ ($p$ odd) or symmetric with respect to the cell interface when $i_R = i_L+1$ ($p$ even), see also \citep{WhiteAdcroft.2008}.
In the former case, the leftmost and rightmost edge values from within the cell are written in terms of the $w^+_{i,k}$ and $w^-_{i,k}$ coefficients using the same zone-centered polynomial. 
In this way, adjacent interface values will automatically be discontinuous, i.e., $Q^+_{i} \ne Q^-_{i+1}$ as they are obtained by different polynomials.
This is the common approach adopted, for example, in WENO schemes.
Conversely, for a face-centered polynomial, left and right states can be initialized to the same unique value $Q^+_i = Q^-_{i+1}$ which can be expressed in terms of the $w^+_{i,k}$ coefficients alone. 
The original PPM is based on this approach.

\item The final interface values (\ref{eq:pm_states}) depend only on the order of the polynomial and the stencil $\{i_L,\, i_R\}$ but not on the particular value of $\xi^c_i$ used to write Eq. (\ref{eq:polynomial}).
The same conclusion holds also for the weight coefficients and one may as well set $\xi^c_i=0$ in Eq. (\ref{eq:linear_system_w}) and (\ref{eq:beta_coeff}).

\item A simplification occurs when the Jacobian is a simple power of $\xi$, that is, $\partial_\xi\vol=\xi^m$ where $m=0,1,2$ correspond to Cartesian, cylindrical and spherical radial coordinates, respectively.
In this case the $\beta$ coefficients (computed with $\xi_c = 0$) simplify to 
\begin{equation}
  \beta_{i+s,n} = \frac{m+1}{n+m+1}
                \frac{\xi_{i+s+\HALF}^{n+m+1} - \xi^{n+m+1}_{i+s-\HALF}}
                     {\xi^{m+1}_{i+s+\HALF} - \xi^{m+1}_{i+s-\HALF}} \,,
\end{equation}
and Eq. (\ref{eq:linear_system_w}) can be solved analytically. 
 
On a uniform grid in the cylindrical radial coordinate (for instance) one obtains, for $i_L=i_R=1$,
\begin{equation}\label{eq:weights_cyl_p}
w^+_{i,-1} = -\frac{( 2i-3)(2i^2 - 1)}{12(i^2-i-1)(2i-1)} \,;\quad
w^+_{i,0}  =  \frac{ ( 10i^2 - 9i - 11)}{12(i^2-i-1)} \,;\quad
w^+_{i,1}  =  \frac{ ( 2i + 1)( 4i^2 - 9i + 4)}{12(i^2-i-1)(2i-1)}
\end{equation}
\begin{equation}\label{eq:weights_cyl_m}
w^-_{i,-1} =  \frac{ ( 2i - 3)(4i^2 + i - 1)}{12(i^2-i-1)(2i-1)}    \,;\quad
w^-_{i,0}  =  \frac{ (10i^2 - 11i - 10)}{12(i^2-i-1)} \,;\quad
w^-_{i,1}  = -\frac{ ( 2i + 1)(2i^2 - 4i + 1)}{12(i^2-i-1)(2i-1)}
\end{equation}
where $i=R_{i+\HALF}/\Delta R$ is the grid index starting at $i=1$ (first active computational zone).
Notice that the weights are defined solely in terms of the grid index and do not explicitly depend on the grid spacing or radial distance.
This feature makes the implementation on adaptively refined grid easier as the coefficients need not be recomputed when a new grid is created. 

In the limit of small curvature, $i\to\infty$, Eqs. (\ref{eq:weights_cyl_p}) and (\ref{eq:weights_cyl_m}) reproduce the well-known Cartesian weights for third-order accurate reconstruction,
\begin{equation}
 w^{+}_{i,-1} = w^{-}_{i,1} \to -\frac{1}{6}
 \,;\quad
 w^+_{i,0} = w^-_{i,0} \to \frac{5}{6}
 \,;\quad
 w^+_{i,1} = w^-_{i,-1} \to \frac{1}{3}\,.
\end{equation}

The complete expressions of the interpolation weights $w^{\pm}_{i,k}$ for third-, fourth- and fifth-order spatial accuracy ($p=3,4,5$) in the radial coordinate are given in Appendix \ref{app:interpolation_weights} for Cartesian, cylindrical and spherical geometries.

\item For the spherical meridional coordinate ($\partial_\xi\vol=\sin\xi$) one has
\begin{equation}
   \beta_{i+s,n} = \frac{1}{\cos\xi_{i_{s-}} - \cos\xi_{i_{s+}}}
    \sum_{k=0}^{n} k!\,\binom{n}{k}\,
    \left[  \xi^{n-k}_{i_{s-}}\cos\left(\xi_{i_{s-}} + \frac{k\pi}{2}\right)
          - \xi^{n-k}_{i_{s+}}\cos\left(\xi_{i_{s+}} + \frac{k\pi}{2}\right)
    \right] \,
\end{equation}                   
where $i_{s\pm}$ is a short-hand for $i+s\pm\HALF$.
However, in this case, the final analytical expressions are quite lengthy and it is advisable to approach the solution of Eq. (\ref{eq:linear_system_w}) numerically using, for example, straightforward LU decomposition.

\item Eq. (\ref{eq:linear_system_w}) retains its validity also when the grid spacing in not uniform.
In these cases, it is also more convenient to resort to direct numerical inversion.

\item Eq. (\ref{eq:linear_system_w}) can be solved to compute the point-value of $Q(\xi)$ not only at the cell interfaces but also at any other point inside the cell.
In particular, setting the constant column vector on the right hand side to $(1,\, 0,\, \dots,\, 0)^T$ (together with $\xi_i=0$ in Eq. \ref{eq:beta_coeff}) allows to find the expansion coefficients needed to approximate the function value at the cell center.
This is useful in the context of nonlinear systems of equations where the reconstruction can be performed from the volume averages of primitive variables rather than conservative ones, see Section \ref{sec:nonlinear_systems}.

\end{enumerate}

\section{Limiting techniques and monotonicity constraints}
\label{sec:limiting}
%
%
%

\subsection{Second-order piecewise linear reconstruction (PLM)}
\label{sec:linear}
%
%
%

Second-order accurate reconstructions can be recovered by fitting a linear polynomial ($p=2$) through either $(\av{Q}_i,\av{Q}_{i+1})$ or $(\av{Q}_i, \av{Q}_{i-1})$ which amounts to solving Eq. (\ref{eq:linear_system_w}) with $\{i_L,\,i_R\}=\{0,1\}$ or $\{i_L,\,i_R\}=\{1,0\}$, respectively.
The final result can be cast in terms of the forward ($\rm f$) and backward ($\rm b$) difference approximations to the derivative:
\begin{equation}\label{eq:linear_FWD_and_BCK}
  Q^{\pm,[\rm f]}_{i} = \av{Q}_i + \Delta Q^F_i\, 
             \frac{\xi_{i\pm\HALF} - \bar{\xi}_{i}}{\Delta\xi_i}
\,;\qquad
  Q^{\pm,[\rm b]}_{i} = \av{Q}_i + \Delta Q^B_i\, 
             \frac{\xi_{i\pm\HALF} - \bar{\xi}_{i}}{\Delta\xi_i}\,,
\end{equation}
where
\begin{equation}\label{eq:fwd_and_bck_slopes}
  \Delta  Q^F_i = \Delta\xi_i\left(
    \frac{\av{Q}_{i+1} - \av{Q}_{i}}{\bar{\xi}_{i+1} - \bar{\xi}_{i}} \right)
     \,;\qquad
  \Delta Q^B_i = \Delta\xi_i\left(
    \frac{\av{Q}_{i} - \av{Q}_{i-1}}{\bar{\xi}_{i} - \bar{\xi}_{i-1}}\right)
\end{equation}
are the forward and backward approximations to the first derivative while $\bar{\xi}_i$ is the centroid of volume, Eq. (\ref{eq:centroid_of_volume}).
The previous expressions are exact for a linear function even on a non-uniform grid.

\begin{figure}[!h]
 \centering
 \includegraphics[width=0.5\textwidth]{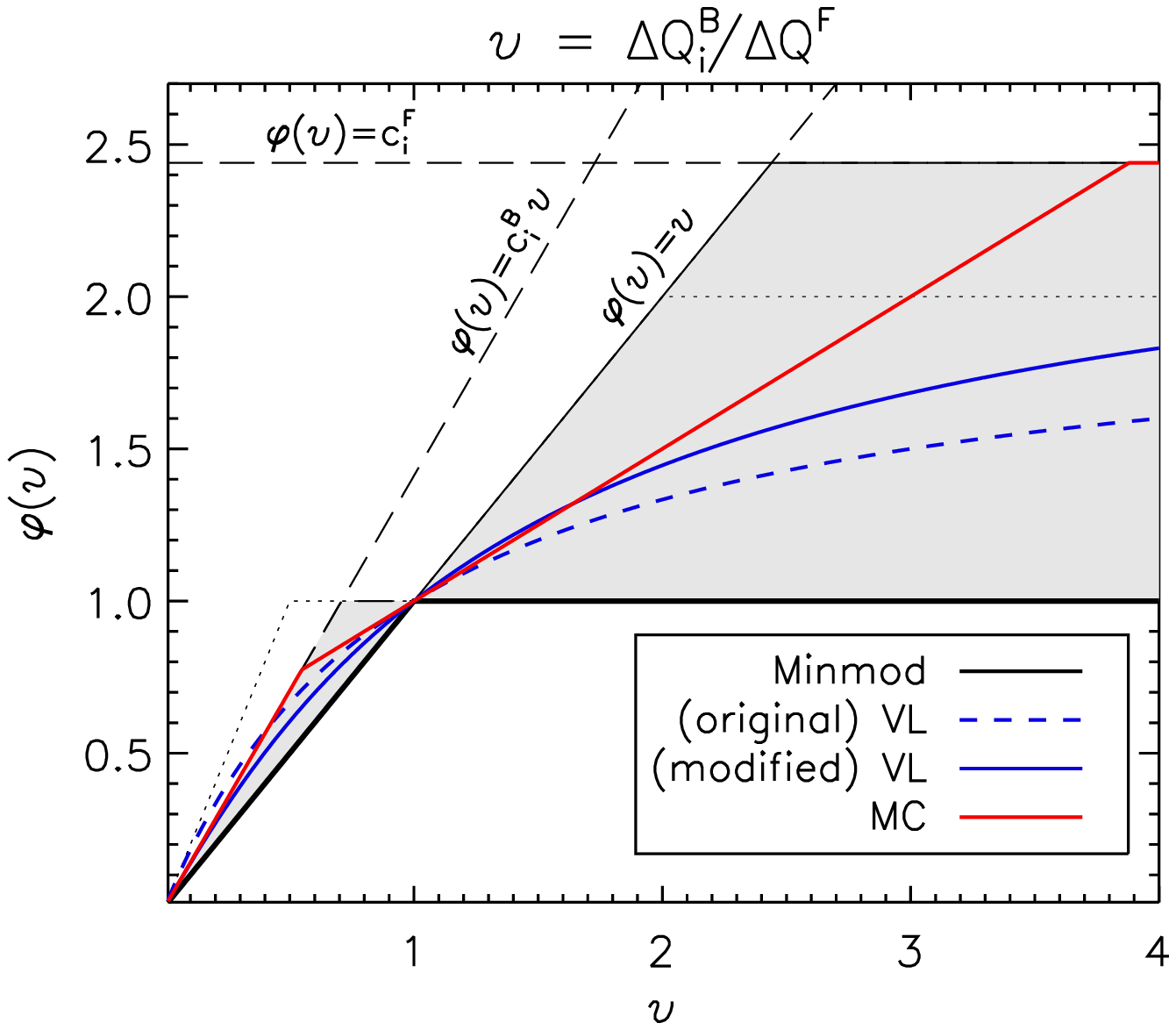}%
 \includegraphics[width=0.5\textwidth]{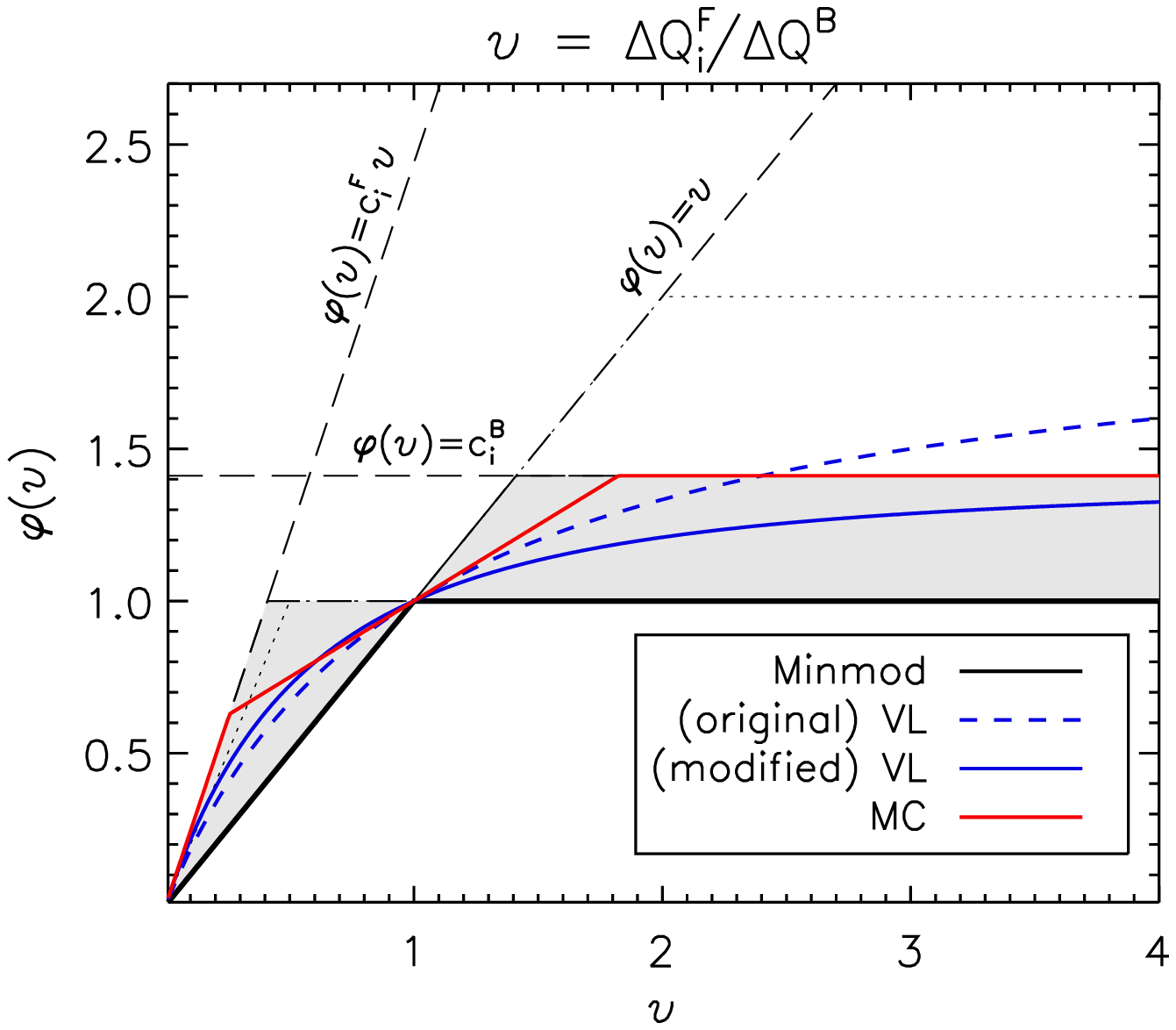}
 \caption{\footnotesize Sweby's diagrams for $\upsilon=\Delta Q^B_i/\Delta Q^F_i$ (left) and $\upsilon=\Delta Q^F_i/\Delta Q^B_i$ (right) for $i=2$ (second active computational zone) on a radial spherical grid. 
 The shaded area gives the TVD region delimited by the constraints imposed in Eq. (\ref{eq:limiter_constraints}).
 The dotted lines give the corresponding limits on a Cartesian grid with uniform spacing.
 The black, blue and red solid lines refer to the Minmod, modified van Leer (VL) and monotonized central (MC) limiter defined, respectively, by Eq. (\ref{eq:MM_lim}), (\ref{eq:VL_lim_modified}) and (\ref{eq:MC_lim}).
 The dashed blue line is the original van Leer limiter.}
 \label{fig:sweby_diagram}
\end{figure}

In order to suppress the appearance of unwanted new extrema, the two slopes can be combined together using a nonlinear slope limiter:
\begin{equation}\label{eq:linear_lim}
 Q^{\pm}_{i} = \av{Q}_i + \overline{\Delta Q}_i
 \,\frac{\xi_{i\pm\HALF} - \bar{\xi}_i}{\Delta\xi_i} \,,
\end{equation}
where the slope $\overline{\Delta Q}_i$ is traditionally written in terms of a nonlinear limiter function $\varphi(\upsilon)$ such that
\begin{equation}\label{eq:limiter_function}
  \overline{\Delta Q}_i = \Delta Q^F_i\varphi(\upsilon) 
  \qquad\mathrm{where}\qquad
  \upsilon = \frac{\Delta Q^B_i}{\Delta Q^F_i} \,.
\end{equation}
Sweby \cite{Sweby.1984} (see also \cite{Toro.1997,LeVeque.2002}) has shown that, in order for the method to be Total Variation Diminishing (TVD), the limiter function must satisfy the following conditions: i) the final slope should be expressed as a weighted average of the forward and backward derivatives, ii) zero gradient must be assigned near local extrema, iii) the reconstruction remains symmetric when $\Delta Q^F_i$ and $\Delta Q^B_i$ are swapped and iv) the reconstructed values always lie between the bounds set by neighboring cells:
\begin{equation}\label{eq:monotonicity}
 \min\big(\av{Q}_{i}, \av{Q}_{i\pm1}\big) \le 
   Q^{\pm}_{i} \le \max\big(\av{Q}_{i}, \av{Q}_{i\pm1}\big)  \,.
\end{equation}
Although conditions i), ii) remain unaltered in curvilinear geometry, the symmetry and monotonicity conditions iii) and iv) are slightly different due the fact that the forward and backward undivided gradients are weighted over the distance between adjacent volume centroids.
These considerations lead to the following modified constraints for the function $\varphi(\upsilon)$ when $\upsilon \ge 0$:
\begin{equation}\label{eq:limiter_constraints}
  \left\{\begin{array}{lcl}
  \min(1,\upsilon) \le \varphi(\upsilon) \le \max(1,\upsilon) & 
      \\ \noalign{\medskip}
\DS  \varphi\left(\frac{1}{\upsilon}\right) = 
      \frac{\varphi(\upsilon)}{\upsilon} &
      \\ \noalign{\medskip}
  \varphi(\upsilon) \le \min\left(c^F_i,\, c^B_i\upsilon\right)
 \end{array}\right.
  \qquad\mathrm{where}\qquad
  c^F_i = \frac{\bar{\xi}_{i+1} - \bar{\xi}_i}{\xi_{i+\HALF} - \bar{\xi}_i}
  \,;\qquad
  c^B_i = \frac{\bar{\xi}_{i} - \bar{\xi}_{i-1}}{\bar{\xi}_i - \xi_{i-\HALF}}\,,
\end{equation}
while $\varphi(\upsilon) = 0$ when $\upsilon < 0$.
Note that, in order to preserve the correct symmetry when $\upsilon\to1/\upsilon$, the coefficients $c^F_i$ and $c^B_i$ must also be interchanged: $(c^F_i,\, c^B_i)\to(c^B_i,\, c^F_i)$.
The corresponding Sweby's diagrams are represented in Fig \ref{fig:sweby_diagram} for $\upsilon = \Delta Q^B/\Delta Q^F$ (left panel) and $\upsilon = \Delta Q^B/\Delta Q^F$ (right panel) for $i=2$ in spherical coordinates.
The permitted TVD region is shown as the shaded area and it becomes askew with respect to the uniform Cartesian grid case (dotted lines).
In the limit of vanishing curvature, $\bar{\xi}_i \to \xi_i$, one has that the coefficients $c^F_i = c^B_i \to 2$ and the monotonicity condition in Eq. (\ref{eq:limiter_constraints}) reduces to the usual condition $\varphi(\upsilon) \le \min(2,2\upsilon)$.

Several limiter functions that satisfy Eq. (\ref{eq:limiter_constraints}) can be devised and three among the most popular limiters are here extended to the curvilinear case.
The simplest (albeit most diffusive) choice is the minmod limiter \cite{Ziegler.2011}:
\begin{equation}\label{eq:MM_lim}
\varphi^{MM}(\upsilon) = \max\Big[0,\,\min(1,\upsilon)\Big]
  \qquad\mathrm{or}\qquad 
  \overline{\Delta Q}_i = 
  {\rm Minmod}\left(\Delta Q^F_i,\, \Delta Q^B_i\right)\,,
\end{equation}
where the Minmod function returns the argument with the smallest absolute value if they all have the same sign and zero otherwise:
\begin{equation}
  {\rm Minmod}\,(a,\,b) = \frac{{\rm sgn}(a) + {\rm sgn}(b)}{2} \min(|a|,\,|b|)\,.
\end{equation}


Another popular choice due to \cite{vanLeer.1974_II} is the van Leer (or harmonic mean) limiter:
\begin{equation}\label{eq:VL_lim}
  \varphi(\upsilon) = \frac{\upsilon + |\upsilon|}{1 + |\upsilon|}
  \qquad\mathrm{or}\qquad
  \overline{\Delta Q}_i = \left\{\begin{array}{ll}  
  \DS  \frac{2\Delta Q^F_i\Delta Q^B_i}
            {\Delta Q^F_i + \Delta Q^B_i} &
      \qquad{\rm if}\quad \Delta Q^F_i\Delta Q^B_i > 0\,,
  \\ \noalign{\medskip}
    0   & \qquad{\rm otherwise}\,.
 \end{array}\right.
\end{equation}
Unfortunately Eq. (\ref{eq:VL_lim}) does not necessarily satisfy the monotonicity constraint given by the third in Eq. (\ref{eq:limiter_constraints}) and may overrun the TVD region as shown in Fig \ref{fig:sweby_diagram}.
Here the following modified van Leer (VL) limiter is proposed:
\begin{equation}\label{eq:VL_lim_modified}
  \varphi^{VL}(\upsilon) = \left\{\begin{array}{ll}
 \DS \frac{\upsilon(c^F_i\upsilon + c^B_i)}{\upsilon^2 + (c^F_i+c^B_i-2)\upsilon+1}
     & \qquad\mathrm{for}\quad \upsilon \ge 0  \,,
  \\ \noalign{\medskip}
   0  & \qquad\mathrm{for}\quad \upsilon < 0 \,.
 \end{array}\right.
\end{equation}
The previous expression is a monotonically increasing function for $c^B_i>0,\,\upsilon>0$, has the correct asymptotic behavior as $\upsilon\to\infty$ and it reduces to the conventional harmonic mean (\ref{eq:VL_lim}) when $c^F_i=c^B_i\to 2$.
Moreover, Eq. (\ref{eq:VL_lim_modified}) is continuously differentiable for $\upsilon \ge 0$, lies inside the allowed TVD region and preserve the correct symmetry when $\upsilon\to 1/\upsilon$ and $(c^F_i,\,c^B_i)\to (c^B_i,\, c^F_i)$.

Finally, the arithmetic average between the two slopes can be considered yielding
\begin{equation}\label{eq:MC_lim}
  \varphi^{MC}(\upsilon) = \max\left[0,\min\left(
   \frac{1+\upsilon}{2},\, c^F_i, \, c^B_i\upsilon\right)\right] \,.
\end{equation}
In the limit of vanishing curvature Eq. (\ref{eq:MC_lim}) reproduces the well known monotonized central (MC) limiter \cite{vanLeer.1977_IV}.
Beware that the arithmetic average of the two slopes does not give, in curvilinear geometry, a second-order accurate approximation to the derivative neither at the cell center nor at the barycenter.
However, it still proves to have smaller numerical dissipation.

Unless otherwise stated, the linear reconstruction scheme given by Eq. (\ref{eq:linear_lim}) with the MC limiter given by Eq. (\ref{eq:MC_lim}) will be referred to as the piecewise linear method (PLM) while the uncorrected version (i.e. Cartesian-like)  will be denoted with PLM$_0$.

\subsection{Third order WENO reconstruction (WENO$_3$)}
\label{sec:WENO3}
%
%
%

The third-order WENO reconstruction employs the information available on a three-point stencil $(i-1,i,i+1)$ to reconstruct left and right interface values through a convex linear combination of second-order accurate values:
\begin{equation}
  Q^{\pm}_{i} =   \omega^{\pm}_{i,0} Q^{\pm,[\rm f]}_{i} 
                + \omega^{\pm}_{i,1} Q^{\pm,[\rm b]}_{i}  \,.
\end{equation}
where $Q^{\pm,[\rm f]}_{i}$ and $Q^{\pm,[\rm b]}_{i}$ are, respectively, the linearly reconstructed values based on the forward and backward derivatives given by Eqs. (\ref{eq:linear_FWD_and_BCK}) and (\ref{eq:fwd_and_bck_slopes}) while $\omega^{\pm}_{i,0}$ and $\omega^{\pm}_{i,1}$ are nonlinear weights assigned to the two stencils $\{\xi_{i-1},\,\xi_{i}\}$ and $\{\xi_{i},\,\xi_{i+1}\}$.
The nonlinear weights should adapt to the relative smoothness of the solution on each candidate stencil \cite{JiaShu.1996} and are functions of the cell averages involved.
A possible choice are the standard weight functions defined in the original formulation by \cite{JiaShu.1996} although the corresponding WENO reconstruction may degenerate to second-order near local extrema.
A better and less dissipative approach is to employ the new weight functions of \cite{YamCar.2009} (see also \cite{MigTzeBod.2010}) conveniently adapted to the case of a curvilinear mesh system by defining 
\begin{equation}\label{eq:WENO3_nonlinear_weights}
 \omega^{\pm}_{i,0} = \frac{\alpha^{\pm}_{i,0}}
                           {\alpha^{\pm}_{i,0} + \alpha^{\pm}_{i,1}}
  \,;\qquad
 \omega^{\pm}_{i,1} = \frac{\alpha^{\pm}_{i,1}}
                           {\alpha^{\pm}_{i,0} + \alpha^{\pm}_{i,1}}
  \,;\qquad
 \alpha^{\pm}_{i,k} = d^{\pm}_{i,k} \left(1 +
   \frac{\left|\Delta Q^F_i - \Delta Q^B_i\right|^2}
        {\beta_{i,k} + \left|Q^{\rm ref}_i\right|^2}\right)
\end{equation}
for $k=0,1$.
In the expression above $d^{\pm}_{i,k}$ are the linear weights, $\beta_{i,0} = \left(\Delta Q^F_i\right)^2$, $\beta_{i,1} = \left(\Delta Q^B_i\right)^2$ are the smoothness indicators while $Q^{\rm ref}_i$ is a reference value ensuring that, for smooth solutions, one has $\omega^\pm_{i,k} - d^\pm_{i,k} = O(\Delta\xi_i^2)$, see \cite{YamCar.2009}.
Numerical experiments presented in this paper have suggested that $Q^{\rm ref}_i$ can be chosen as
\begin{equation}
 Q^{\rm ref}_i = \frac{C^{\rm ref}}{N} \max
                 \left(|\av{Q}_{i-1}|,\,|\av{Q}_i|,\, |\av{Q}_{i+1}|\right)\,,
\end{equation}
where $C^{\rm ref}$ is positive constant and $N$ is the number of grid zones in the $\xi$ direction.
In the numerical tests presented here, $C^{\rm ref}=20$ is used although strongly nonlinear problems involving discontinuous waves may benefit from using a lower value.

It should be noted that, in Cartesian geometry, the linear coefficients are grid-independent and equal to  $d^+_{i,0} = 2/3$, $d^+_{i,1} = 1/3$ while the previous expressions become identical to those of \cite{YamCar.2009}.
Furthermore, there is no need to define $d^-_{i,k}$ in the Cartesian case since the reconstruction of $Q_{i}^{-}$ is mirror symmetric with respect to the cell central point (i.e., $d^-_{i,0} = d^+_{i,1}$, $d^-_{i,1} = d^+_{i,0}$).
These properties are lost in a curvilinear coordinate system whenever $\partial^2\vol/\partial\xi^2\ne 0$ because adjacent zones have now different volumes and there is no translational invariance.
Therefore two sets of coefficients must be defined in each cell to retrieve the leftmost and rightmost interface values. 
However, one can still take advantage of the normalization condition $d^{\pm}_{i,0} + d^{\pm}_{i,1} = 1$.

For a third-order reconstruction the linear weights can be found by matching the coefficients of $\av{Q}_{i\pm1}$ in Eq. (\ref{eq:pm_states}, using $i_L=i_R=1$) with those obtained from Eq. (\ref{eq:linear_FWD_and_BCK}) using Eq. (\ref{eq:fwd_and_bck_slopes}):
\begin{equation}\label{eq:WENO3_linear_weights}
 d^{+}_{i,0} = w^+_{i,1}\frac{\bar{\xi}_{i+1}-\bar{\xi}_i}{\xi_{i+\HALF}-\bar{\xi}_i}
 \,;\qquad
 d^{-}_{i,0} = w^-_{i,1}\frac{\bar{\xi}_{i+1}-\bar{\xi}_i}{\xi_{i-\HALF}-\bar{\xi}_i}
%
\end{equation}
while the remaining weights are simply given by $d^{\pm}_{i,1} = 1 - d^{\pm}_{i,0}$.
Equations (\ref{eq:WENO3_nonlinear_weights}) and (\ref{eq:WENO3_linear_weights}) are also valid on non-equidistant meshes as long as $w^{\pm}_{i,1}$ are consistently computed from Eq. (\ref{eq:linear_system_w}).
For a uniform mesh spacing, instead, $w^{\pm}_{i,1}$ are respectively given by Eqs. (\ref{eq:iL1_iR1_cyl}) and (\ref{eq:iL1_iR1_sph}) in cylindrical and spherical coordinates while the volume centroids $\bar{\xi}$ are computed using Eq. (\ref{eq:centroid_of_volume}).
Finally, note that in the limit of vanishing curvature, $i\to\infty$, the linear weights defined above tend to the Cartesian limits $d^{+}_{0} = 2/3$, $d^{-}_{0}=1/3$ of the classical third-order WENO scheme.

\subsection{Piecewise parabolic method (PPM)}
\label{sec:PPM}
%
%
%

In the Piecewise Parabolic Method (PPM, \cite{ColWoo.1984}) a parabolic interpolant is uniquely determined by the cell average $\av{Q}_i$ and by the left and right extrapolated edge values:
\begin{equation}\label{eq:PPM_parabola}
 Q_i(\xi) = Q^{-}_{i} + y\Big[Q^{+}_{i} - Q^{-}_{i} + (1 - y)Q_{6,i}\Big]\,,
\end{equation}
where $y = (\xi - \xi_{i-\HALF})/\Delta\xi_i$ and $Q_{6,i}$ is a parabolic coefficient giving a measure of the second-derivative.
The left and right interface values $Q^{-}_{i}$ and $Q^{+}_{i}$ should be initially computed using a third-order (or higher) accurate approximation and then further modified to satisfy monotonicity constraints.

In the original formulation \cite{ColWoo.1984}, the authors suggested that left and right interface states could be obtained in curvilinear coordinates by interpolating the variables in the volume coordinate (rather than the radial coordinate) using the same formalism adopted on a Cartesian mesh.
However, as noted by \cite{BloLuf.1993}, this approach suffers from two major disadvantages.
First of all, the resulting interface values are only first-order accurate and the error becomes increasingly large near the coordinate origin.
In spherical coordinates, for instance, a parabolic profile would be approximated by $q(r) \approx a_0 + a_1r^3 + a_2r^6$ rather than $q(r) \approx a_0 + a_1r + a_2 r^2$.
Secondly, reconstruction in the volume coordinate has to be performed on a non-equidistant mesh even if the original radial grid is uniformly spaced.

To overcome these shortcomings, the authors in \cite{BloLuf.1993} suggested to interpolate the primitive function
\begin{equation}
  F_Q(\xi) = \int^\xi Q(\zeta) \zeta^m d\zeta\,,
\end{equation}
so that the point values of $Q$ could be obtained  by straightforward differentiation, $Q(\xi) = dF_Q(\xi)/(\xi^md\xi)$.
In cylindrical geometry ($m=1$), for example, this approach is formally equivalent to obtaining the interface values of $RQ(R)$ rather than $Q(R)$ (see also \cite{SkiOst.2010}) resulting in the loss of one order of accuracy. 
In spherical geometry ($m=2$), the situation worsens as the resulting profile  would be $Q(r) = a_{-2}/r^2 + a_{-1}/r + a_0 + a_1 r$ thus decreasing by 2 the order of accuracy.
Furthermore this procedure cannot be used, in this form, at the origin since the resulting states would clearly be singular.

In the present approach, the (unlimited) interface values are directly computed using Eq. (\ref{eq:pm_states}) to the desired order of accuracy to produce regular and well-behaved interface values also when $\xi = 0$.
The unlimited left and right interface values are then corrected to ensure that the resulting parabolic profile is bounded between neighboring cell averages and monotone.
This is achieved through the following steps.
\begin{enumerate}
 \item Interface values must be constrained to lie between adjacent cell averages.
       This is obtained by resetting $Q^\pm_i$ to the maximum or the minimum of the two averages if the original estimate falls outside this range:
 \begin{equation}
   Q^\pm_i \;\to \;
           \min\left[Q^\pm_i,\,\max\left(\av{Q}_i,\, \av{Q}_{i\pm1}\right)\right]
   \,;\qquad
   Q^\pm_i \; \to \;
           \max\left[Q^\pm_i,\,\min\left(\av{Q}_i,\, \av{Q}_{i\pm1}\right)\right]
 \end{equation}
 Note that we do not use the conventional van-Leer limiting as in the original PPM formulation \cite{ColWoo.1984}.
 \item Monotonicity is then enforced by requiring that no extrema in the distribution given by Eq. (\ref{eq:PPM_parabola}) appear for $0\le y\le 1$.
 This involves two modifications.
 First, the distribution is flattened whenever $\av{Q}_i$ is a local maximum or minimum.
 Second, when an extremum takes place close to $y=0$ or $y=1$, the interface value on the opposite edge of the zone is modified so that $Q(y)$ has zero derivative at $y=1$ or $y=0$, respectively.
 From Eq. (\ref{eq:PPM_parabola}) this condition is verified when $|Q^+_i-Q^-_i|\le|Q_{6,i}|$.

 These criteria lead to the following redefinition of the parabolic limiter \cite{ColWoo.1984} in curvilinear geometry as
\begin{equation}\label{eq:PPM_parabolic_limiter}
  Q^\pm_i \to \av{Q}_i + \left\{\begin{array}{cll}
   0 & \quad{\rm if}\quad & \DS \delta Q^{+}_i \delta Q^{-}_i \ge 0  \,,
  \\ \noalign{\bigskip}
\DS  -\frac{h^\mp_i + 1}{h^\pm_i -1}\delta Q^{\mp}_i
            & \quad{\rm if}\quad & 
\DS       \big|\delta Q^{\pm}_i\big| \ge
          \frac{h^\mp_i + 1}{h^\pm_i -1}\big|\delta Q^{\mp}_i\big| 
          \quad\mathrm{and}\quad \delta Q^{+}_i \delta Q^{-}_i < 0 
  \\ \noalign{\bigskip}
  \delta Q^{\pm}_i & \quad{\rm otherwise}
\end{array}\right.
\end{equation}
 where $\delta Q^\pm_i = Q^\pm_i-\av{Q}_i$ and $h^\pm_i$ are geometrical factors derived below.
 Note that the first of the two conditions impairs the order of accuracy to first order at smooth extrema (this deficiency has been recently overcome by \cite{ColSek.2008, Corquodale_Colella.2011}).
 The test condition for the second case follows from the definition of the $Q_{6,i}$ coefficient which can be found upon integrating Eq. (\ref{eq:PPM_parabola}) over the cell volume.
 The result can be cast in the following general expression:
 \begin{equation}
   Q_{6,i} = - \left(h^+_i\delta Q^+_i + h^-_i\delta Q^-_i\right)\,,
\quad\mathrm{with}\quad
   h^{\pm}_i = \mp\Delta\xi_i\frac{\DS
      \int_{\xi_{i-\HALF}}^{\xi_{i+\HALF}}
      \left(\xi-\xi_{i\mp\HALF}\right) \,d\vol}
     {\DS\int_{\xi_{i-\HALF}}^{\xi_{i+\HALF}} 
            \left(\xi-\xi_{i-\HALF}\right)\left(\xi-\xi_{i+\HALF}\right) \,d\vol}\,.
 \end{equation}
%
where $d\vol = (\partial\vol/\partial\xi)\,d\xi$.
Specializing to Cartesian, cylindrical and spherical coordinates one obtains
\begin{equation}
 h^\pm_{i} = \left\{\begin{array}{ll}
\DS   3
  \qquad & \mathrm{Cartesian},\,\xi=x\,  \\ \noalign{\bigskip}
\DS   3 \pm\frac{\Delta R_i}{2R_i}
  \qquad & \mathrm{cylindrical},\,\xi=R\,  \\ \noalign{\bigskip}
\DS 3 + \frac{2\Delta r_i(\pm 10r_i + \Delta r_i)}{20r_i^2+\Delta r_i^2}
  \qquad & \mathrm{spherical},\,\xi=r\,   \\ \noalign{\bigskip}
\DS 
  \pm\frac{\Delta\theta_i\left(
           \Delta\tilde{\mu}_i + \Delta\theta_i\cos\theta_{i\pm\HALF}\right)}
          {\Delta\theta_i(\sin\theta_{i-\HALF}+\sin\theta_{i+\HALF})- 2\Delta\mu_i}
  \qquad & \mathrm{spherical}\,,\xi=\theta\,,   \\ \noalign{\bigskip}
\end{array}\right.  
\end{equation}
where $\Delta\mu_i = \cos\theta_{i-\HALF} - \cos\theta_{i+\HALF}$, 
$\Delta\tilde{\mu}_i = \sin\theta_{i-\HALF} - \sin\theta_{i+\HALF}$.
\end{enumerate}

Although traditionally a $4^{\rm th}$-order accurate approximation is employed to construct the right-extrapolated edge values, three different schemes are considered here based on third-, fourth- and fifth-order approximations to the interface values and labeled with PPM$_3$, PPM$_4$ and PPM$_5$, respectively.
For the sake of comparison, the fourth-order scheme which does not employ any geometrical correction (i.e. Cartesian-like) will be denoted with PPM$_0$.

\section{Extension to nonlinear systems}
\label{sec:nonlinear_systems}
%
%
%

The formalism introduced the previous sections can be extended to nonlinear systems of conservation laws in a component-wise manner by considering more than one equation of the form (\ref{eq:prototype}).
In this case, $\vec{U}$ defines an array of conservative variables while $Q$ should be regarded as one component of this set, i.e.,  $Q\in\vec{U}$.

In what follows, two crucial aspects concerning the employment of high order methods in curvilinear geometry are discussed.
First, in Section \ref{sec:rec_prim} it is shown how higher than second-order reconstruction schemes can be employed on a different set of variables (i.e. non-conservative).
This is relevant when the employment of, say, primitive variables (density, velocity and pressure) is preferred over the conservative ones (density momentum and energy).
Second, in \S\ref{sec:source_terms} the numerical integration of geometrical source terms arising when using differential operators in curvilinear coordinates is discussed.
In particular, it is shown how high (second- and third-) order quadrature rules can be derived in cylindrical and spherical coordinates.

\subsection{Reconstruction from volume averages of primitive variables}
\label{sec:rec_prim}
%
%
%

The reconstruction of the interface states from volume averages may not be necessarily carried on the conserved variables but on a different set, say $\vec{V}=V(\vec{U})$ where $V$ is a nonlinear variable transformation.
In the case of gas dynamics, for instance, conserved fluid variables include density $\rho$, momentum $\rho\vec{v}$ and total energy $E$ whereas primitive variables are customary chosen as $\vec{V}=(\rho, \vec{v}, p)$ where $\vec{v}$ and $p$ denote velocity and thermal pressure, respectively.

The reason behind introducing such a transformation is based on a consolidated experience suggesting that interpolation of primitive variables rather than conservative ones leads to less oscillatory and better-behaved results.
At the second order level this poses no difficulty since it is a well known result that (see Eq. (\ref{eq:volume_point_err})
\begin{equation}
  \av{\vec{V}}_i = V(\av{\vec{U}}_i) + O(\Delta x^2) \,,
\end{equation}
and therefore one may interchange the volume averages of the conserved and primitive variables indifferently.
However, wih order higher than 2, this operation becomes inaccurate and one has to be more careful when computing $\av{\vec{V}}_i$.
Here we follow the approach of \cite{Corquodale_Colella.2011} and extend it to the case of curvilinear coordinates.
The method can be summarized through the following steps.
\begin{enumerate}
 \item Start from volume averages of conservative variables, $\av{\vec{U}}_i$.
 \item Using Taylor expansion, form point values of conserved variables using a three-point stencil
 \begin{equation}\label{eq:Upoint_values}
   \vec{U}_i = w_{i,-1}\av{\vec{U}}_{i-1} + 
               w_{i,0}\av{\vec{U}}_{i}    + 
               w_{i,1}\av{\vec{U}}_{i+1}  + \epsilon\,,
 \end{equation}
 where $w_{i,-1}$, $w_{i,0}$ and $w_{i,1}$ are geometry-dependent coefficients
 and $\epsilon$ is the leading error term.

 In Cartesian coordinates, for instance, one has the simple expressions
 \begin{equation}\label{eq:cartesian_weights}
  w_{i,\pm1} = -\frac{1}{24}\,,\qquad
  w_{i,0}    = \frac{13}{12}\,,\qquad
  \epsilon   = \frac{3}{640}\left(\pd{^4 \vec{U}}{\xi^4}\right)_i\Delta\xi^4\,,
 \end{equation}
 which, upon re-arranging terms, give the fourth-order approximation to $\vec{U}_i$:
 \begin{equation}
   \vec{U}_i = \av{\vec{U}}_i - \frac{\Delta^2\av{\vec{U}}_i}{24} + O(\Delta x^4)\,,
 \end{equation}
 where $\Delta^2\av{\vec{U}}_i = \av{\vec{U}}_{i+1} - 2\av{\vec{U}}_i + \av{\vec{U}}_{i-1}$ is a second-order accurate approximation to the undivided second derivative.
 
 In cylindrical coordinates ($\xi=R$) one finds
 \begin{equation}
   w_{i,\pm1} = -\frac{J\pm1}{24J}\,,\qquad
   w_{i,0}    =  \frac{13}{12}\,,\qquad
   \epsilon   =  \frac{3}{160J}\left(\pd{^3\vec{U}}{R^3}\right)_i\Delta R^3 
                +\frac{3}{640} \left(\pd{^4\vec{U}}{R^4}\right)_i\Delta R^4\,,
 \end{equation}
 where $J \equiv i-1/2=1/2, 3/2, 5/2, ...$ is a half-integer number labeling the zone.
 From the previous expressions one can immediately see that, close to the 
 origin, the approximation is third-order accurate while, in the limit of vanishing curvature ($J\to\infty$), one recovers the Cartesian limit given by Eq. (\ref{eq:cartesian_weights}).

 Finally, in spherical coordinates ($\xi=r$) one obtains, after some algebra,
  \begin{equation} \label{eq:spherical_weights}
  \begin{array}{lcl}
  w_{i,\pm 1} &=&\DS 
     -\frac{(12J^2 \pm 24J + 13)(80J^4 - 288J^2 \pm 216J + 15)}{72\Delta_s}
    \\ \noalign{\medskip}
  w_{i,0} &=&\DS \frac{(12J^2 + 1)(1040J^4 - 2448J^2 + 1815)}{36\Delta_s}\,,
  \end{array}
  \end{equation}
  where $\Delta_s = 320J^6 - 720J^4 + 492J^2 + 45$.
  As before, the leading error term can be written as the sum of two contributions,
  \begin{equation}\label{eq:spherical_error}
    \epsilon = 
    \frac{J}{4}\frac{48J^4 - 120J^2 + 91}{\Delta_s}
    \left(\pd{^3\vec{U}}{r^3}\right)_i \Delta r^3 
 +  \frac{1344J^6 - 2640J^4 + 1516J^2 + 405}{896\Delta_s}
    \left(\pd{^4\vec{U}}{r^4}\right)_i\Delta r^4
  \end{equation}
  showing that, when $r\sim \Delta r$, the approximation is again third-order accurate. 
  On the contrary, in the limit of vanishing curvature, the weights given by Eq. (\ref{eq:spherical_weights}) and the error in Eq. (\ref{eq:spherical_error}) reproduce the corresponding Cartesian limits given by Eq. (\ref{eq:cartesian_weights}) and the solution becomes fourth-order accurate.

  \item Convert point-values of the conserved variables into primitive using the nonlinear change of variables:
  \begin{equation}\label{eq:Vpoint_values}
    \vec{V}_i = V\left(\vec{U}_i\right)
  \end{equation}
   and approximate $\vec{V}(\xi)$ with a parabolic profile inside the $i$-th zone:
 \begin{equation}
   \vec{V}(\xi) = \vec{V}_i 
   + \left[\frac{\vec{V}_{i+1}-\vec{V}_{i-1}}{2\Delta \xi}\right] (\xi-\xi_i)
   + \left[\frac{\vec{V}_{i+1}-2\vec{V}_i + \vec{V}_{i-1}}{\Delta \xi^2}\right]
 \frac{(\xi-\xi_i)^2}{2}
 \end{equation}
 where the terms in square brackets are second-order approximations to the first and second derivative of $\vec{V}$, respectively.

 \item  Form volume averages of primitive variables by straighforward integration.
        The final results reads
 \begin{equation}\label{eq:Vaverage_values}
   \av{\vec{V}}_i = \left\{\begin{array}{ll}
 \DS \frac{1}{24}\vec{V}_{i-1} + \frac{11}{12}\vec{V}_i + \frac{1}{24}\vec{V}_{i+1} 
   & \qquad \mathrm{(Cartesian)}
  \\ \noalign{\bigskip}
 \DS \frac{J-1}{24J}\vec{V}_{i-1} 
   + \frac{11}{12}\vec{V}_i + \frac{J+1}{24J}\vec{V}_{i+1}
   & \qquad \mathrm{(cylindrical)}
  \\ \noalign{\bigskip}
 \DS   \frac{20J^2 - 40J + 3}{40(12J^2 + 1)}\vec{V}_{i-1} 
     + \frac{220J^2 + 17}{20(12J^2 + 1)}\vec{V}_{i} 
     + \frac{20J^2 + 40J + 3}{40(12J^2 + 1)}\vec{V}_{i+1} 
   & \qquad \mathrm{(spherical)}
  \end{array}\right.
 \end{equation}
\end{enumerate}
The previous expressions give third-order accurate approximations to the volume average of the primitive variables close to the origin while, for large $J$, the error becomes $O(\Delta\xi^4)$.

The conversion from volume averages to point values is carried out without any particular form of limiting and it has been succesfully employed in those (1D) problems containing only smooth profiles. 
Still, spurious oscillations may arise in presence of discontinuities since the conversion operation consists, essentially, of subtracting the (unlimited) second derivative from the solution.
Future extension of this work should consider a more careful treatment in order to avoid the onset of unphysical values.
\subsection{Source term integration}
\label{sec:source_terms}
%
%
%

The choice of a curvilinear coordinate system is based on a local vector basis that has no fixed orientation in space but changes from point to point.
In the case of the Euler equations, for instance, this leads to the appearance of additional source terms in the momentum equation arising upon taking the divergence of the momentum flux tensor.
A typical example consists of the pressure and centrifugal terms in the radial momentum equation which, in cylindrical coordinates, reads
\begin{equation}\label{eq:S:radial_mom_cyl}
  S = \frac{p + \rho v_\phi^2}{R}\,,
\end{equation}
where $p$ is the gas pressure, $\rho$ is the density and $v_\phi$ is the azimuthal velocity.
Other terms such as body forces (e.g. gravity) or viscous drag may also be present.

In general, source terms in the FV formalism should be treated as averages over the cell volume as in Eq. (\ref{eq:prototype_integrated}):
\begin{equation}\label{eq:S:integral1}
  \av{S}_i = \frac{1}{\Delta\vol_i}\int_{i-\HALF}^{i+\HALF} S\,\xi^m d\xi\,,
\end{equation}
where $m=0,1,2$ for Cartesian, cylindrical and spherical geometry, respectively.
However, in some cases, the integral may be reduced to a somewhat simpler form by taking advantage of the explicit spatial dependence.
This is the case for geometrical source terms containing $1/\xi$ factors (such as Eq. \ref{eq:S:radial_mom_cyl}) for which Eq. (\ref{eq:S:integral1}) can be written as
\begin{equation}\label{eq:S:integral2}
  \av{S}_i = \frac{1}{\Delta\vol_i}\int_{i-\HALF}^{i+\HALF} \hat{S} \xi^{m-1} d\xi\,,
\end{equation}
where $\hat{S}$  is regular near the origin since does not explicitly depend on $\xi$ and $m\ge 1$.

Eq. (\ref{eq:S:integral1}) or (\ref{eq:S:integral2}) may be computed using different quadrature rules.
To second-order accuracy, for example, one can replace the volume average in Eq. (\ref{eq:S:integral1}) with the integrand evaluated at the corresponding centroid of volume (by the same argument used in Sec. \ref{sec:linear}) yielding
\begin{equation}\label{eq:S:midpoint1}
 \av{S}_i \approx S_{\xi=\bar{\xi}} + {\cal O}(\Delta\xi^2) \,.
\end{equation}

Alternatively, Eq. (\ref{eq:S:integral1}) or (\ref{eq:S:integral2}) may be approximated using a trapezoidal rule based on a linear fit through the leftmost and rightmost interface values.
This yields, in the case of Eq. (\ref{eq:S:integral1}), 
\begin{equation}\label{eq:S:trapezoidal1}
 \av{S}_i \approx  \frac{\xi_{i+\HALF} - \bar{\xi}_i}{\Delta\xi_i}S_{i-\HALF} 
                 + \frac{\bar{\xi}_i - \xi_{i-\HALF}}{\Delta\xi_i}S_{i+\HALF}
                 + {\cal O}(\Delta\xi^2) \,,
\end{equation}
where one may use, for example, $S_{i-\HALF}\equiv S(\vec{V}^{-}_{i})$ and $S_{i+\HALF}\equiv S(\vec{V}^+_{i})$.
The previous expression extends the trapezoidal rule to cylindrical and spherical geometries and is exact for second-order polynomials. 
Note also that, in the limit of vanishing curvature, the weights in Eq. (\ref{eq:S:trapezoidal1}) become equal to $1/2$ (Cartesian case).
Likewise, for Eq. (\ref{eq:S:integral2}), one obtains
\begin{equation}\label{eq:S:trapezoidal2}
 \av{S}_i = \left\{\begin{array}{ll}
\DS    \frac{\hat{S}_{i-\HALF} + \hat{S}_{i+\HALF}}{2R_{i}} 
    & \qquad \mathrm{for}\quad m = 1  \\ \noalign{\bigskip}
\DS    \left(\frac{6r_i - \Delta r_i}{12r_i^2 + \Delta r_i^2}\hat{S}_{i-\HALF} 
           + \frac{6r_i + \Delta r_i}{12r_i^2 + \Delta r_i^2}\hat{S}_{i-\HALF} \right) 
    & \qquad \mathrm{for}\quad m = 2  \\ \noalign{\medskip}
   \end{array}\right\}
      + {\cal O}(\Delta\xi^2)
\end{equation}
Eqs. (\ref{eq:S:midpoint1}), (\ref{eq:S:trapezoidal1}) and (\ref{eq:S:trapezoidal2}) are second-order accurate.

A third-order accurate expression based on Simpson quadrature rule is
\begin{equation}\label{eq:S:simpson12}
 \av{S} = \gamma^{-}_{i} S_{i-\HALF} + \gamma_i S_{i} + \gamma^{+}_{i}S_{i+\HALF}
         + O(\Delta\xi^4) 
  \qquad\mathrm{or}\qquad
 \av{S} =   \hat{\gamma}^{-}_{i}\hat{S}_{i-\HALF} 
          + \hat{\gamma}_i \hat{S}_{i} 
          + \hat{\gamma}^{+}_{i}\hat{S}_{i+\HALF}
         + O(\Delta\xi^4)\,,
\end{equation}
requiring three function evaluations.
The weight coefficients $\gamma$ and $\hat{\gamma}$ for the first or second expression can be found from Eq. (\ref{eq:S:integral1}) or (\ref{eq:S:integral2}), respectively, by fitting a parabola through the interface values and the central point.
The results for $m=1$ and $m=2$ for a source term of the type (\ref{eq:S:integral1}) are found to be
\begin{equation}\label{eq:S:simpson_weights1}
  \Big(\gamma^{-}_{i},\, \gamma_{i},\, \gamma^{+}_{i}\Big) = 
  \left\{\begin{array}{ll}
\DS \left(\frac{1}{6} - \frac{\Delta R}{12R_i},\,
          \frac{2}{3},\,
          \frac{1}{6} + \frac{\Delta R}{12R_i}\right)
  & \qquad\mathrm{for}\quad m = 1\,,
  \\ \noalign{\bigskip}
\DS \left(\frac{20r_i(r_i - \Delta r_i) + 3\Delta r_i^2}{10(12r_i^2 + \Delta r_i^2)}
      ,\, \frac{2}{5}\frac{20r_i^2 + \Delta r_i^2}{12r_i^2 + \Delta r_i^2}
      ,\, \frac{20r_i(r_i + \Delta r_i) + 3\Delta r_i^2}{10(12r_i^2 + \Delta r_i^2)}
     \right) 
   & \qquad\mathrm{for}\quad m = 2\,,
  \end{array}\right.
\end{equation}
while, for Eq. (\ref{eq:S:integral2}), they become:
\begin{equation}\label{eq:S:simpson_weights2}
  \Big(\hat{\gamma}^{-}_{i},\, 
       \hat{\gamma}_{i},\, 
       \hat{\gamma}^{+}_{i}\Big) = 
  \left\{\begin{array}{ll}
\DS \left(\frac{1}{6R_i},\, \frac{2}{3R_i},\, \frac{1}{6R_i}\right)
  & \qquad\mathrm{for}\quad m = 1\,,
  \\ \noalign{\bigskip}
\DS \left(\frac{2r_i - \Delta r_i}{12r_i^2 + \Delta r_i^2}
      ,\, \frac{8r_i}{12r_i^2 + \Delta r_i^2}
      ,\, \frac{2r_i + \Delta r_i}{12r_i^2 + \Delta r_i^2}
     \right) 
   & \qquad\mathrm{for}\quad m = 2\,.
  \end{array}\right.
\end{equation}

Note that Eq. (\ref{eq:S:simpson12}) requires the knowledge of the solution at the cell center which, for a third-order accurate polynomial, can be found from Eq. (\ref{eq:PPM_parabola}) using $y=1/2$.

As a final remark, we point out that the trapezoidal or Simpson quadrature rules approximating Eq. (\ref{eq:S:integral2}) have the advantage of being well-behaved near the coordinate origin making them more suited for source terms containing diverging factors like $1/\xi$.
Conversely, Eqns. (\ref{eq:S:trapezoidal1}) and the first of (\ref{eq:S:simpson12}) may become singular.

%
%
\section{Numerical benchmarks}
\label{sec:numerical_benchmarks}
%
%
%


In this section the accuracy of the proposed reconstruction schemes is measured using selected numerical benchmarks in one and two dimensions.
For the sake of comparison, the governing conservation laws are evolved in time using the explicit third-order TVD Runge-Kutta time stepping \citep{Shu.1988, GotShu.1998}
\begin{equation}\label{eq:rk3}
 \begin{array}{lcl}
 \vec{U}^*     &=& \vec{U}^n + \Delta t^n{\cal L}\left(\vec{U}^n\right) \,,\\ \noalign{\medskip}
 \vec{U}^{**}  &=& \DS \frac{3}{4}\vec{U}^n + \frac{1}{4}\vec{U}^*
                     + \frac{\Delta t^n}{4}{\cal L}\left(\vec{U}^*\right)\,, \\ \noalign{\medskip}
 \vec{U}^{n+1} &=& \DS \frac{1}{3}\vec{U}^n + \frac{2}{3}\vec{U}^{**}
                     + \frac{2}{3}\Delta t^n{\cal L}\left(\vec{U}^{**}\right)\,.
\end{array}
\end{equation}
where $\vec{U}$ is an array of conservative quantities and ${\cal L}$ is a discrete approximation to the right hand side of the conservation laws.
In the scalar case, for instance, ${\cal L}$ corresponds to the right hand side of Eq. (\ref{eq:cons_law_cylindrical}) or Eq. (\ref{eq:cons_law_spherical}) for cylindrical or spherical coordinates, respectively.

The time step $\Delta t^n$ is computed from the Courant-Friedrichs-Lewy (CFL) condition:
\begin{equation}\label{eq:dt}
  \Delta t^n = C_a\left[\max_\vec{i}\left(\frac{1}{D}
   \sum_d\frac{\lambda_{d,\vec{i}}}{\Delta l_{d,\vec{i}}}\right)\right]^{-1}
\end{equation}
where $C_a$ is the Courant number, $D$ is the number of spatial dimensions while $\lambda_{d}$ and $\Delta l_d$ are, respectively, the maximum signal speed and the zone spatial length in the direction $\hvec{e}_d$ .

Numerical benchmarks for a scalar conservation law in one and two dimensions are first presented in Section \ref{sec:scalar_tests} while verification tests for nonlinear systems are discussed in Section \ref{sec:nonlinear_tests}.
Unless otherwise stated, errors for a generic flow quantity $Q$ are computed using the $L_1$ discrete norm defined by
\begin{equation}\label{eq:err_L1}
 \epsilon_1(Q) = \frac{\DS \sum_\vec{i} 
 \left|\av{Q}_\vec{i} - \av{Q}^{\rm ref}_\vec{i}\right|\Delta\vol_\vec{i} \,,}
 {\DS \sum_\vec{i} \Delta\vol_\vec{i}}
\end{equation}
where the summation extends to all grid zones, $\av{Q}^{\rm ref}_\vec{i}$ is the volume average of the reference (or exact) solution and $\Delta\vol_\vec{i}$ is the zone volume.

\subsection{Scalar advection tests}
\label{sec:scalar_tests}
%
%

\subsubsection{Advection equation in cylindrical and spherical radial coordinates}
\label{sec:radial_advection}
%
%

As a first benchmark, the one-dimensional advection equation in cylindrical and spherical geometries is considered:
\begin{equation}\label{eq:radial_advection}
  \pd{Q}{t} + \frac{1}{\xi^m}\pd{}{\xi}\left(\xi^m Q v\right) = 0 \,,
\end{equation}
where $v = \alpha\xi$ is a linear velocity profile, $\alpha = 1$ is a constant and $m=0,1,2$ for Cartesian, cylindrical or spherical geometry (respectively).
Eq. (\ref{eq:radial_advection}) admits the exact solution 
\begin{equation}
  Q^{\rm ref}(\xi,t) = e^{-(m+1)\alpha t} Q\left(\xi e^{-\alpha t}, 0\right) \,,
\end{equation}
where $Q(\xi,0)$ is the initial condition.
For the present test, a Gaussian profile is employed:
\begin{equation}\label{eq:rad_adv_initial_profile}
  Q(\xi,0) = e^{-a^2(\xi - b)^2} \,,
\end{equation}
where $a$ and $b$ are constants. 
The computational grid spans the interval $\xi\in[0,2]$ with $N_\xi$ regularly spaced zones and initial condition given by
Eq. (\ref{eq:rad_adv_initial_profile}) is integrated over the corresponding zone-volume $\Delta\vol_i$ using a five-point Gaussian quadrature rule to correctly initialize $\av{Q}_i$ at $t=0$.
Boundary conditions are symmetric at the origin ($\xi=0$) while zero-gradient is imposed at $\xi=2$.
Two different sets of parameters are considered, namely, $\{a=10, b=0\}$ (set A) corresponding to a monotonically decreasing profile and $\{a=16, b=1/2\}$ (case B) yielding a maximum at $\xi = 1/2$.
Computations are carried out until $t=1$ using a CFL number of $0.9$ while the interface flux (Eq. \ref{eq:interface_flux}) is computed using upwinding:
\begin{equation}\label{eq:upwind_Riemann_solver}
  \tilde{F}_{i+\HALF} = \frac{1}{2}\left[
    v_{i+\HALF}\left(Q^{-}_{i+1} + Q^+_i\right)
  - \left|v_{i+\HALF}\right|\left(Q^{-}_{i+1} - Q^+_{i}\right)\right]
\end{equation}

\begin{figure}[!h]
 \centering
 \includegraphics[width=0.4\textwidth]
                 {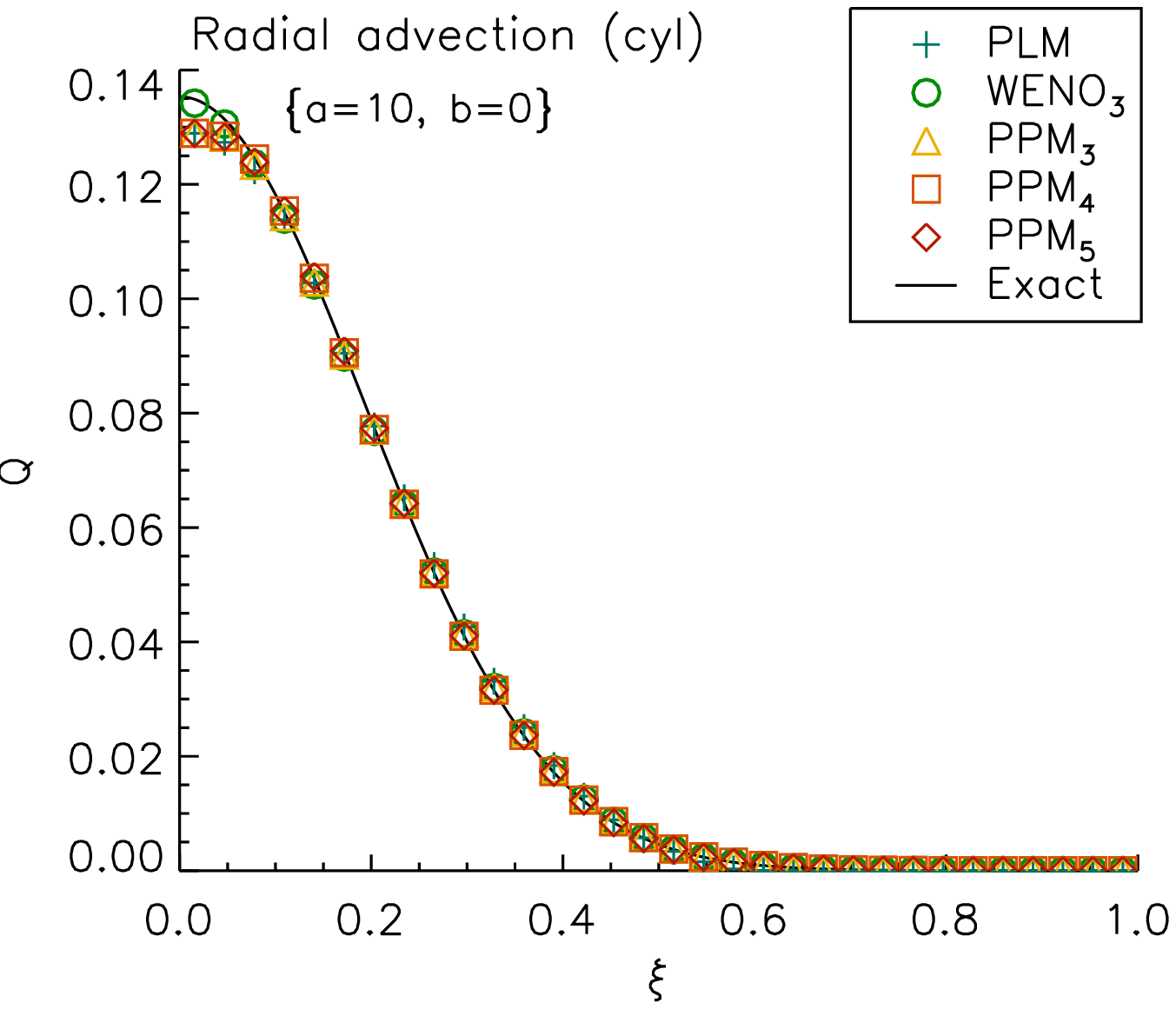}%
 \includegraphics[width=0.4\textwidth]
                 {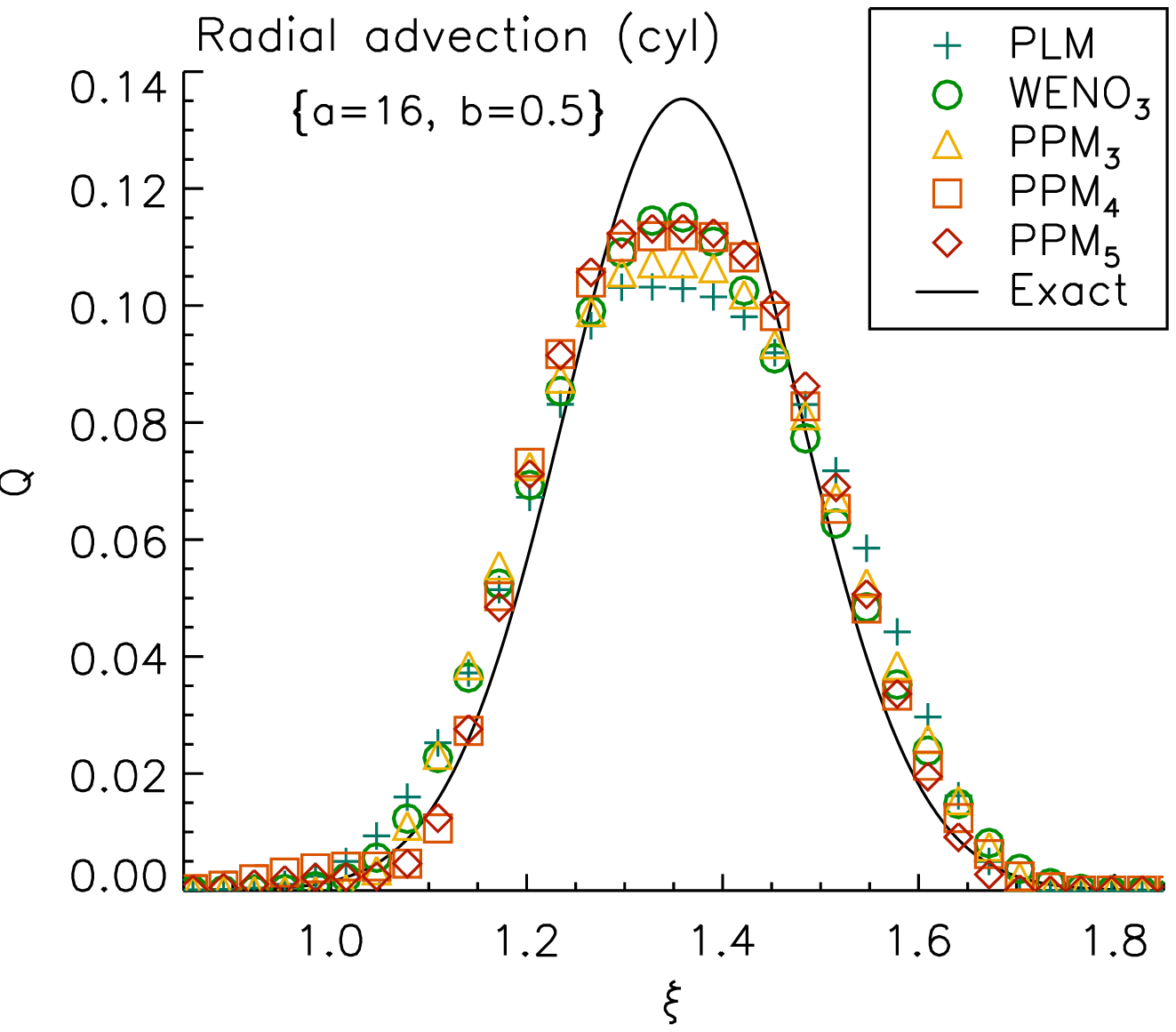}

 \vspace{12pt}

 \includegraphics[width=0.4\textwidth]
                 {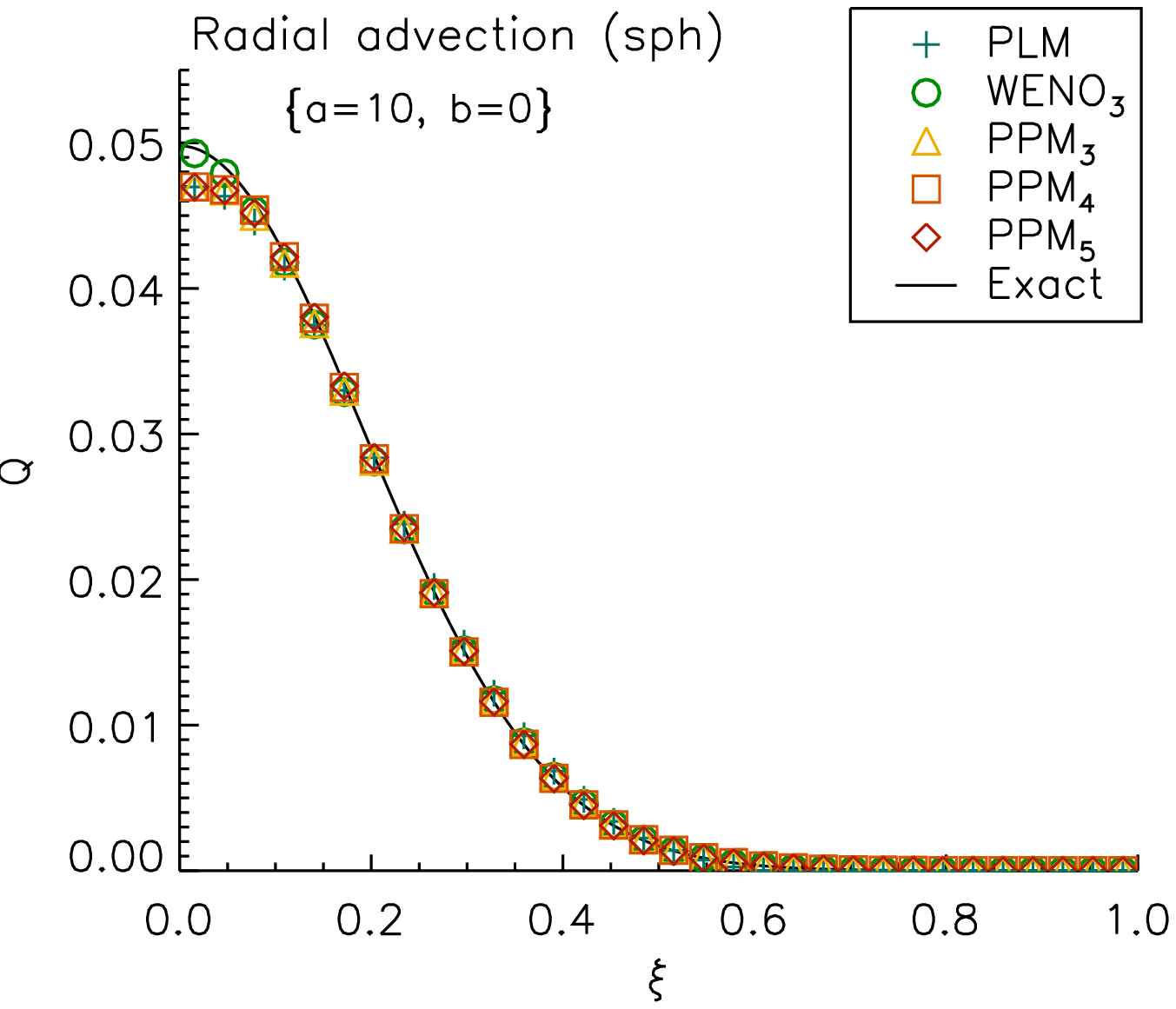}%
 \includegraphics[width=0.4\textwidth]
                 {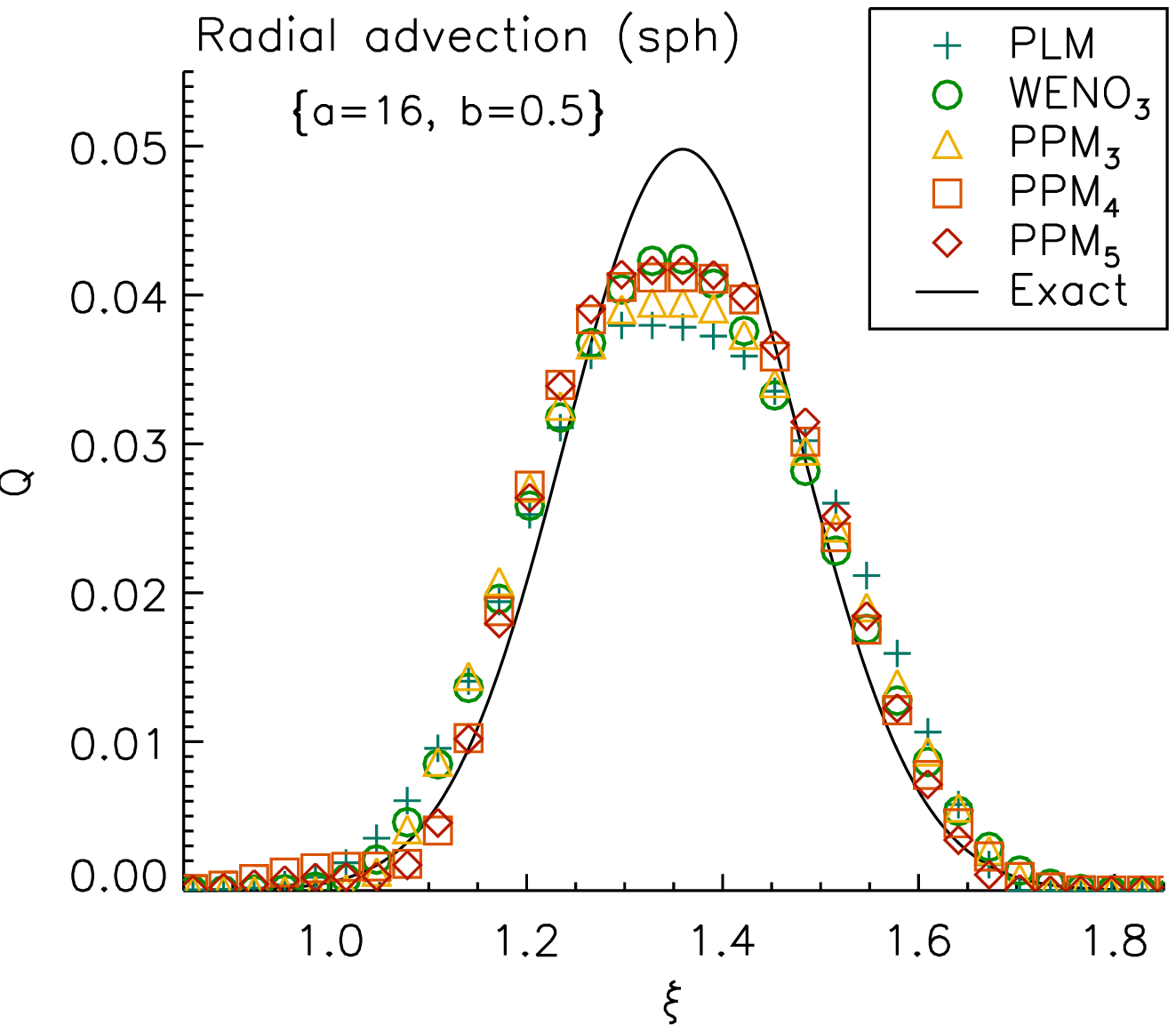}
 \caption{\footnotesize Spatial profiles at $t=1$ for the radial advection test problem using $N=64$ grid zones in cylindrical (top panels) and spherical coordinates (bottom panels).
 Left and right panels refer, respectively, to computations carried out with $\{a=10,\, b=0\}$ and $\{a=16,\, b=1/2\}$.
 For the sake of clarity, a smaller portion of the computational domain is shown.}
 \label{fig:rad_adv_pro}
\end{figure}

\begin{figure}[!h]
 \centering
 \includegraphics[width=0.4\textwidth]
                 {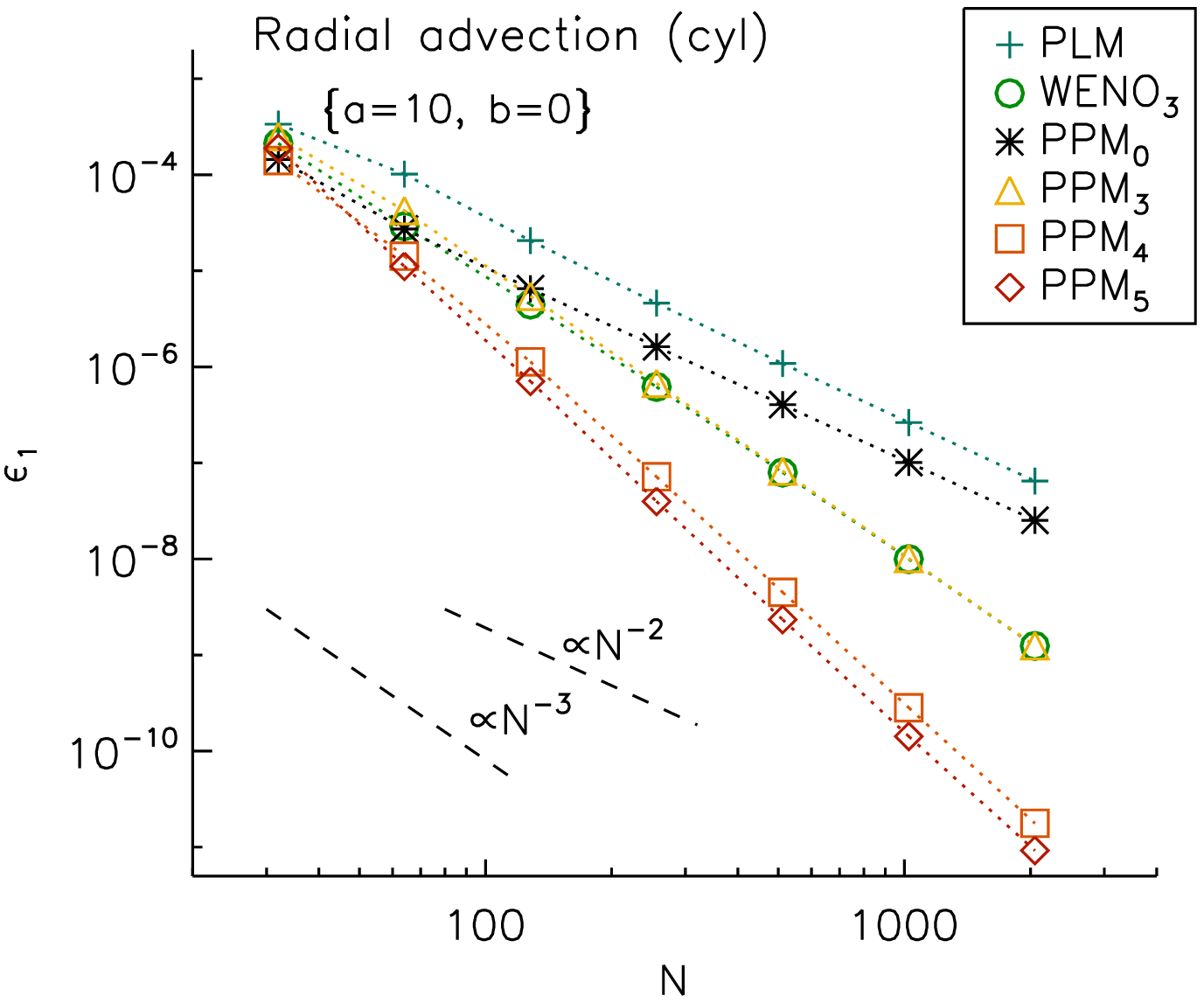}%
 \includegraphics[width=0.4\textwidth]
                 {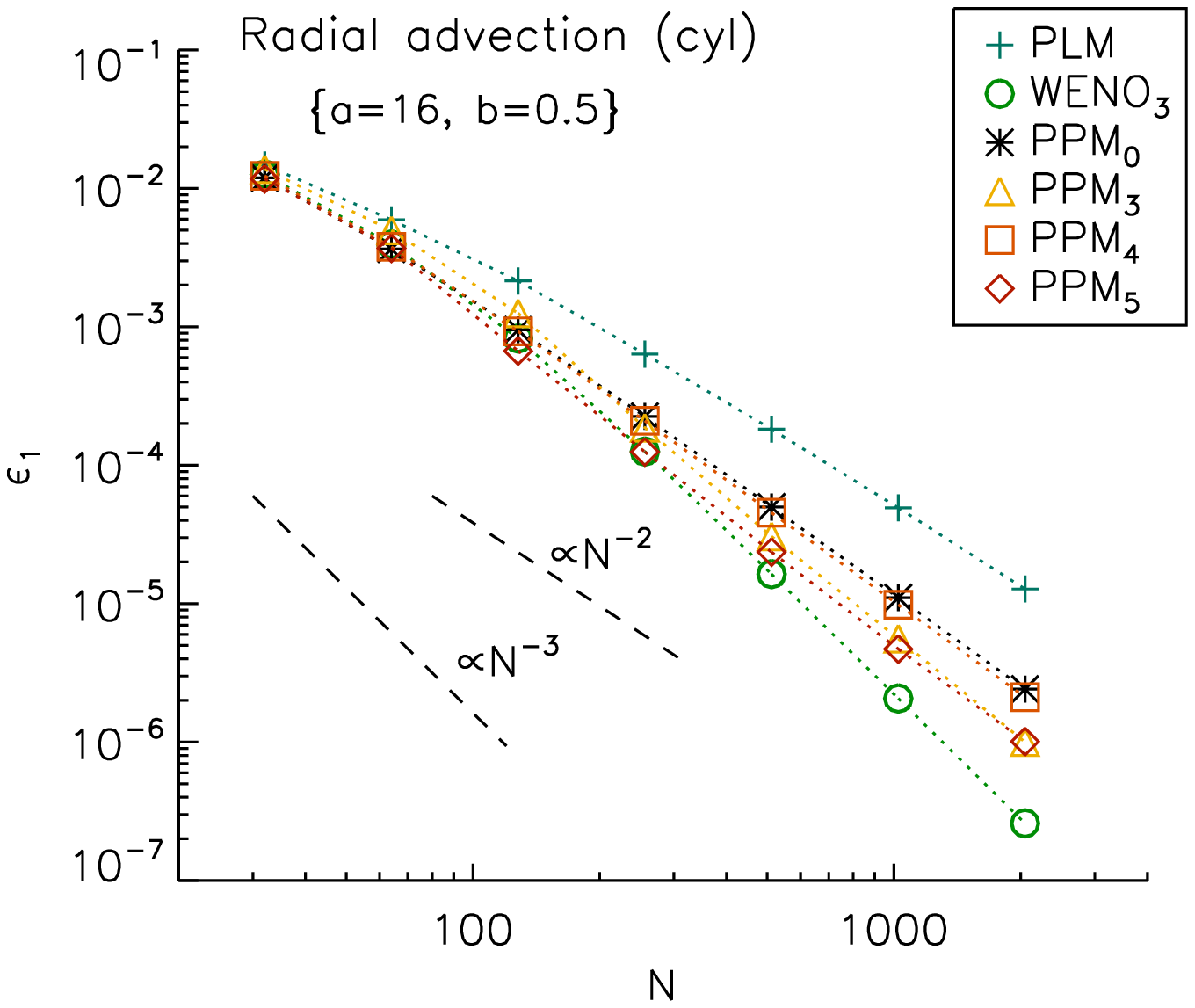}

 \vspace{12pt}

 \includegraphics[width=0.4\textwidth]
                 {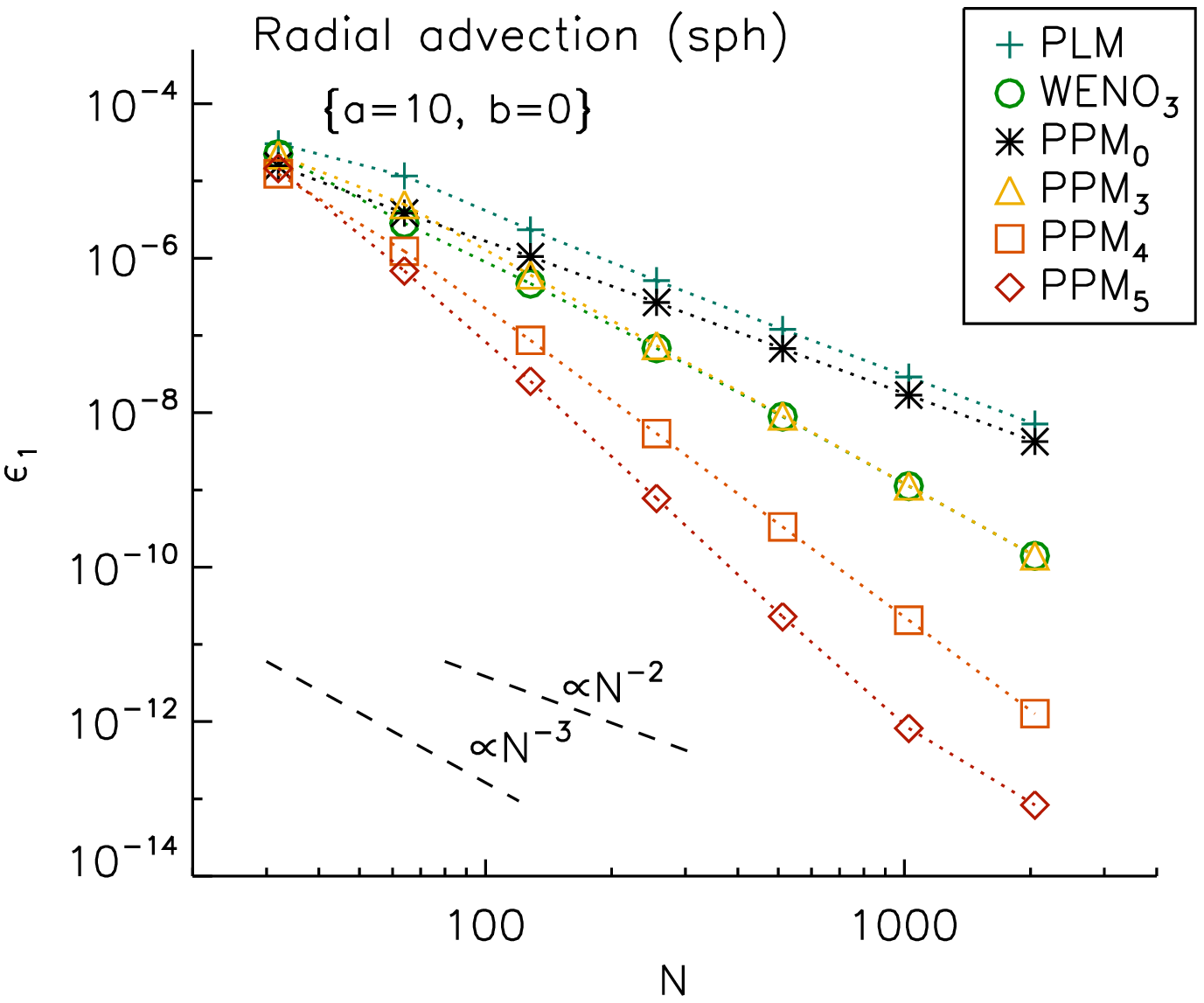}%
 \includegraphics[width=0.4\textwidth]
                 {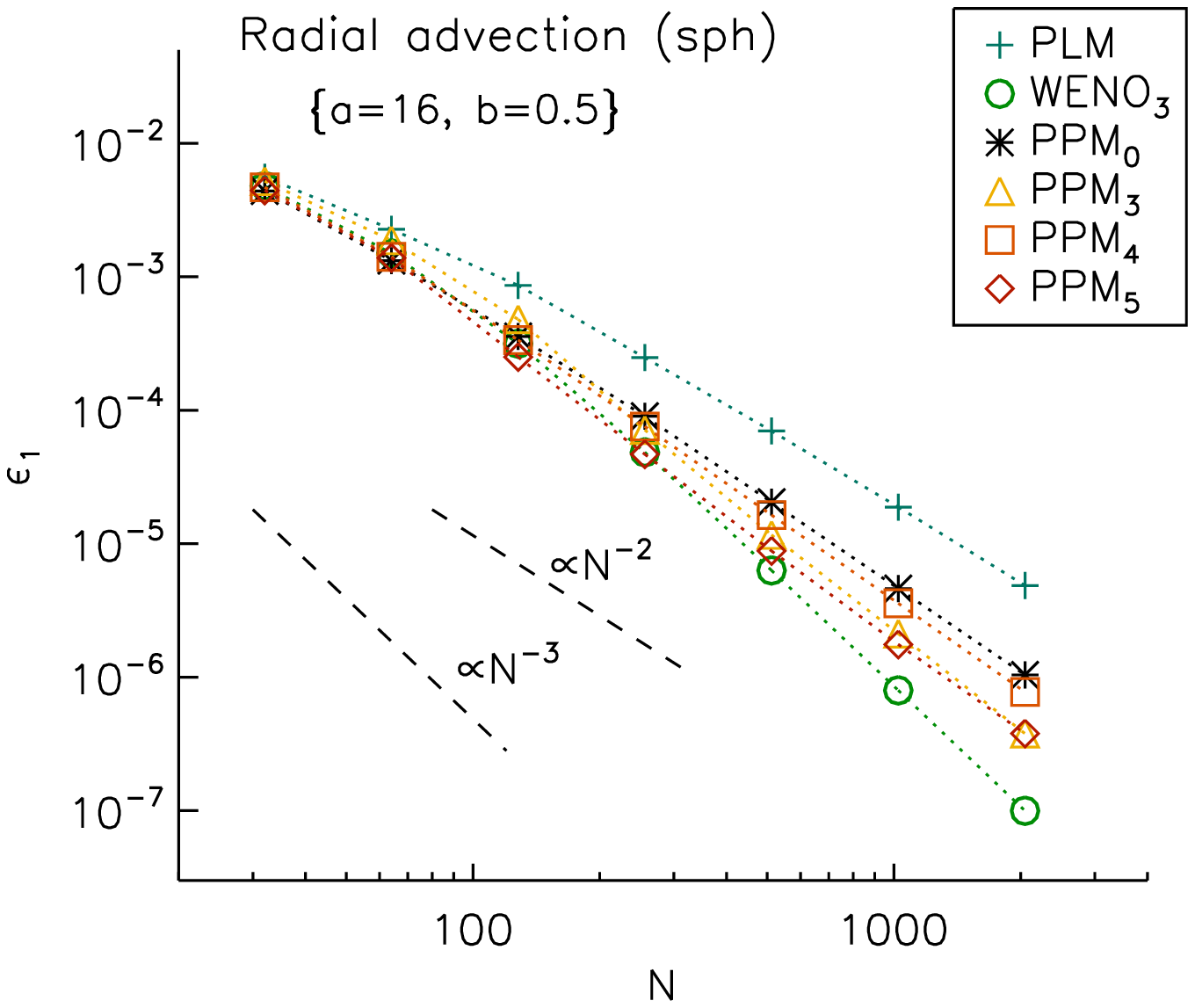}
 \caption{\footnotesize $L_1$ norm errors for the radial advection problem at $t=1$ in cylindrical coordinates (top panels) and spherical coordinates (bottom panels) as function of the resolution. 
 Left and right panels refer, respectively, to computations performed with $\{a=10,\, b=0\}$ and $\{a=16,\, b=1/2\}$.
 Third- and second-order ideal scalings are given by the dashed lines.}
 \label{fig:rad_adv_err}
\end{figure}

Fig. \ref{fig:rad_adv_pro} shows the spatial profiles obtained using $N = 64$ zones at $t=1$ in cylindrical ($\xi=R$, top panel) and spherical coordinates ($\xi=r$, bottom) for the two sets, respectively.
For the monotonically decreasing profile (Case A), all schemes yields comparable errors with the exception of WENO$_3$ which shows superior performance as it does not suffer from clipping in proximity of the maximum located at the coordinate origin.
The non-monotonic case (B) tests more severely the accuracy of the proposed interpolation methods.
Here, WENO$_3$ is still the best shape-preserving method (at this resolution) followed by PPM$_5$, PPM$_4$, PPM$_3$ and, lastly, by PLM which shows the largest numerical diffusion.

A resolution study is presented in Fig \ref{fig:rad_adv_err} with the corresponding errors and orders of convergence (in $L_1$ norm) being sorted in Table \ref{tab:rad_adv_err}.
For set A (left panels in Fig \ref{fig:rad_adv_err}), WENO$_3$ and PPM$_3$ show third-order accuracy, PPM$_4$ and PPM$_5$ converge somewhat faster ($\sim N^{-4}$) while PLM converges as $N^{-2}$, as expected.
On the contrary, the original PPM scheme without any geometric correction (PPM$_0$, cross symbols) is only $2^{\rm nd}$-order accurate owing to the interpolation errors generated close to $\xi=0$.
At the maximum resolution ($N = 2048$ zones), the error obtained with PPM$_4$ is $\gtrsim 10^3$ times smaller than the traditional PPM scheme.

For parameter set B (with a maximum in the initial profile), only WENO$_3$ is truly $3^{\rm rd}$-order accurate whereas the orders of accuracy of the different PPM versions decrease to $2$ owing to the well-known clipping phenomenon near local extrema.
Here the difference between geometrically-corrected schemes such as PPM$_4$ and the traditional PPM$_0$ are less evident.

\begin{table}[!ht]
\caption{\footnotesize $L_1$ norm errors and orders of convergence for the radial advection test in cylindrical (columns 3-6) and spherical (columns 7-10) coordinates at $t=1$ for the selected reconstruction schemes.
The errors are given for different sets of the constants $\{a,\,b\}$ used to define the initial Gaussian profile, Eq. (\ref{eq:rad_adv_initial_profile}).}
\label{tab:rad_adv_err}
\centering
\footnotesize
\begin{tabular*}{\textwidth}{@{\extracolsep{\fill}} lr rrrr rrrr}\hline
        &   &  \multicolumn{4}{c}{Cylindrical} 
            &  \multicolumn{4}{c}{Spherical} \\
  \cline{3-6} \cline{7-10} 
        &   &  \multicolumn{2}{c}{$\{a=10,b=0\}$} 
            &  \multicolumn{2}{c}{$\{a=16,b=1/2\}$} 
            &  \multicolumn{2}{c}{$\{a=10,b=0\}$} 
            &  \multicolumn{2}{c}{$\{a=16,b=1/2\}$} \\
  \cline{3-4} \cline{5-6} \cline{7-8} \cline{9-10}
Method & $N_r$  & $\epsilon_1\left(Q\right)$  & ${\cal O}_{L_1}$
                & $\epsilon_1\left(Q\right)$  & ${\cal O}_{L_1}$ 
                & $\epsilon_1\left(Q\right)$  & ${\cal O}_{L_1}$
                & $\epsilon_1\left(Q\right)$  & ${\cal O}_{L_1}$ \\
 \noalign{\smallskip} \hline\noalign{\smallskip}
\hline PLM
 & 32 & 3.36E-004 & - & 1.48E-002 & - & 3.04E-005 & - & 5.58E-003 & -\\ \noalign{\smallskip}
 & 64 & 1.02E-004 & 1.73 & 5.93E-003 & 1.32 & 1.16E-005 & 1.39 & 2.27E-003 & 1.30\\ \noalign{\smallskip}
 & 128 & 2.07E-005 & 2.30 & 2.15E-003 & 1.47 & 2.34E-006 & 2.31 & 8.62E-004 & 1.40\\ \noalign{\smallskip}
 & 256 & 4.61E-006 & 2.16 & 6.35E-004 & 1.76 & 5.13E-007 & 2.19 & 2.48E-004 & 1.80\\ \noalign{\smallskip}
 & 512 & 1.08E-006 & 2.09 & 1.82E-004 & 1.80 & 1.20E-007 & 2.10 & 6.99E-005 & 1.83\\ \noalign{\smallskip}
 & 1024 & 2.63E-007 & 2.04 & 4.92E-005 & 1.89 & 2.90E-008 & 2.05 & 1.88E-005 & 1.90\\ \noalign{\smallskip}
 & 2048 & 6.48E-008 & 2.02 & 1.28E-005 & 1.95 & 7.13E-009 & 2.02 & 4.86E-006 & 1.95\\ \noalign{\smallskip}
\hline WENO$_3$
 & 32 & 2.12E-004 & - & 1.26E-002 & - & 2.22E-005 & - & 4.79E-003 & -\\ \noalign{\smallskip}
 & 64 & 2.91E-005 & 2.87 & 3.94E-003 & 1.68 & 2.84E-006 & 2.97 & 1.50E-003 & 1.68\\ \noalign{\smallskip}
 & 128 & 4.46E-006 & 2.71 & 8.24E-004 & 2.26 & 4.70E-007 & 2.60 & 3.15E-004 & 2.25\\ \noalign{\smallskip}
 & 256 & 6.18E-007 & 2.85 & 1.26E-004 & 2.71 & 6.83E-008 & 2.78 & 4.84E-005 & 2.70\\ \noalign{\smallskip}
 & 512 & 7.95E-008 & 2.96 & 1.63E-005 & 2.94 & 8.90E-009 & 2.94 & 6.31E-006 & 2.94\\ \noalign{\smallskip}
 & 1024 & 9.97E-009 & 2.99 & 2.06E-006 & 2.99 & 1.12E-009 & 2.99 & 7.97E-007 & 2.99\\ \noalign{\smallskip}
 & 2048 & 1.24E-009 & 3.00 & 2.58E-007 & 3.00 & 1.40E-010 & 3.00 & 9.97E-008 & 3.00\\ \noalign{\smallskip}
\hline PPM$_3$
 & 32 & 2.51E-004 & - & 1.38E-002 & - & 2.18E-005 & - & 5.20E-003 & -\\ \noalign{\smallskip}
 & 64 & 4.23E-005 & 2.57 & 4.98E-003 & 1.47 & 4.95E-006 & 2.14 & 1.88E-003 & 1.47\\ \noalign{\smallskip}
 & 128 & 5.43E-006 & 2.96 & 1.25E-003 & 1.99 & 6.08E-007 & 3.03 & 4.80E-004 & 1.97\\ \noalign{\smallskip}
 & 256 & 6.74E-007 & 3.01 & 1.87E-004 & 2.74 & 7.41E-008 & 3.04 & 7.12E-005 & 2.75\\ \noalign{\smallskip}
 & 512 & 8.24E-008 & 3.03 & 3.06E-005 & 2.61 & 9.10E-009 & 3.02 & 1.16E-005 & 2.61\\ \noalign{\smallskip}
 & 1024 & 1.01E-008 & 3.02 & 5.58E-006 & 2.45 & 1.13E-009 & 3.01 & 2.12E-006 & 2.46\\ \noalign{\smallskip}
 & 2048 & 1.25E-009 & 3.01 & 9.99E-007 & 2.48 & 1.40E-010 & 3.01 & 3.77E-007 & 2.49\\ \noalign{\smallskip}
\hline PPM$_4$
 & 32 & 1.41E-004 & - & 1.23E-002 & - & 1.24E-005 & - & 4.68E-003 & -\\ \noalign{\smallskip}
 & 64 & 1.42E-005 & 3.31 & 3.78E-003 & 1.70 & 1.23E-006 & 3.33 & 1.41E-003 & 1.73\\ \noalign{\smallskip}
 & 128 & 1.13E-006 & 3.65 & 9.27E-004 & 2.03 & 8.65E-008 & 3.83 & 3.36E-004 & 2.07\\ \noalign{\smallskip}
 & 256 & 7.18E-008 & 3.97 & 2.09E-004 & 2.15 & 5.37E-009 & 4.01 & 7.56E-005 & 2.15\\ \noalign{\smallskip}
 & 512 & 4.50E-009 & 4.00 & 4.54E-005 & 2.20 & 3.32E-010 & 4.02 & 1.64E-005 & 2.20\\ \noalign{\smallskip}
 & 1024 & 2.83E-010 & 3.99 & 9.82E-006 & 2.21 & 2.05E-011 & 4.01 & 3.57E-006 & 2.20\\ \noalign{\smallskip}
 & 2048 & 1.77E-011 & 4.00 & 2.11E-006 & 2.22 & 1.27E-012 & 4.02 & 7.75E-007 & 2.21\\ \noalign{\smallskip}
\hline PPM$_5$
 & 32 & 1.89E-004 & - & 1.17E-002 & - & 1.46E-005 & - & 4.43E-003 & -\\ \noalign{\smallskip}
 & 64 & 1.11E-005 & 4.09 & 3.69E-003 & 1.67 & 6.86E-007 & 4.41 & 1.38E-003 & 1.68\\ \noalign{\smallskip}
 & 128 & 7.07E-007 & 3.98 & 6.70E-004 & 2.46 & 2.56E-008 & 4.74 & 2.51E-004 & 2.47\\ \noalign{\smallskip}
 & 256 & 3.98E-008 & 4.15 & 1.25E-004 & 2.42 & 7.79E-010 & 5.04 & 4.68E-005 & 2.42\\ \noalign{\smallskip}
 & 512 & 2.34E-009 & 4.09 & 2.36E-005 & 2.41 & 2.28E-011 & 5.09 & 8.83E-006 & 2.41\\ \noalign{\smallskip}
 & 1024 & 1.42E-010 & 4.04 & 4.69E-006 & 2.33 & 8.19E-013 & 4.80 & 1.76E-006 & 2.33\\ \noalign{\smallskip}
 & 2048 & 9.23E-012 & 3.95 & 1.01E-006 & 2.22 & 8.32E-014 & 3.30 & 3.78E-007 & 2.22\\ \noalign{\smallskip}
\hline
\end{tabular*}
\end{table}
\normalsize

\subsubsection{Advection equation in the meridional spherical coordinate}
\label{sec:mer_adv}

\begin{figure}
 \centering
 \includegraphics[width=0.4\textwidth]
                 {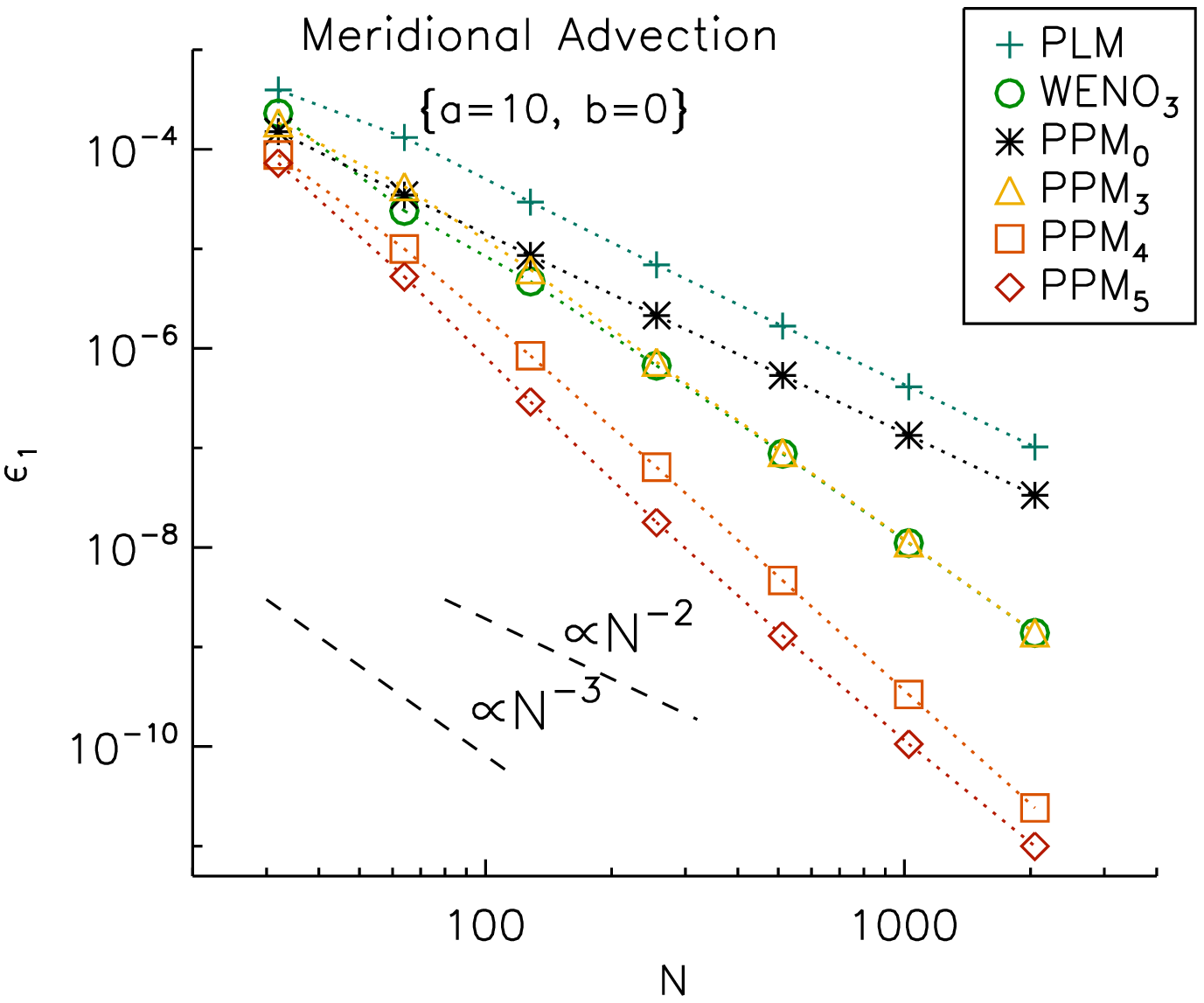}%
 \includegraphics[width=0.4\textwidth]
                 {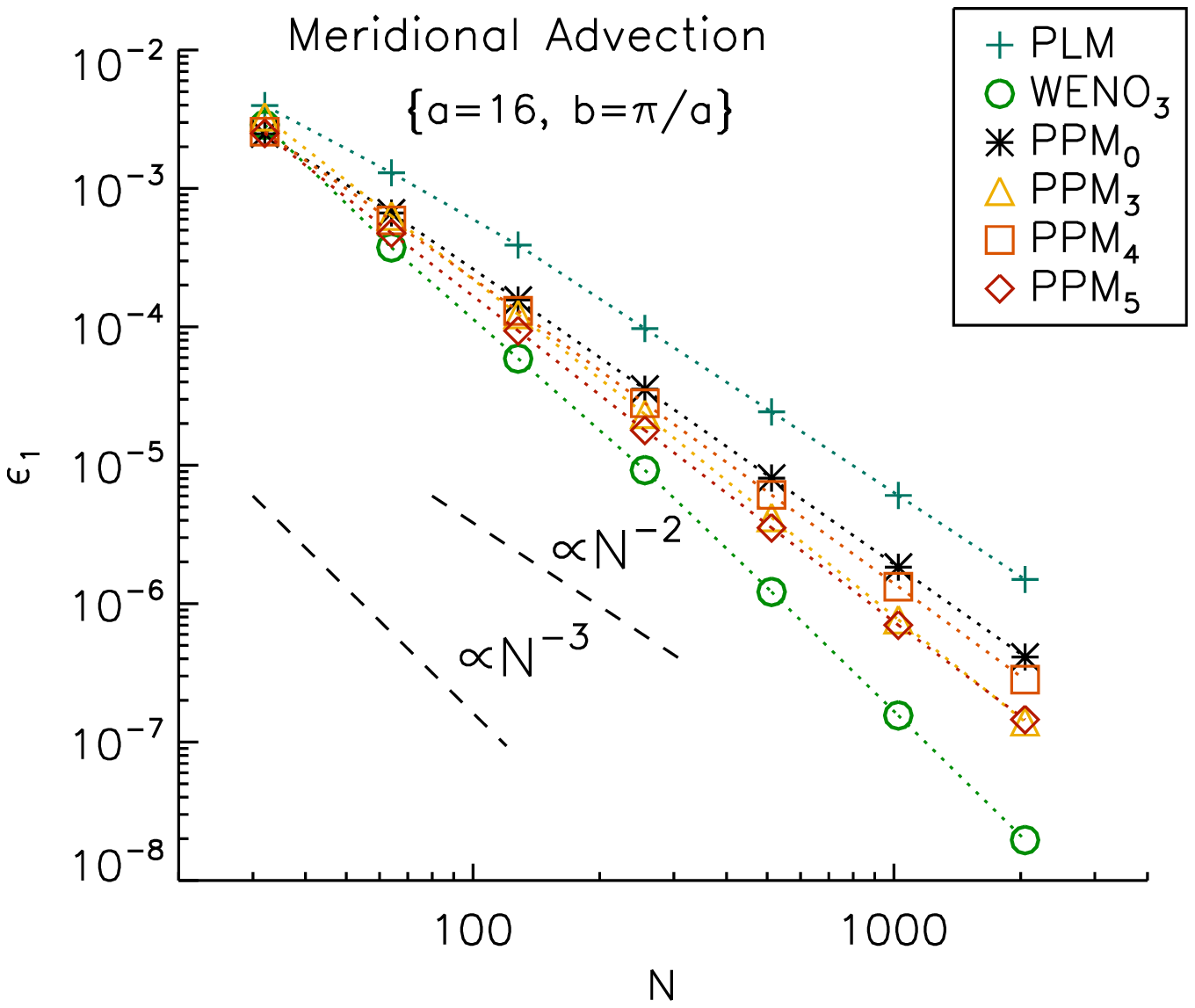}
 \caption{\footnotesize Errors in $L_1$ norm vs. resolution for the meridional linear advection test at $t=1$ using $\{a=10,\, b=0\}$ (left panel) and $\{a=16,\,b=\pi/a\}$ (right panel).
 The ideal second- and third-order scaling are shown as dashed lines.}
 \label{fig:mer_adv_err}
\end{figure}

In order to assess the accuracy of the proposed reconstruction schemes in the meridional spherical coordinate $\theta$ we consider, in analogy with the previous test, the solution of
\begin{equation}\label{eq:meridional_advection}
 \pd{Q}{t} + \frac{1}{\sin\theta}\pd{}{\theta}
             \left(\sin\theta Q v\right)=0 \,,
\end{equation}
where $v = \alpha\theta$ is a linear velocity profile.
Eq. (\ref{eq:meridional_advection}) can be solved exactly, yielding
\begin{equation}
 Q^{\rm ref}(\theta,t) = e^{-\alpha t}\frac{\sin(e^{-\alpha t}\theta)}{\sin\theta}
                 Q\left(e^{-\alpha t}\theta,0\right)\,.
\end{equation}
Without loss of generality, $\alpha=1$ is considered in the following.

A one-dimensional computational grid spanning the interval $\theta\in[0,\pi/2]$ is set up with $N$ zones. 
At $t=0$ the following initial condition is prescribed:
\begin{equation}\label{eq:mer_adv_init}
  Q(\theta, 0) = \left\{\begin{array}{ll}
  \DS \left[\frac{1 + \cos\left( a(\theta-b)\right)}{2}\right]^2 & \qquad\textrm{for}\quad \DS   |\theta-b| < \frac{\pi}{a} \,,\\ \noalign{\medskip}
   0 & \qquad\textrm{otherwise} \,,
 \end{array}\right.
\end{equation}
where $a$ and $b$ are constants.
Note that Eq. (\ref{eq:mer_adv_init}) is continuous at $\theta = b\pm \pi/a$ up to the third derivative.
The initial solution values are given by the volume average of Eq. (\ref{eq:mer_adv_init}) over the cell and integration stops at $t=1$ using a CFL number $C_a = 0.9$ while the interface flux is computed using Eq. (\ref{eq:upwind_Riemann_solver}).

Two different choices of $a$ and $b$ are considered: the first one corresponding to a monotonically decreasing profile ($\{a=10,\, b=0\}$), and the second one resulting in a non-monotone function ($\{a=16,\, b=a/\pi\}$) with a maximum at $\theta = b$.
Results for different resolutions $N = 2^n$ ($n=5,\dots,10$), shown in Fig \ref{fig:mer_adv_err} and Table \ref{tab:mer_adv_err}, confirm the general trend already established for the radial advection test, see \S\ref{sec:radial_advection}.
Overall, PPM and WENO schemes have comparable errors and converge to the exact solution with third- (or higher-) order accuracy in the monotonic profile case.
The situation is somehow different for the non-monotonic profile (right panel in Fig \ref{fig:mer_adv_err}) where only WENO$_3$ achieves the expected order of accuracy.
The loss of accuracy for the PPM schemes is due, once again, to the clipping at extrema.
By comparing the performance of the original PPM$_0$ scheme to that of the others, one can again conclude that geometrical corrections are particularly important close to the coordinate origin. 

\begin{table}[!ht]
\caption{\footnotesize $L_1$ norm errors and orders of convergence for different reconstruction schemes for the meridional advection test at $t=1$ using $\{a=10,\, b=0\}$ (columns 3 and 4) and $\{a=16,\, b=\pi/a\}$ (columns 5 and 6).}
\label{tab:mer_adv_err}
\centering
\footnotesize
\begin{tabular*}{\textwidth}{@{\extracolsep{\fill}} lr rrrr}\hline
        &   &  \multicolumn{2}{c}{$a=10,\,b=0$}  
            &  \multicolumn{2}{c}{$a=16,\,b=\pi/a$} \\
  \cline{3-4} \cline{5-6}
Method & $N_\theta$  & $\epsilon_1\left(Q\right)$  & ${\cal O}_{L_1}$
                     & $\epsilon_1\left(Q\right)$  & ${\cal O}_{L_1}$ \\
 \noalign{\smallskip} \hline\noalign{\smallskip}
\hline PLM
 & 32 & 3.95E-004 & - & 3.97E-003 & -\\ \noalign{\smallskip}
 & 64 & 1.31E-004 & 1.59 & 1.30E-003 & 1.61\\ \noalign{\smallskip}
 & 128 & 2.95E-005 & 2.15 & 3.90E-004 & 1.74\\ \noalign{\smallskip}
 & 256 & 6.92E-006 & 2.09 & 9.74E-005 & 2.00\\ \noalign{\smallskip}
 & 512 & 1.67E-006 & 2.05 & 2.43E-005 & 2.00\\ \noalign{\smallskip}
 & 1024 & 4.12E-007 & 2.02 & 6.05E-006 & 2.01\\ \noalign{\smallskip}
 & 2048 & 1.02E-007 & 2.01 & 1.50E-006 & 2.02\\ \noalign{\smallskip}
\hline WENO$_3$
 & 32 & 2.29E-004 & - & 2.88E-003 & -\\ \noalign{\smallskip}
 & 64 & 2.41E-005 & 3.25 & 3.74E-004 & 2.94\\ \noalign{\smallskip}
 & 128 & 4.71E-006 & 2.36 & 5.90E-005 & 2.67\\ \noalign{\smallskip}
 & 256 & 6.71E-007 & 2.81 & 9.22E-006 & 2.68\\ \noalign{\smallskip}
 & 512 & 8.75E-008 & 2.94 & 1.22E-006 & 2.92\\ \noalign{\smallskip}
 & 1024 & 1.11E-008 & 2.98 & 1.56E-007 & 2.97\\ \noalign{\smallskip}
 & 2048 & 1.39E-009 & 3.00 & 1.96E-008 & 2.99\\ \noalign{\smallskip}
\hline PPM$_3$
 & 32 & 1.85E-004 & - & 3.30E-003 & -\\ \noalign{\smallskip}
 & 64 & 4.23E-005 & 2.13 & 6.32E-004 & 2.38\\ \noalign{\smallskip}
 & 128 & 6.19E-006 & 2.77 & 1.23E-004 & 2.36\\ \noalign{\smallskip}
 & 256 & 7.30E-007 & 3.08 & 2.34E-005 & 2.39\\ \noalign{\smallskip}
 & 512 & 9.03E-008 & 3.01 & 4.20E-006 & 2.48\\ \noalign{\smallskip}
 & 1024 & 1.12E-008 & 3.01 & 7.66E-007 & 2.45\\ \noalign{\smallskip}
 & 2048 & 1.39E-009 & 3.01 & 1.40E-007 & 2.45\\ \noalign{\smallskip}
\hline PPM$_4$
 & 32 & 8.76E-005 & - & 2.57E-003 & -\\ \noalign{\smallskip}
 & 64 & 9.97E-006 & 3.13 & 5.87E-004 & 2.13\\ \noalign{\smallskip}
 & 128 & 8.44E-007 & 3.56 & 1.30E-004 & 2.17\\ \noalign{\smallskip}
 & 256 & 6.41E-008 & 3.72 & 2.81E-005 & 2.22\\ \noalign{\smallskip}
 & 512 & 4.66E-009 & 3.78 & 6.10E-006 & 2.20\\ \noalign{\smallskip}
 & 1024 & 3.33E-010 & 3.81 & 1.33E-006 & 2.20\\ \noalign{\smallskip}
 & 2048 & 2.42E-011 & 3.78 & 2.82E-007 & 2.23\\ \noalign{\smallskip}
\hline PPM$_5$
 & 32 & 7.28E-005 & - & 2.50E-003 & -\\ \noalign{\smallskip}
 & 64 & 5.25E-006 & 3.79 & 4.75E-004 & 2.40\\ \noalign{\smallskip}
 & 128 & 2.91E-007 & 4.17 & 9.36E-005 & 2.34\\ \noalign{\smallskip}
 & 256 & 1.79E-008 & 4.03 & 1.79E-005 & 2.39\\ \noalign{\smallskip}
 & 512 & 1.30E-009 & 3.78 & 3.53E-006 & 2.34\\ \noalign{\smallskip}
 & 1024 & 1.06E-010 & 3.61 & 7.00E-007 & 2.33\\ \noalign{\smallskip}
 & 2048 & 9.97E-012 & 3.41 & 1.45E-007 & 2.27\\ \noalign{\smallskip}
\hline
\end{tabular*}
\end{table}
\normalsize

\subsubsection{Advection of a cosine bell}
%
%
%

The solid-body rotation of a cosine bell profile on the surface of the sphere is considered \cite{Williamson_etal.1992}.
The problem is solved in the $(\theta,\,\phi)$ coordinates with velocity given by the non-deformational field
\begin{equation}
  \vec{v} = (v_r,\,v_\theta,\,v_\phi) = 
  \left(0,\, -\frac{1}{\sin\theta}\pd{\Psi}{\phi},\,\pd{\Psi}{\theta}\right) \,,
\end{equation}
where $\Psi = -u_0(\cos\theta\cos\alpha - \cos\phi\sin\theta\sin\alpha)$ is the horizontal stream function while $\alpha$ gives the inclination angle between the axis of rotation and the polar axis of the spherical coordinate system ($\theta=0$).
Setting $\alpha = 0$, for instance, gives a purely azimuthal velocity field $\vec{v}=(0,\, 0, \, u_0\sin\theta)$ (advection parallel to the equator).
Conversely, for $\alpha=\pm\pi/2$, the axis of rotation coincides with the $x$ direction so that $\vec{v}=\pm(0,\, u_0\sin\phi, \, u_0\cos\theta\cos\phi)$ and any initial profile lying in the $yz$ plane is transported across the poles.

The initial cosine bell to be advected is given by 
\begin{equation}
  Q = \left\{\begin{array}{ll}
    \DS \frac{1}{2}\left[1 + \cos\left(\pi \sigma/\sigma_0\right)\right] 
  & \quad \textrm{if}\quad \sigma < \sigma_0 \\ \noalign{\medskip}
     0   & \quad\textrm{otherwise} \,,
  \end{array}\right.
\end{equation}
where $\sigma$ is the great circle distance between the point $(\theta,\phi)$ on the sphere and the center $(\theta_c,\phi_c)$:
\begin{equation}
  \sigma = \cos^{-1}\left[\cos\theta_c\cos\theta + \sin\theta_c\sin\theta
                     \cos(\phi - \phi_c)\right] \,.
\end{equation}
Here, $\sigma_0 = 1/3$ while the initial position of the cosine bell is centered around $\theta_c = \pi/2$, $\phi_c = 3\pi/2$.
In what follows only advection through the poles ($\alpha=\pi/2$) is considered which results in a particularly challenging test owing to the grid singularities at $\theta=0,\,\pi$ where latitudinal fluxes have zero value ($\sin\theta=0$) and transport is accomplished only by the contribution of longitudinal fluxes.
Computations are performed at three different resolutions: $32\times 64$ (low, $\Delta\theta=\Delta\phi=5.625^\circ$), $64\times128$ (mid, $\Delta\theta=\Delta\phi=2.8125^\circ$) and $128\times 256$ (high,  $\Delta\theta=\Delta\phi=1.40625^\circ$) computational zones.

\begin{figure}[!h]
 \centering
 \includegraphics[width=0.4\textwidth]{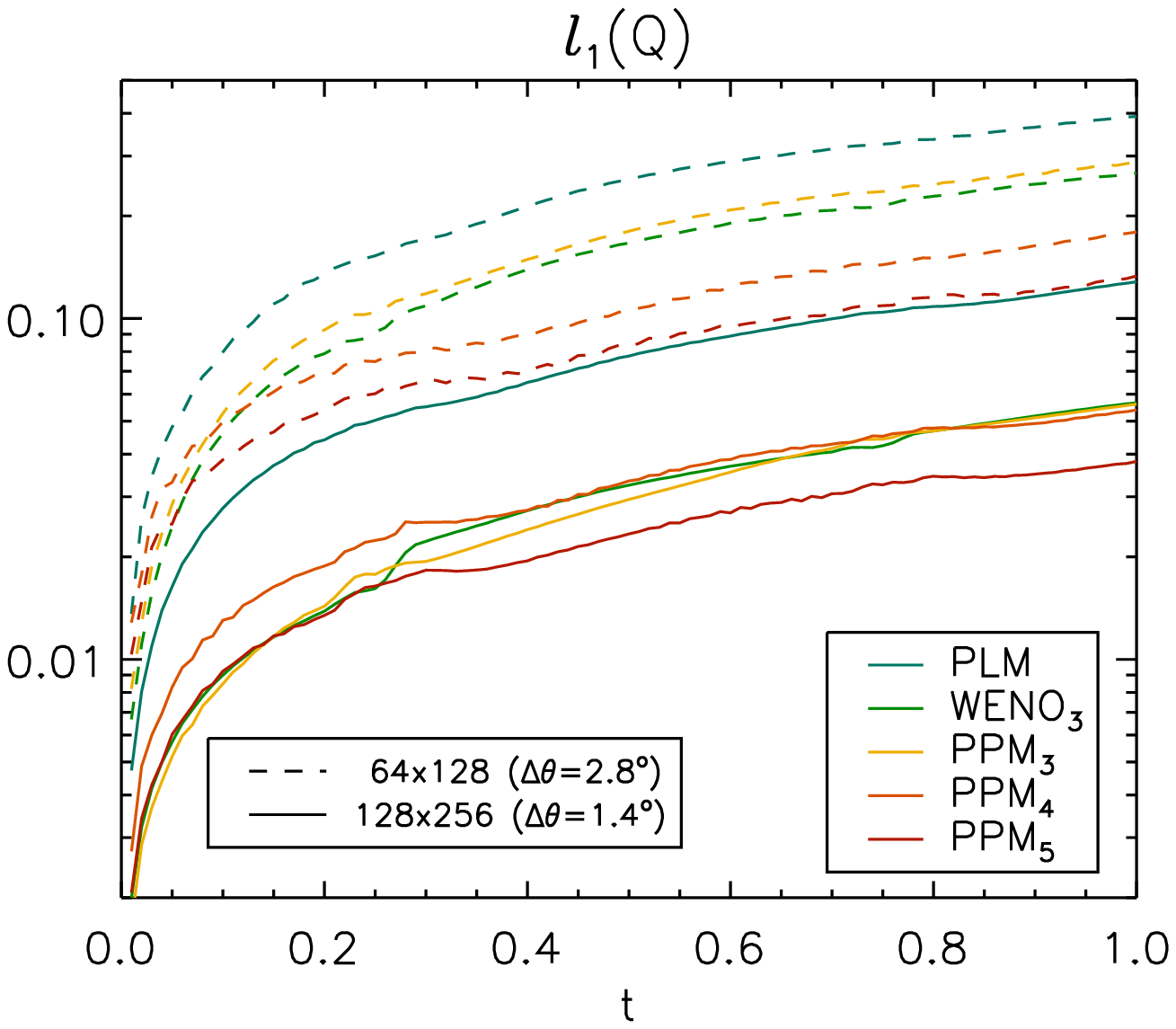}%
 \includegraphics[width=0.4\textwidth]{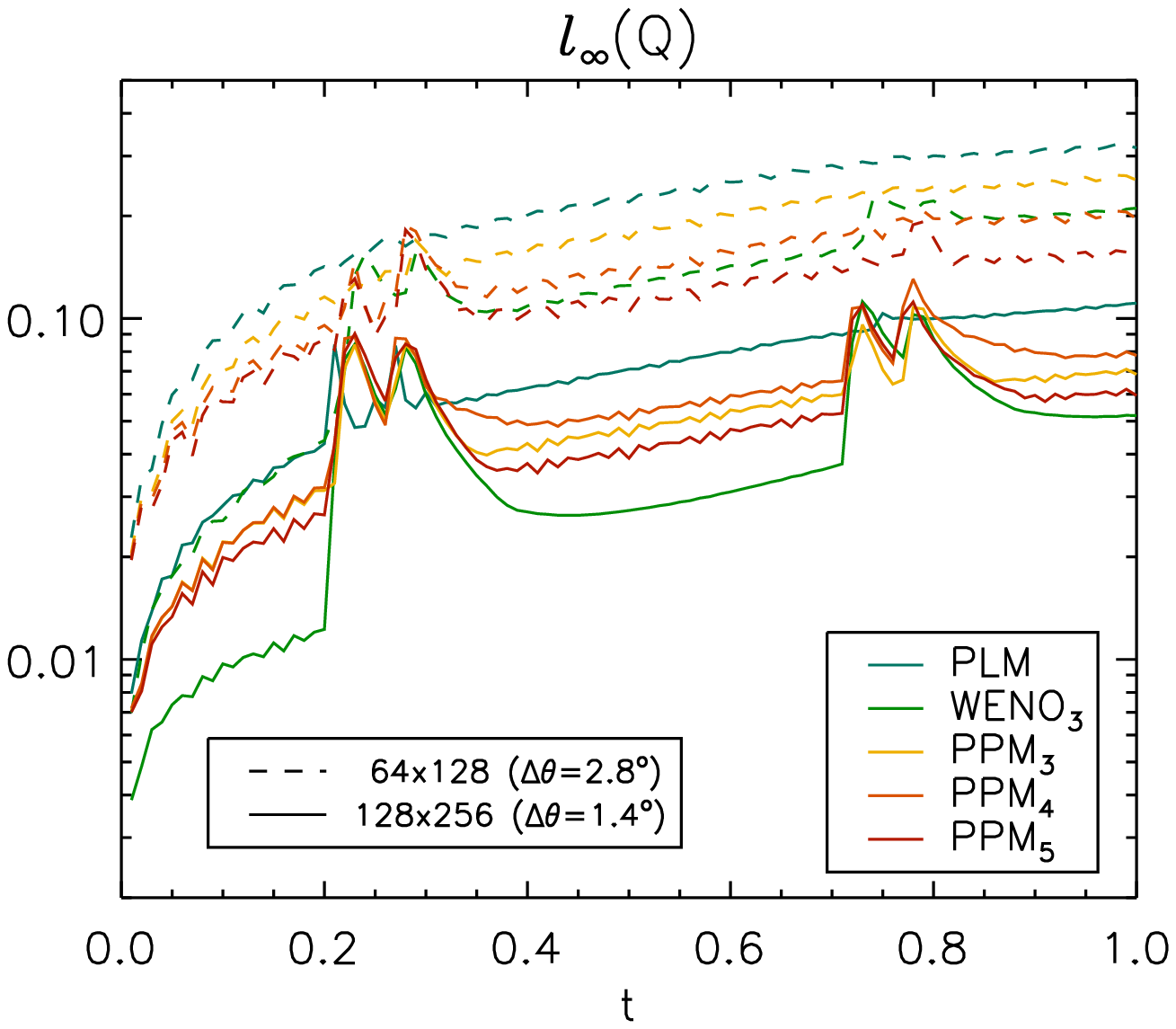}
 \caption{\footnotesize Time history of the normalized errors (left: $l_1$ errors, right: $l_\infty$ errors) for the cosine bell test using $64\times 128$  (dashed lines) and $128\times 256$ (solid lines) grid zones.}
 \label{fig:cosine_bell_err}
\end{figure}

In order to ease the comparison with previous results we plot, in Fig. \ref{fig:cosine_bell_err}, the time history of the $l_1$ and $l_\infty$ errors during the first revolution at the resolutions of $64\times 128$ (dashed line) and $128\times256$ (solid line) grid zones.
The final error values (including also the $l_2$ error) can be inspected from Table \ref{tab:cosine_bell_err}.
Here $l_1$, $l_2$ and $l_\infty$ are the normalized errors computed as in \cite{Williamson_etal.1992}:
\begin{equation}
 l_1(Q) = \frac{\sum_\vec{i} |Q_\vec{i}-Q^{\rm ref}_\vec{i}|\Delta\vol_\vec{i}}
               {\sum_\vec{i} |Q^{\rm ref}_\vec{i}|\Delta\vol_\vec{i}}
  \,;\quad
 l_2(Q) = \sqrt{\frac{\sum_\vec{i} |Q_\vec{i}-Q^{\rm ref}_\vec{i}|^2\Delta\vol_\vec{i}}
               {\sum_\vec{i} |Q^{\rm ref}_\vec{i}|^2\Delta\vol_\vec{i}}}
  \,;\quad
 l_\infty(Q) = \frac{\max_\vec{i}(|Q_\vec{i}-Q^{\rm ref}_\vec{i}|)}
                    {\max_\vec{i}(|Q^{\rm ref}_\vec{i}|)}\,,
\end{equation}
where $\vec{i} = (j,k)$ is a 2D integer vector spanning the computational zones in $(\theta,\phi)$.

\begin{figure}[!h]
 \centering
 \includegraphics[width=0.4\textwidth]{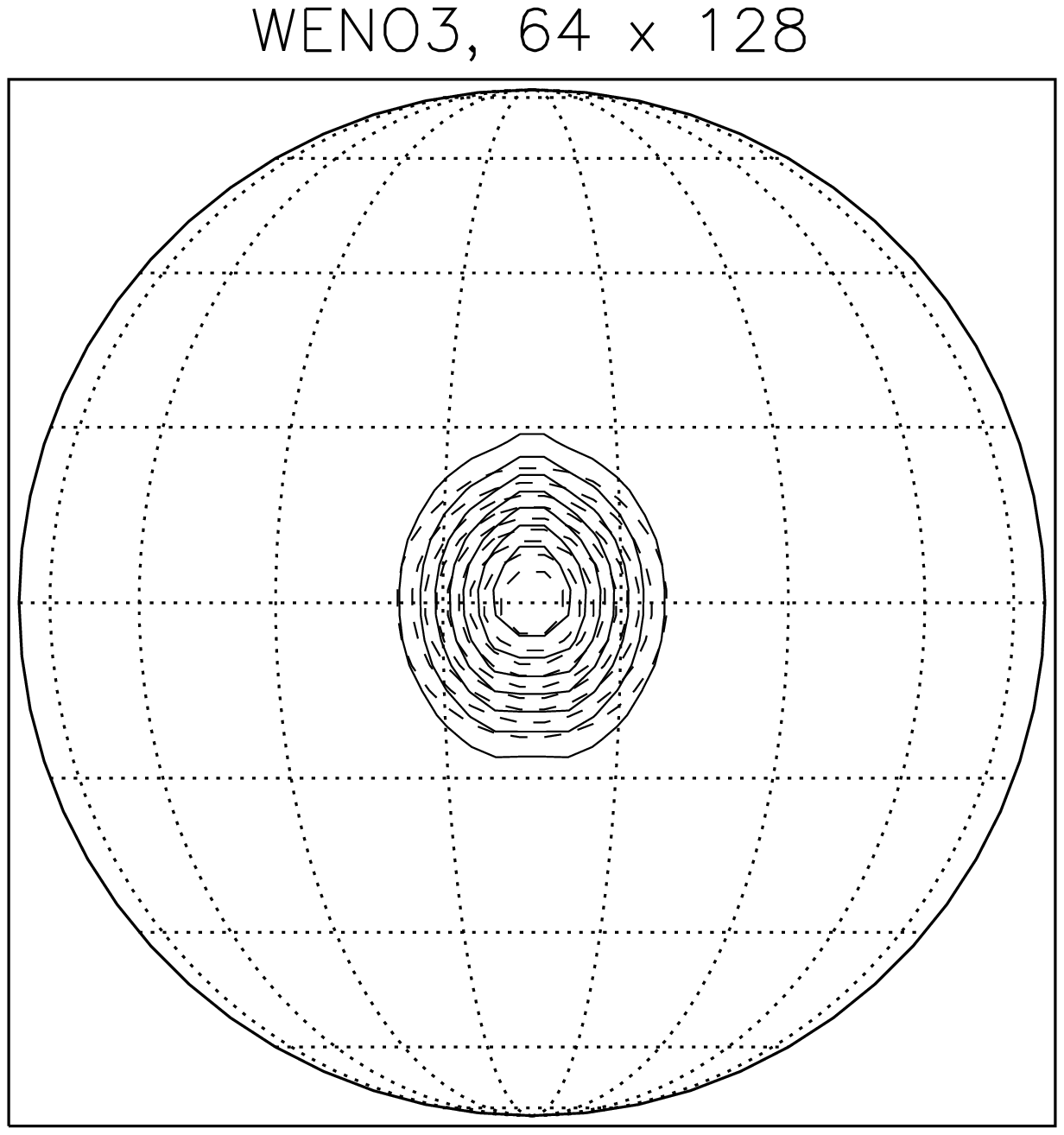}
 \includegraphics[width=0.4\textwidth]{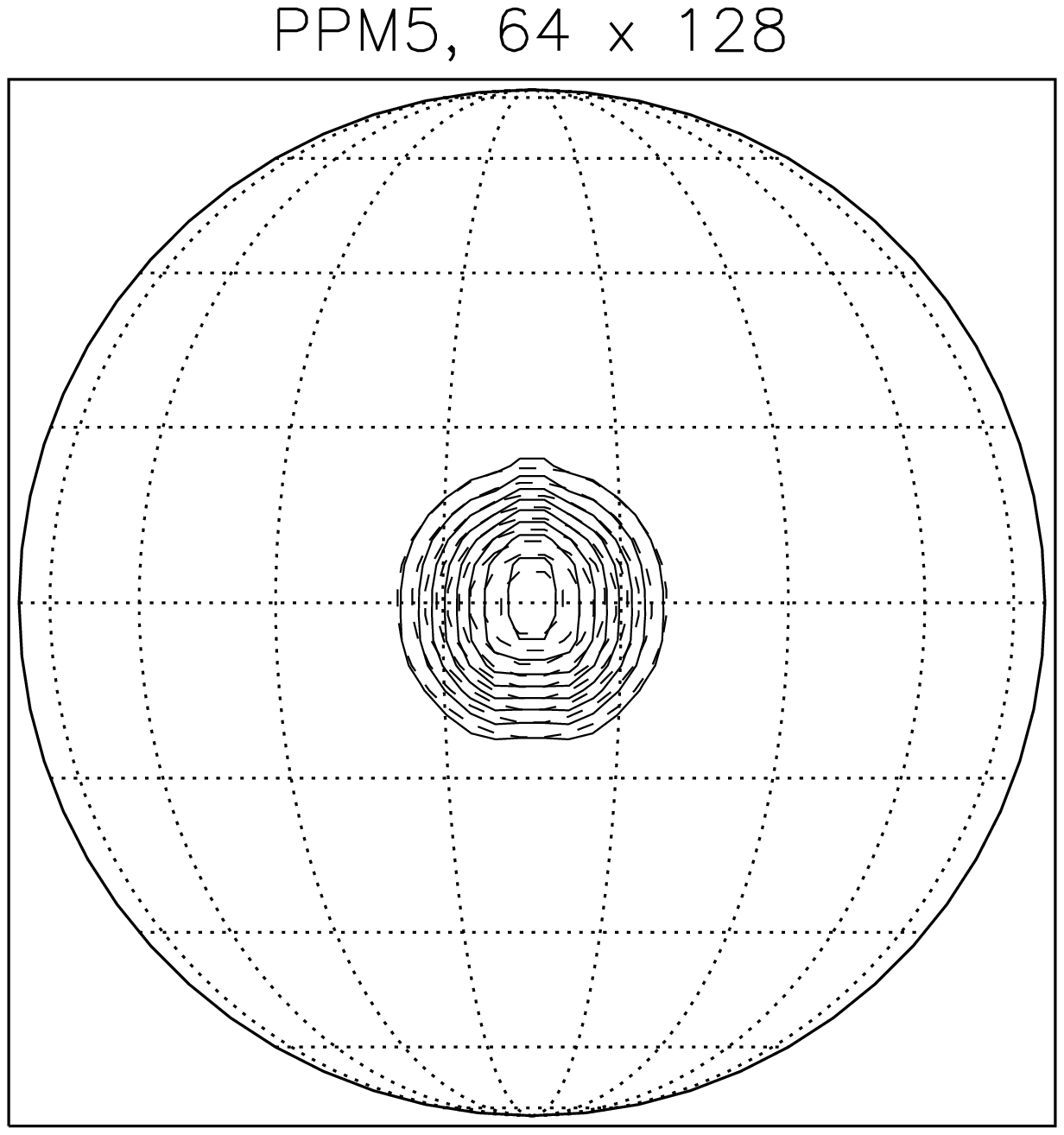}
 \includegraphics[width=0.4\textwidth]{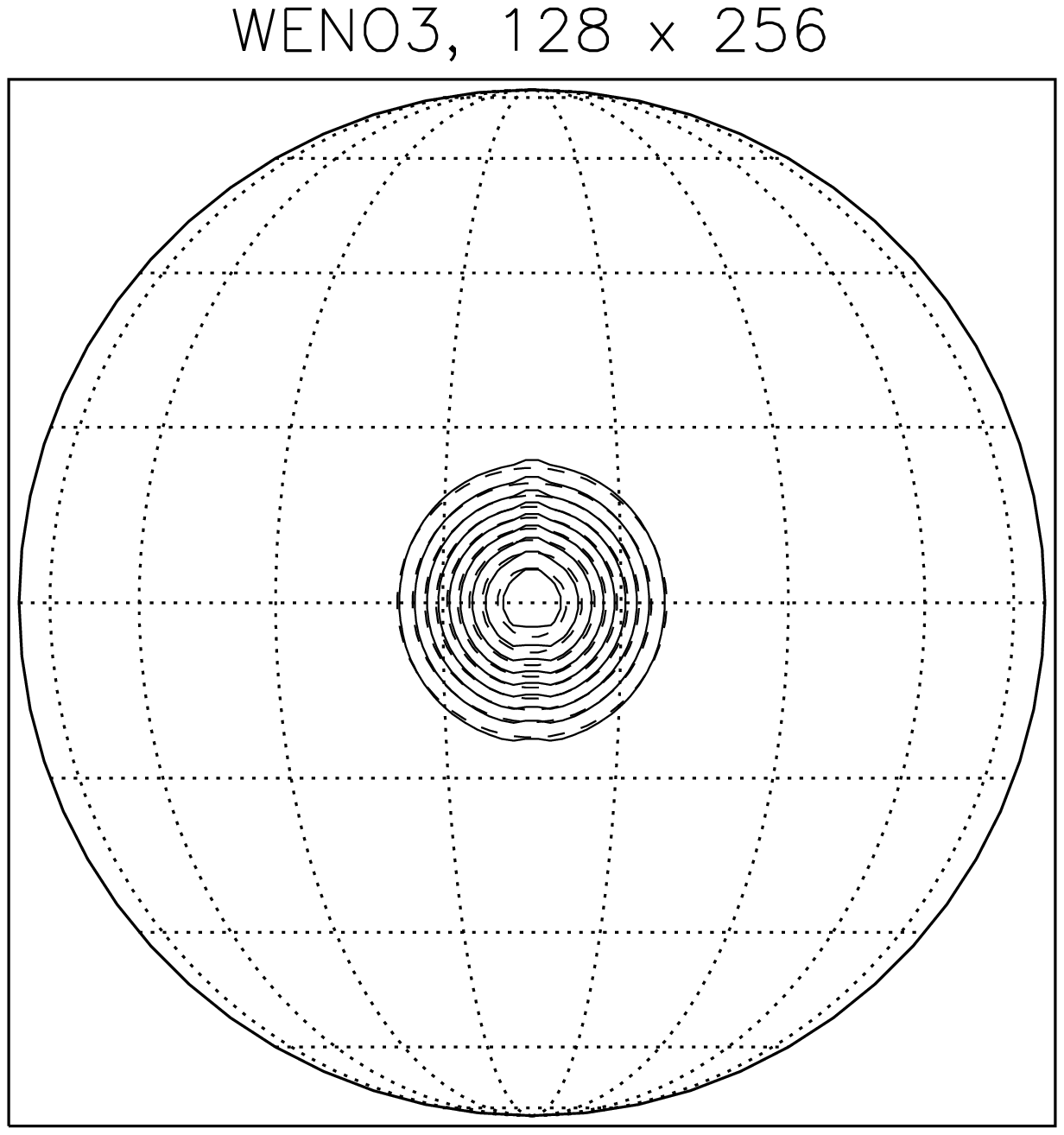}
 \includegraphics[width=0.4\textwidth]{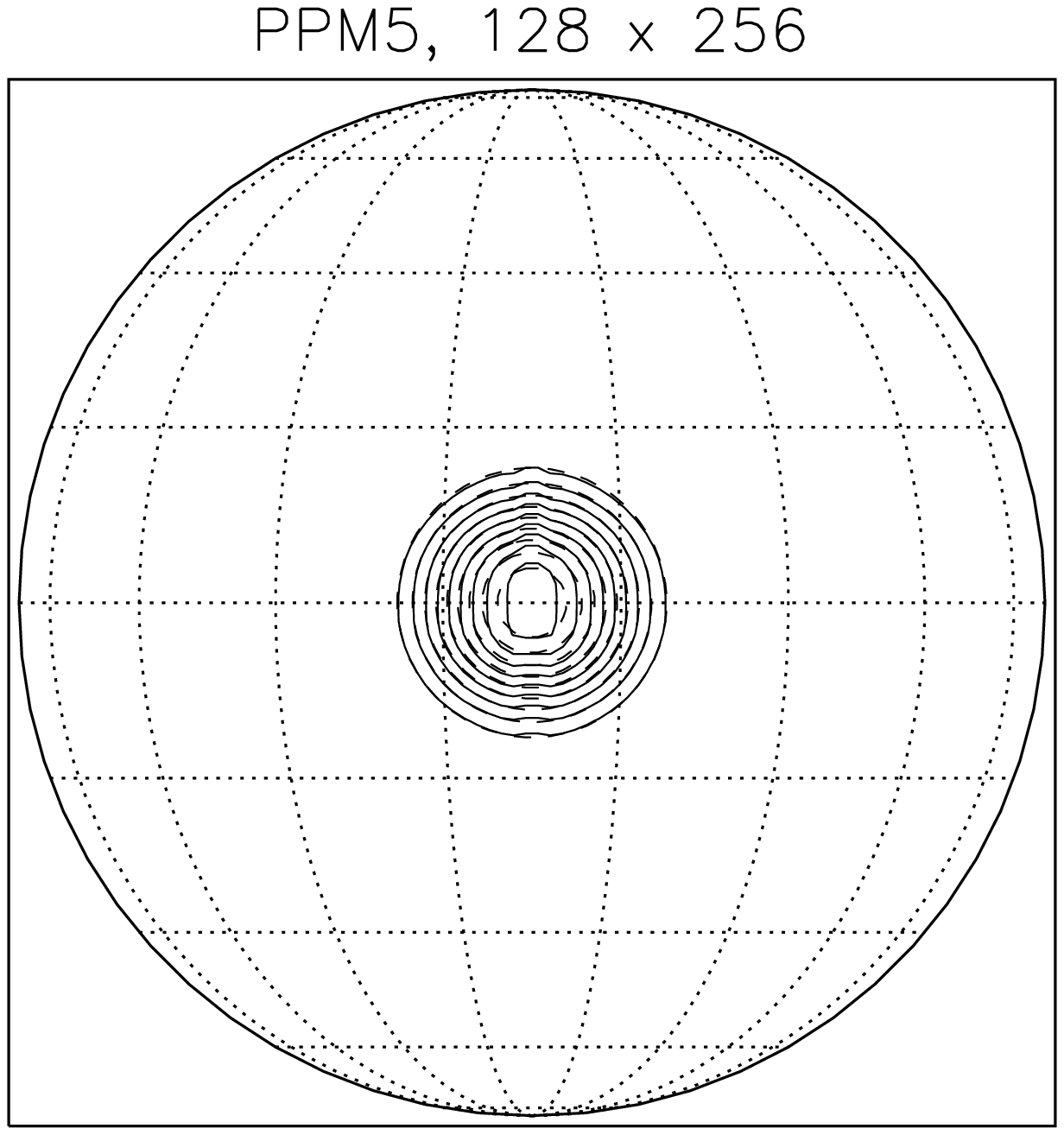}
 \caption{\footnotesize Orthographic projections of the computed solution (solid contours) and exact solution (dotted contour) for the cosine bell advection test after one revolution. 
 Panels to the left and to the right show the results obtained with WENO$_3$ and PPM$_5$, respectively, at the resolutions of $64\times128$ (top) and $128\times256$ (bottom) grid zones.}
 \label{fig:cosine_bell_ortho}
\end{figure}

Errors steadily increase with time and show sudden peaks (in $l_\infty$) in proximity of pole crossing at $t\approx 1/4$ and $t\approx 3/4$.
Our results indicate that PPM$_5$ gives the best accuracy followed by PPM$_4$ and then by WENO$_3$ and PPM$_3$ with comparable errors.
It is worth noticing that PLM requires twice the resolution to match the accuracy obtained with PPM$_5$.
The corresponding orthographic projections of the solutions are shown in Fig \ref{fig:cosine_bell_ortho} for WENO$_3$ (left panels) and PPM$_5$ (right panels) for the middle and high resolutions together with the exact solution.
The distortion of the cosine bell is significantly reduced at the largest resolution ($128\times 256$).
Our results compare favourably to those obtained by other authors, see, for instance, \cite{Rasch.1994, HubNik.2003}.

Notice that all presented schemes do not achieve higher than second-order accuracy owing to the fact that, in 2D, the flux integral computed at a zone interface (e.g. Eqns. \ref{eq:flux_average} and \ref{eq:interface_flux}) is approximated using a midpoint quadrature rule.

\small
\begin{table}[!ht]
\caption{\footnotesize Errors measurements for the cosine bell test after one revolution for the selected reconstruction schemes using three different grid sizes.}
\label{tab:cosine_bell_err}
\centering
\footnotesize
\begin{tabular*}{\textwidth}{@{\extracolsep{\fill}} lr rrr}\hline
Method & $N_\theta\times N_\phi$  
       &  $l_1\left(Q\right)$ & $l_2\left(Q\right)$ 
       &  $l_\infty\left(Q\right)$
  \\
 \noalign{\smallskip} \hline\noalign{\smallskip}
\hline PLM
 & 32 $\times $ 64 & 8.67E-001 & 6.02E-001 & 6.08E-001 \\ \noalign{\smallskip}
 & 64 $\times $ 128 & 3.92E-001 & 3.13E-001 & 3.17E-001 \\ \noalign{\smallskip}
 & 128 $\times $ 256 & 1.28E-001 & 1.07E-001 & 1.11E-001 \\ \noalign{\smallskip}
\hline WENO$_3$
 & 32 $\times $ 64 & 7.41E-001 & 5.12E-001 & 5.15E-001 \\ \noalign{\smallskip}
 & 64 $\times $ 128 & 2.67E-001 & 2.12E-001 & 2.11E-001 \\ \noalign{\smallskip}
 & 128 $\times $ 256 & 5.66E-002 & 4.43E-002 & 5.19E-002 \\ \noalign{\smallskip}
\hline PPM$_3$
 & 32 $\times $ 64 & 7.68E-001 & 5.59E-001 & 5.65E-001 \\ \noalign{\smallskip}
 & 64 $\times $ 128 & 2.88E-001 & 2.42E-001 & 2.55E-001 \\ \noalign{\smallskip}
 & 128 $\times $ 256 & 5.61E-002 & 5.21E-002 & 6.86E-002 \\ \noalign{\smallskip}
\hline PPM$_4$
 & 32 $\times $ 64 & 6.38E-001 & 4.81E-001 & 4.93E-001 \\ \noalign{\smallskip}
 & 64 $\times $ 128 & 1.80E-001 & 1.50E-001 & 1.96E-001 \\ \noalign{\smallskip}
 & 128 $\times $ 256 & 5.40E-002 & 4.93E-002 & 7.80E-002 \\ \noalign{\smallskip}
\hline PPM$_5$
 & 32 $\times $ 64 & 5.77E-001 & 4.49E-001 & 4.63E-001 \\ \noalign{\smallskip}
 & 64 $\times $ 128 & 1.33E-001 & 1.14E-001 & 1.48E-001 \\ \noalign{\smallskip}
 & 128 $\times $ 256 & 3.81E-002 & 3.65E-002 & 5.95E-002 \\ \noalign{\smallskip}
\hline
\end{tabular*}
\end{table}
\normalsize

\clearpage
\subsection{Nonlinear systems of equations}
\label{sec:nonlinear_tests}
%
%
%

The performance of the reconstruction schemes is now investigated on problems requiring the solution of nonlinear system of conservation laws.
Although conservative variables are evolved in time, the reconstruction is performed on the volume averages of primitive variables following the procedure outlined in Section \ref{sec:nonlinear_systems}.
This choice has shown to yield less oscillatory results in some of the problems below (in particular, see Section \ref{sec:radial_wind}).
Unless otherwise stated, the interface flux  (\ref{eq:interface_flux}) is approximated with a simple Rusanov Lax-Friedrichs scheme \cite{Rusanov.1961} with local speed estimate.
The time step in Eq. (\ref{eq:dt}) is computed using the maximum characteristic speed of the system. 
 
\subsubsection{Homologous dust collapse}

\begin{figure}[!h]
 \centering
 \includegraphics[width=0.4\textwidth]{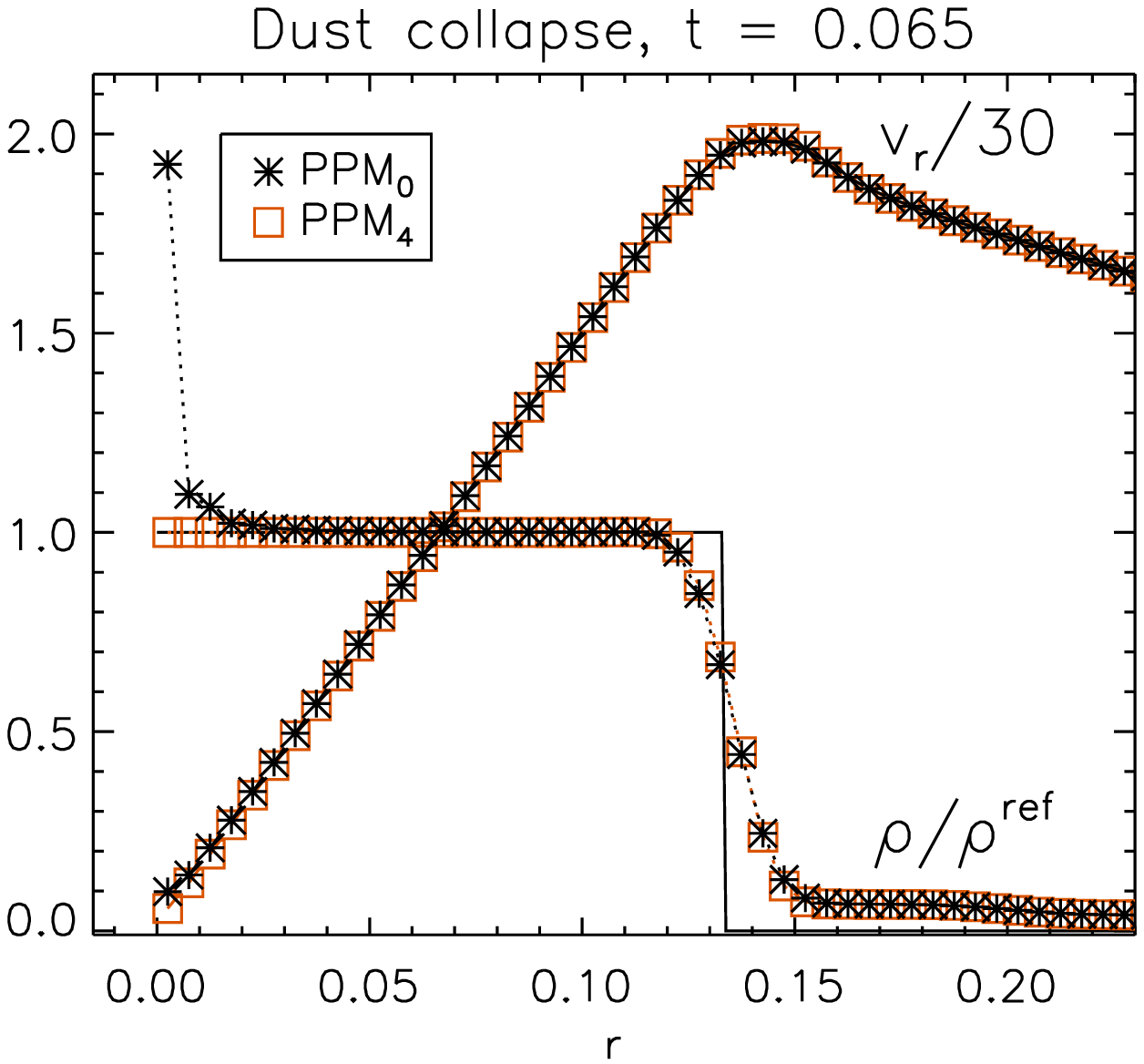}
 \includegraphics[width=0.4\textwidth]{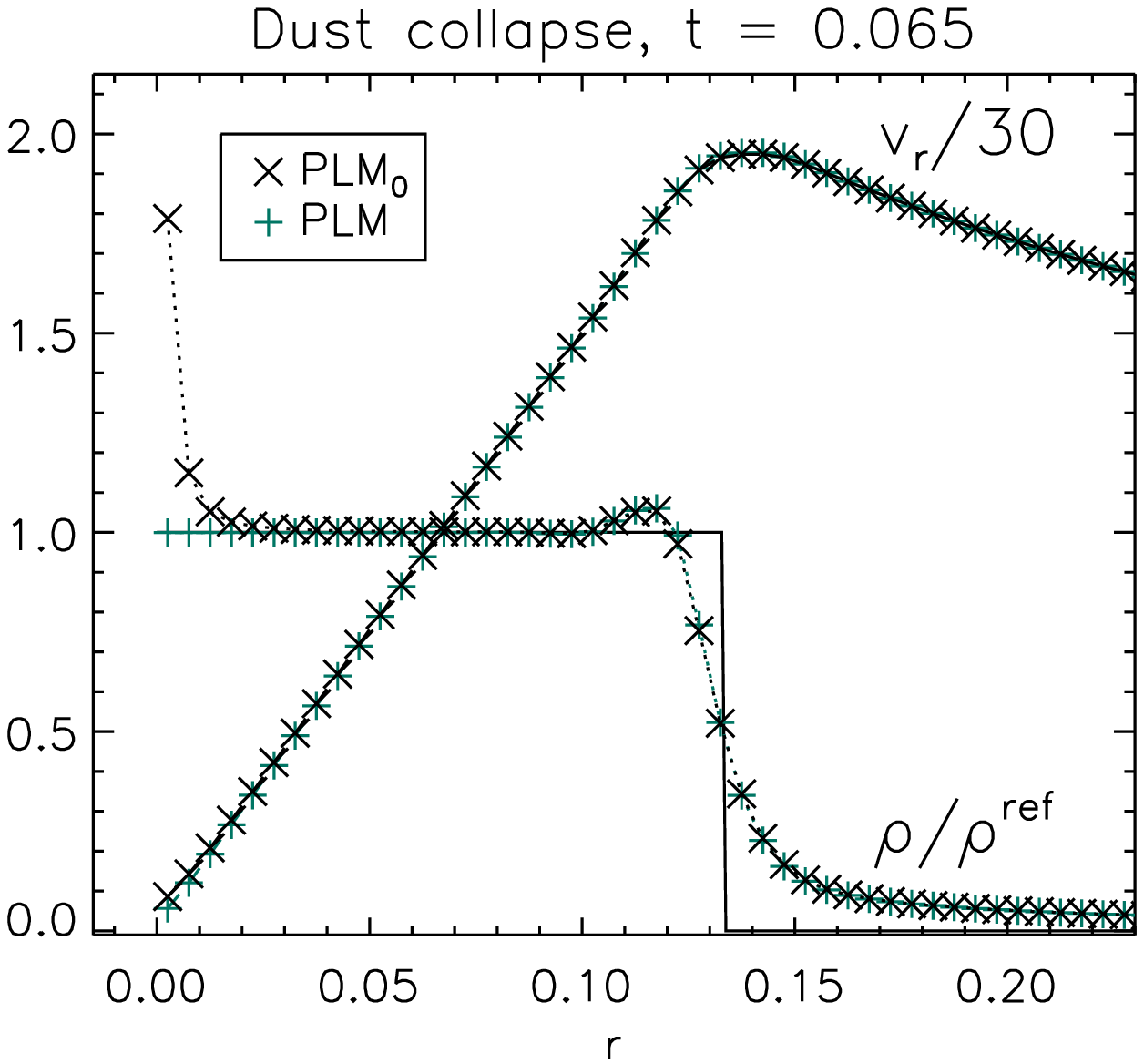}
 \caption{\footnotesize Numerical solutions of the homologous pressureless collapse at $t=0.065$ showing density and velocity.
 Left panel: comparison between PPM$_5$ and PPM$_0$. Right panel: comparison between PLM and PLM$_0$.
 In all panels the reference solution for the density is shown as a solid line.}
 \label{fig:dust_collapse}
\end{figure}

The spherically symmetric collapse of a dust sphere under the influence of its own gravity (\cite{Colgate_White.1966}) is a valuable benchmark demonstrating the importance of geometrical corrections in a finite difference or FV scheme in curvilinear coordinates \cite{MonchMull.1989, BossMyhill.1992}.

The problem consists of a unit sphere initially at rest filled with constant density $\rho=1$ obeying the laws of a pressure-less fluid (dust) and complemented by the Poisson equation for the gravitational potential:
\begin{equation}\label{eq:Dust1D}
\left\{\begin{array}{lcl}
\DS  \pd{\rho}{t} + \frac{1}{r^2}\pd{}{r}\left(\rho v_r r^2\right) & = & 0   \,,
\\ \noalign{\bigskip}
\DS  \pd{(\rho v_r)}{t} + \frac{1}{r^2}\pd{}{r}\left(\rho v^2_r r^2\right) 
 & = & \DS -\rho\Psi'   \,,
\\ \noalign{\bigskip}
\DS \frac{1}{r^2}\pd{}{r}(r^2\Psi')  & = & 4\pi G\rho  \,,
\end{array}\right.
\end{equation}
where $\Psi' = \partial_r\Psi$ is the gravitational acceleration, $\Psi$ is the gravitational potential and $G$ is the gravitational constant.

The problem has an analytical solution \citep{Colgate_White.1966} demonstrating  that the collapse proceeds while preserving a linear velocity profile ($v_r\propto -r$) and a uniform constant density inside a sphere of radius $r(t)$ given by the relation
\begin{equation}\label{eq:dust_radius}
  \left(\frac{8\pi G}{3}\rho_0\right)^{1/2}\; t = 
  \left(\frac{r(t)}{r_0}\right)^{1/2}\left(1-\frac{r(t)}{r_0}\right)^{1/2} 
   + \sin^{-1}\left(1-\frac{r(t)}{r_0}\right)^{1/2}\,.
\end{equation}
where $\rho_0$ and $r_0$ are the initial gas density and radius.
As pointed out by \cite{MonchMull.1989}, a numerical code with second- (or higher-) order spatial accuracy should reproduce the profiles exactly.

Eqs. (\ref{eq:Dust1D}) are solved until $t=0.065$ on the domain $r\in[0,1]$ using $200$ zones and a CFL number $C_a=0.4$.
Reflective boundary conditions are applied at $r=0$ while at the sphere boundary ($r=1$) density has zero gradient and the velocity is set to scale as $\sim 1/\sqrt{r}$, appropriate for a free-fall trajectory.
Numerical values can be scaled to actual physical units by introducing a reference density $\rho_0$, length $L_0$ and velocity $V_0$ and modifying the constant in the Poisson solver by letting $4\pi G\to 4\pi G\rho_0(L_0/V_0)^2$.
Following \cite{MonchMull.1989}, we use $\rho_0 = 10^{9}\,{\rm gr/cm}^3$, $L_0 = 6.5\times 10^8\,{\rm cm}$ and $V_0 = L_0/t_0$ where $t_0 = 1\,{\rm sec}$ is our time reference unit.

The gravitational acceleration term $\Psi'$ is readily obtained at cell interfaces by integrating the Poisson equation:
\begin{equation}\label{eq:dust:Psi'}
  \Psi'_{i+\HALF} = \frac{4\pi G}{r_{i+\HALF}^2}
  \sum_{l=1}^i \rho_l\Delta\vol_l   \,.
\end{equation}
For the purpose of the test, it is sufficient to compute the source term in the momentum equation using second-order accuracy.
Since the gravitational acceleration is known at cell interface, a natural choice is to use the trapezoidal rule (\ref{eq:S:trapezoidal1}) giving
\begin{equation}
 -\frac{1}{\Delta\vol_i}\int_{i-\HALF}^{i+\HALF} \rho\Psi' r^2\, dr \approx 
  -\left( \frac{r_{i+\HALF} - \bar{r}_i}{\Delta r}\rho^{-}_{i}\,\Psi'_{i-\HALF} 
         +\frac{\bar{r}_i - r_{i-\HALF}}{\Delta r}\rho^{+}_{i}\,\Psi'_{i+\HALF} \right)
 \,,
\end{equation}
where $\Psi'_{i+\HALF}$ is obtained from (\ref{eq:dust:Psi'}).

Fig. \ref{fig:dust_collapse} compares the results obtained with the fourth-order PPM and linear (PLM) schemes with the corresponding uncorrected versions.
The density reference solution (overplotted) can be found at any time $t$ by solving Eq. (\ref{eq:dust_radius}) for $r(t)$ and then using mass conservation: $\rho^{\rm ref}(t) = \rho_0 r^3_0/r^3(t)$.
The density plateau and the linear velocity profiles are correctly preserved close to the origin for the corrected schemes.
Results obtained with the remaining schemes behave similarly to PPM$_4$ and are not shown to avoid cluttered plots.
On the contrary, the uncorrected versions (PPM$_0$ and PLM$_0$) show a systematic accumulation of mass near the center and a deviation from linearity in the velocity profile (although to a less degree) owing to an incorrect numerical discretization.
Our results favourably compare to those obtained by other investigators using second-order finite difference schemes, e.g., \cite{MonchMull.1989, BossMyhill.1992, Castro.2010}.

\subsubsection{Radial wind test problem}
\label{sec:radial_wind}

\begin{figure}[!ht]
 \centering
 \includegraphics[width=0.4\textwidth]{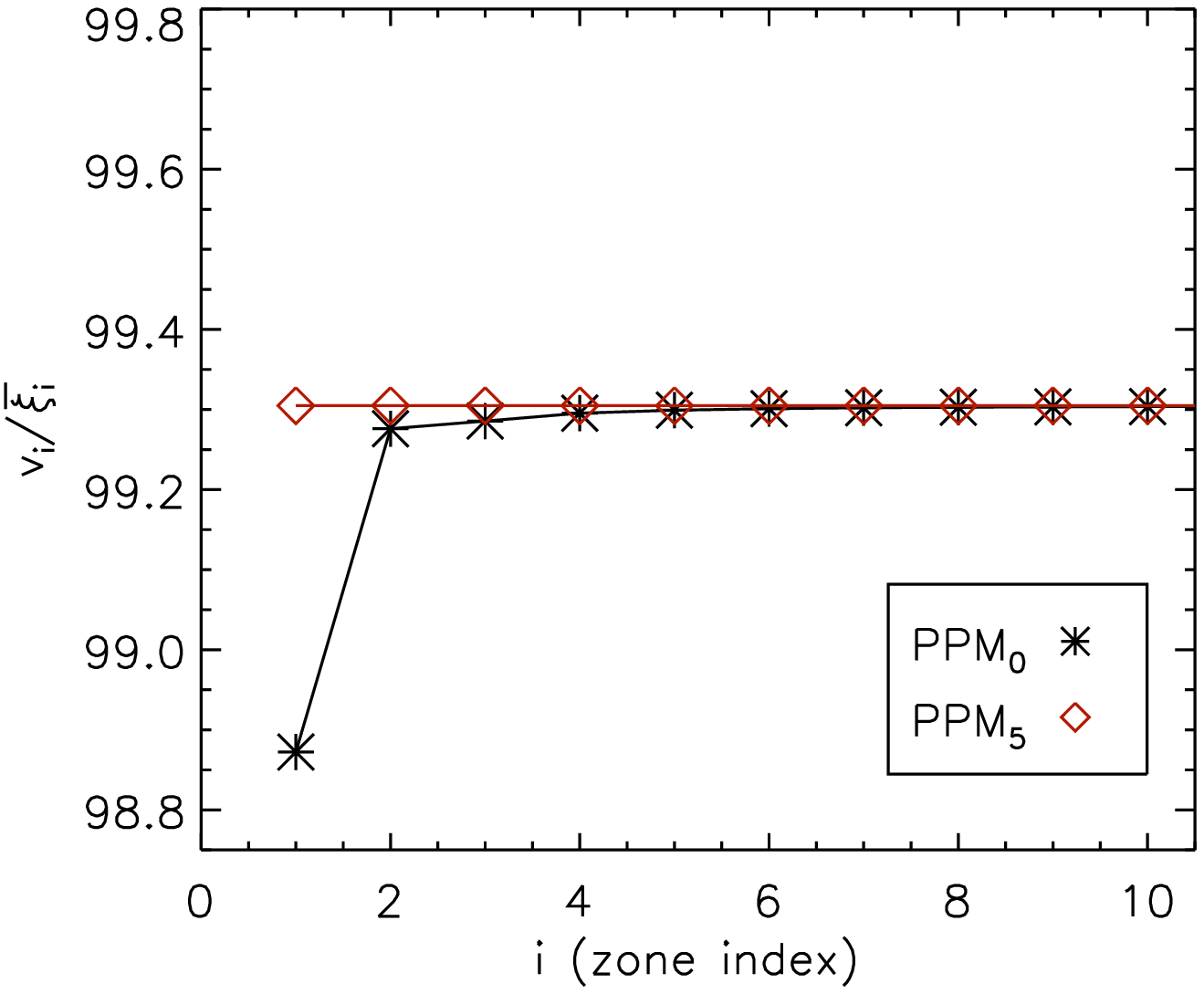}%
 \includegraphics[width=0.4\textwidth]{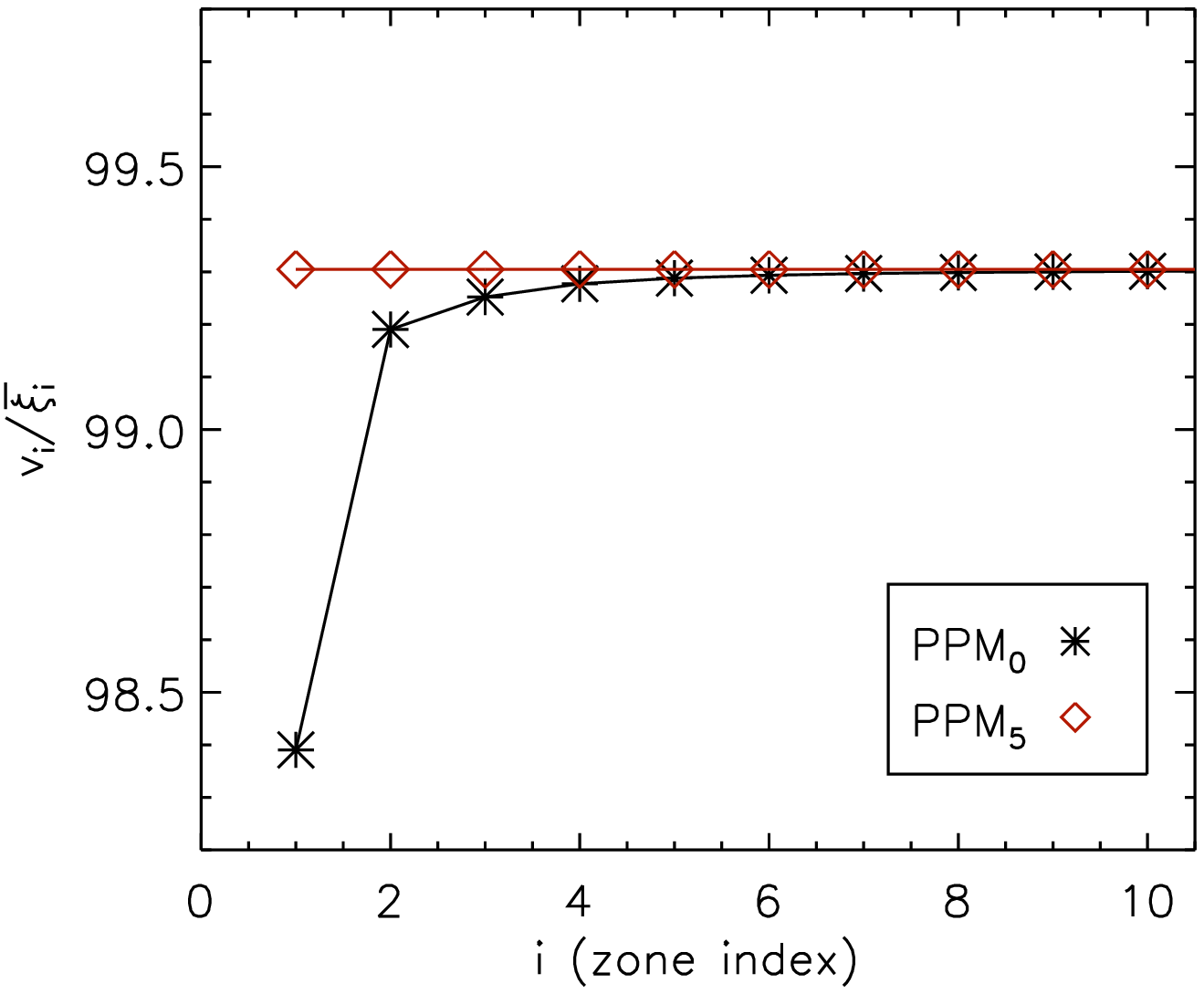}
 \caption{\footnotesize Numerical solution for the isothermal radial wind problem with constant density after one time step using $\alpha_0=100$ in cylindrical (left panel) and spherical (right panel) coordinates.
 The profiles show $v/\bar{\xi}$ for the original PPM$_0$ scheme with no geometrical correction (stars) and the PPM$_5$ scheme (diamonds).}
 \label{fig:rad_wind_one_step}
\end{figure}

\begin{figure}[!h]
 \centering
 \includegraphics[width=0.4\textwidth]{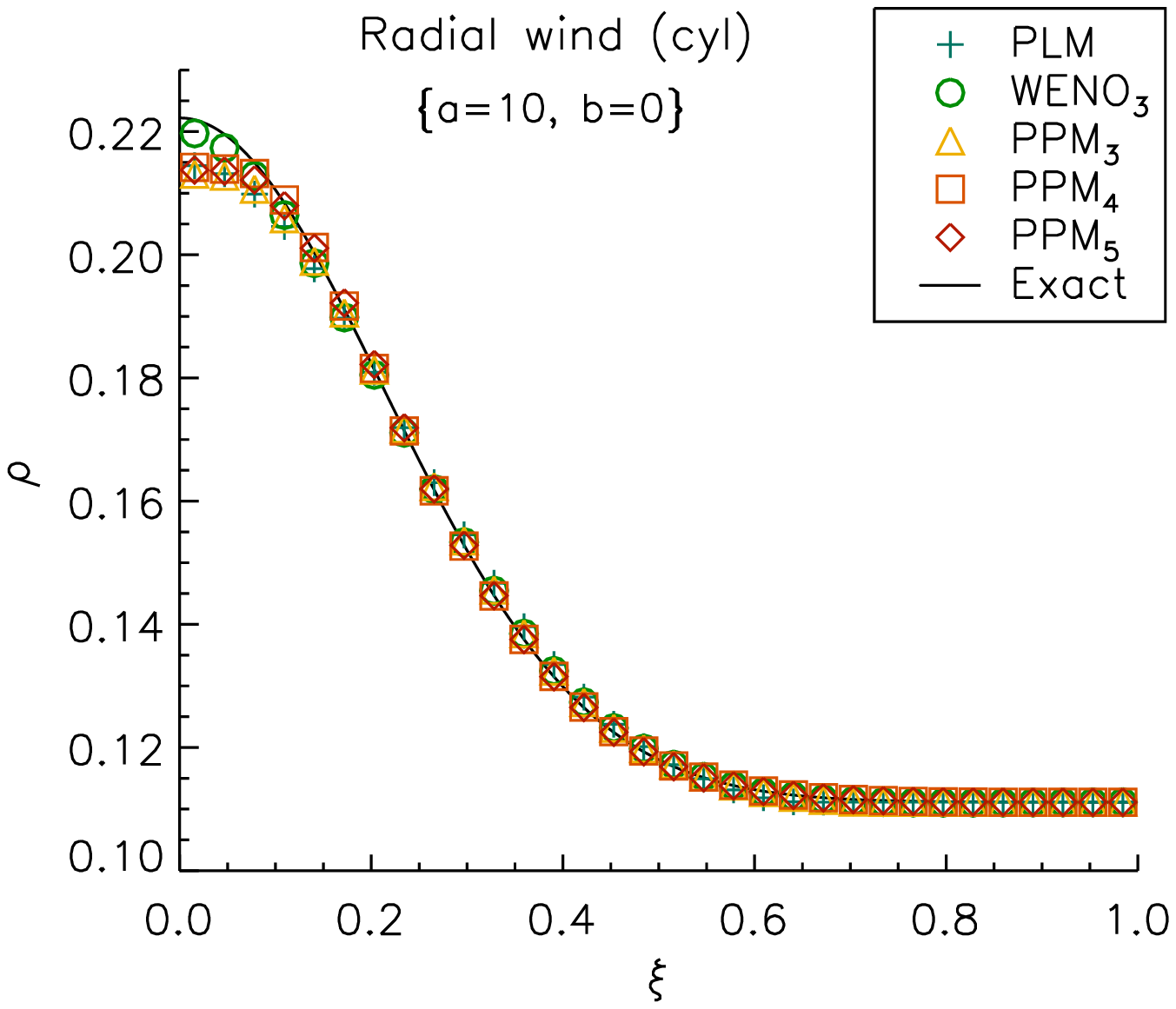}%
 \includegraphics[width=0.4\textwidth]{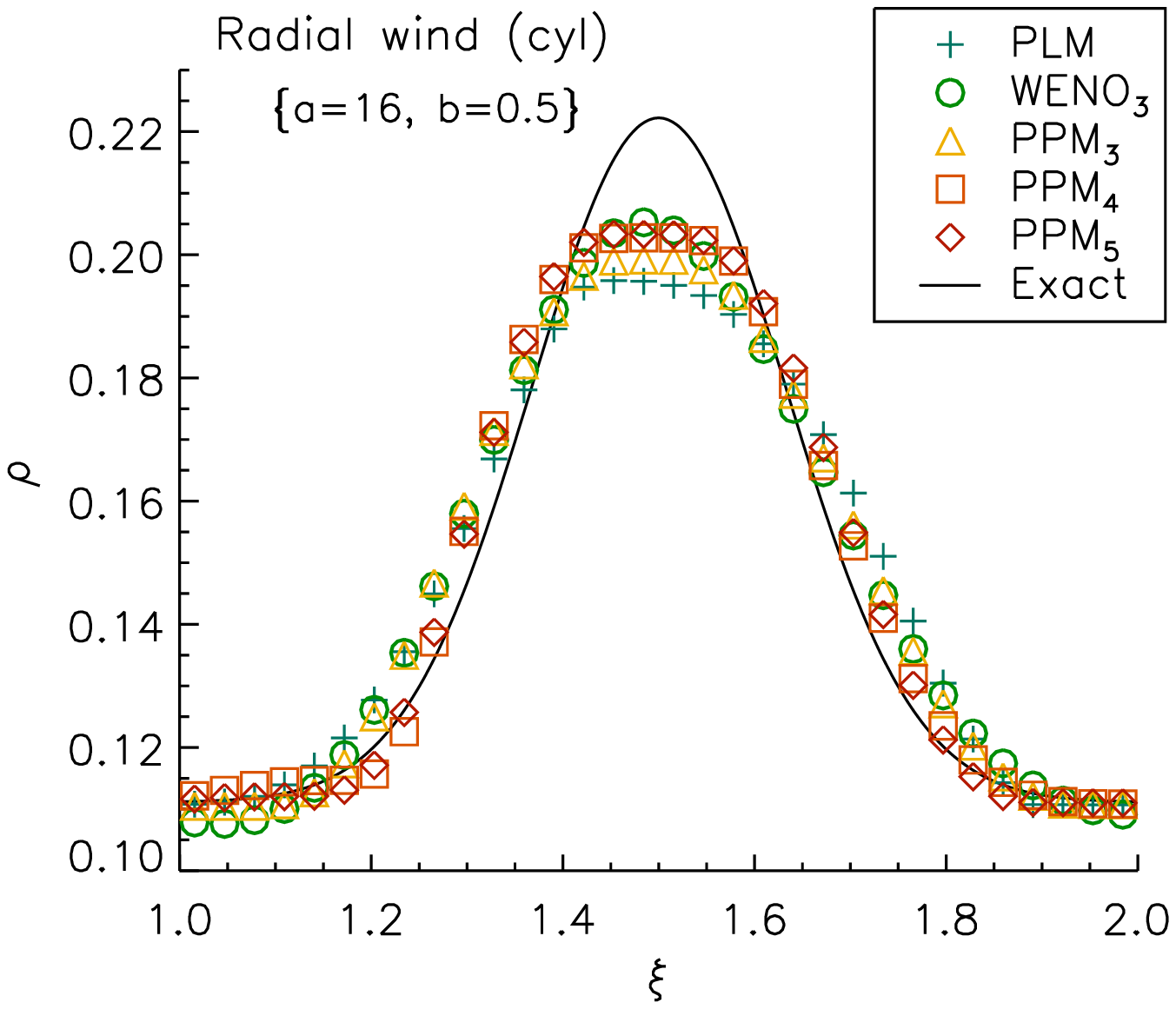}

 \vspace{12pt}

 \includegraphics[width=0.4\textwidth]{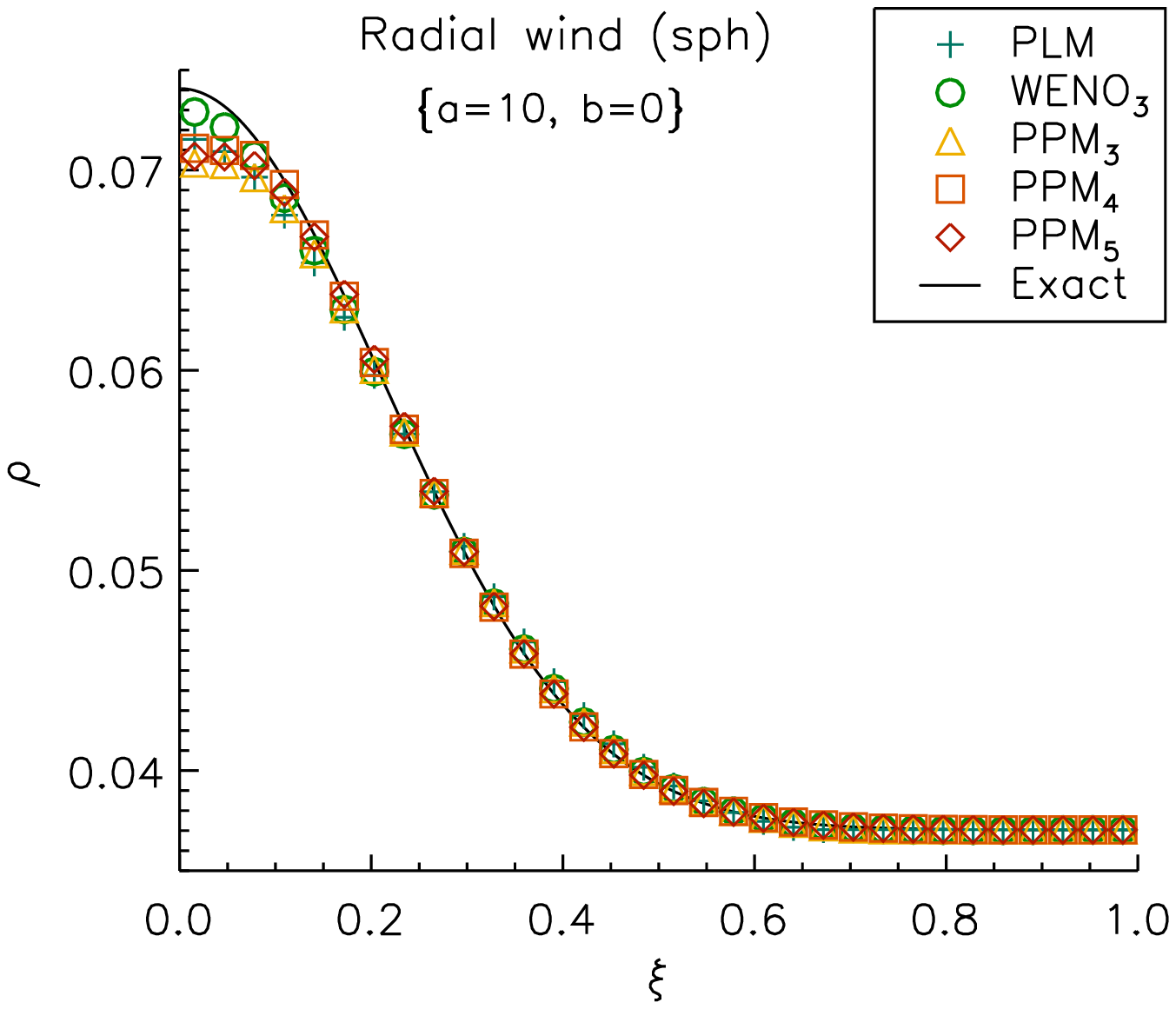}%
 \includegraphics[width=0.4\textwidth]{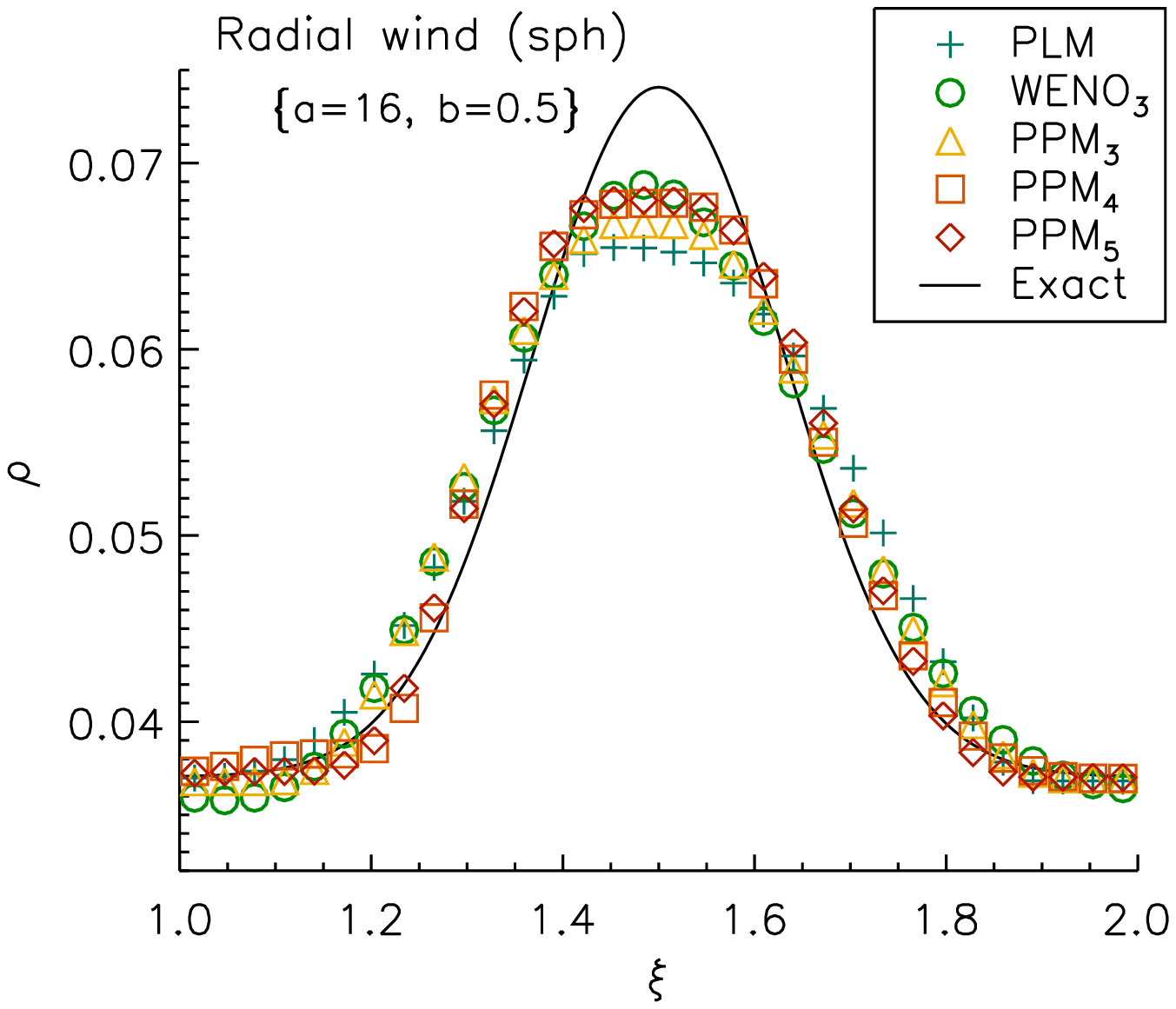}
 \caption{\footnotesize Spatial profiles at $t=0.4$ for the radial wind problem using $N=64$ grid zones in cylindrical (top panels) and spherical coordinates (bottom panels).
 Left and right panels refer, respectively, to computations carried out with $a=10$, $b=0$ (left) and $a=16$, $b=1/2$ (right). 
 For the sake of clarity, only a smaller portion of the computational domain is shown.}
 \label{fig:rad_wind_pro}
\end{figure}

The radial advection problem presented in Section \ref{sec:radial_advection} can be generalized to the Euler equations of gas dynamics in one dimension,
\begin{equation}\label{eq:Euler1D}
  \pd{}{t}\left(\begin{array}{l}
  \rho      \\ \noalign{\medskip}
  \rho v  \\ \noalign{\medskip}
  E        
\end{array}\right)
 +
 \frac{1}{\xi^m} \pd{}{r}
\left(\begin{array}{c}
   \rho v \xi^m \\ \noalign{\medskip}
   (\rho v^2 + p)\xi^m \\ \noalign{\medskip}
   (E+p)v \xi^m
\end{array}\right)
  = 
\left(\begin{array}{c}
   0    \\ \noalign{\medskip}
   mp/\xi \\ \noalign{\medskip}
   0
\end{array}\right) \,,
\end{equation}
where $\rho$ is the mass density, $v$ is the radial velocity, $p$ is the gas pressure, $E$ is the total energy density and $m=0,1,2$ for Cartesian, cylindrical or spherical (radial) coordinates.
For an isothermal flow, the energy equation is discarded whereas for an adiabatic equation of state one has
\begin{equation}\label{eq:Energy}
  E = \frac{p}{\Gamma-1} + \frac{1}{2}\rho v^2\,,
\end{equation}
where $\Gamma=5/3$ is assumed.
 
The initial condition consists of a self-similar radial outflow with spatially varying density, a linear velocity profile and a constant pressure:
\begin{equation}
  \rho(\xi,0) = \rho_0(\xi)\,;\qquad
  v(\xi,0)  = \alpha_0 \xi\,;\qquad
  p(\xi,0)    = \frac{1}{\Gamma}  \,,
\end{equation}
where $\rho_0(\xi)$ is an arbitrary function and $\alpha_0$ is a constant.
This problem has an exact analytical solution which can be written as
\begin{equation}\label{eq:rad_wind_exact_sol}
  \rho^{\rm ref}(\xi,t) = \left(\frac{\alpha(t)}{\alpha_0}\right)^{1+m}
                                 \rho_0\left(\xi\frac{\alpha(t)}{\alpha_0}\right)
  \,;\qquad
  v^{\rm ref}(\xi,t)  = \alpha(t)\xi
   \,;\qquad
  p^{\rm ref}(\xi,t)    = \frac{1}{\Gamma}
                   \left(\frac{\alpha(t)}{\alpha_0}\right)^{\Gamma(m+1)}   \,,
\end{equation}
while $\alpha(t) = \alpha_0/(1 + \alpha_0 t)$.

Computations are carried out on the interval $0\le \xi \le 2$ using $N$ equally-spaced zones using a  Courant number $C_a = 0.9$.
A five-point Gaussian quadrature rule is used to assign the initial volume averages for density, momentum and energy density.
At $\xi=0$ axisymmetric boundary conditions apply while at the outer edge density, pressure and $v/\xi$ have zero gradient.

As a first benchmark, Eqs. (\ref{eq:Euler1D}) are solved in cylindrical and spherical coordinates with $N=100$, $\alpha_0=100$, $\rho_0(\xi) = 1$ and an isothermal Equation of state for a direct comparison with the results of \cite{BloLuf.1993}.
The spatial profiles of $v/\bar{\xi}$ are plotted in Fig. \ref{fig:rad_wind_one_step} after one integration step $\Delta t = 7\times 10^{-5}$ using the traditional PPM$_0$ and the PPM$_5$ schemes (here $\bar{\xi}$ is the centroid of volume defined by Eq. \ref{eq:centroid_of_volume}).
For the sake of clarity, only the first computational zones are shown while the results produced with the other schemes are identical to PPM$_5$ and have been omitted.
Since density is constant and velocity remains linear at all times (see Eq. \ref{eq:rad_wind_exact_sol}) the error should be set only by the temporal accuracy of the scheme and not by the spatial reconstruction for second- or higher-order methods.
This expectation is indeed fulfilled by the proposed geometrically corrected  methods whereas the original PPM scheme shows significant deviations close to the origin.
The same conclusions have been drawn in \cite{BloLuf.1993}.

\begin{figure}[!h]
 \centering
 \includegraphics[width=0.4\textwidth]{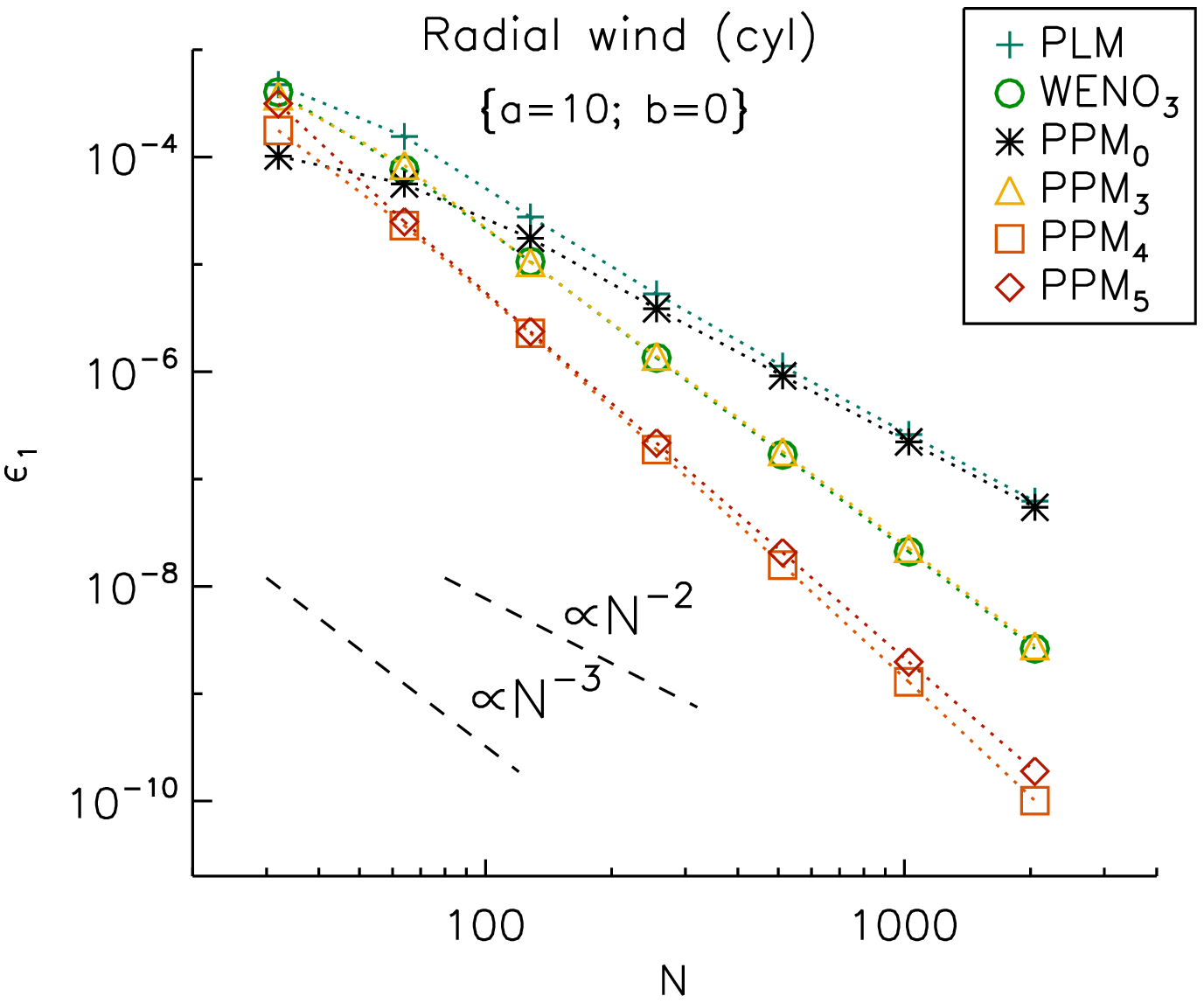}%
 \includegraphics[width=0.4\textwidth]{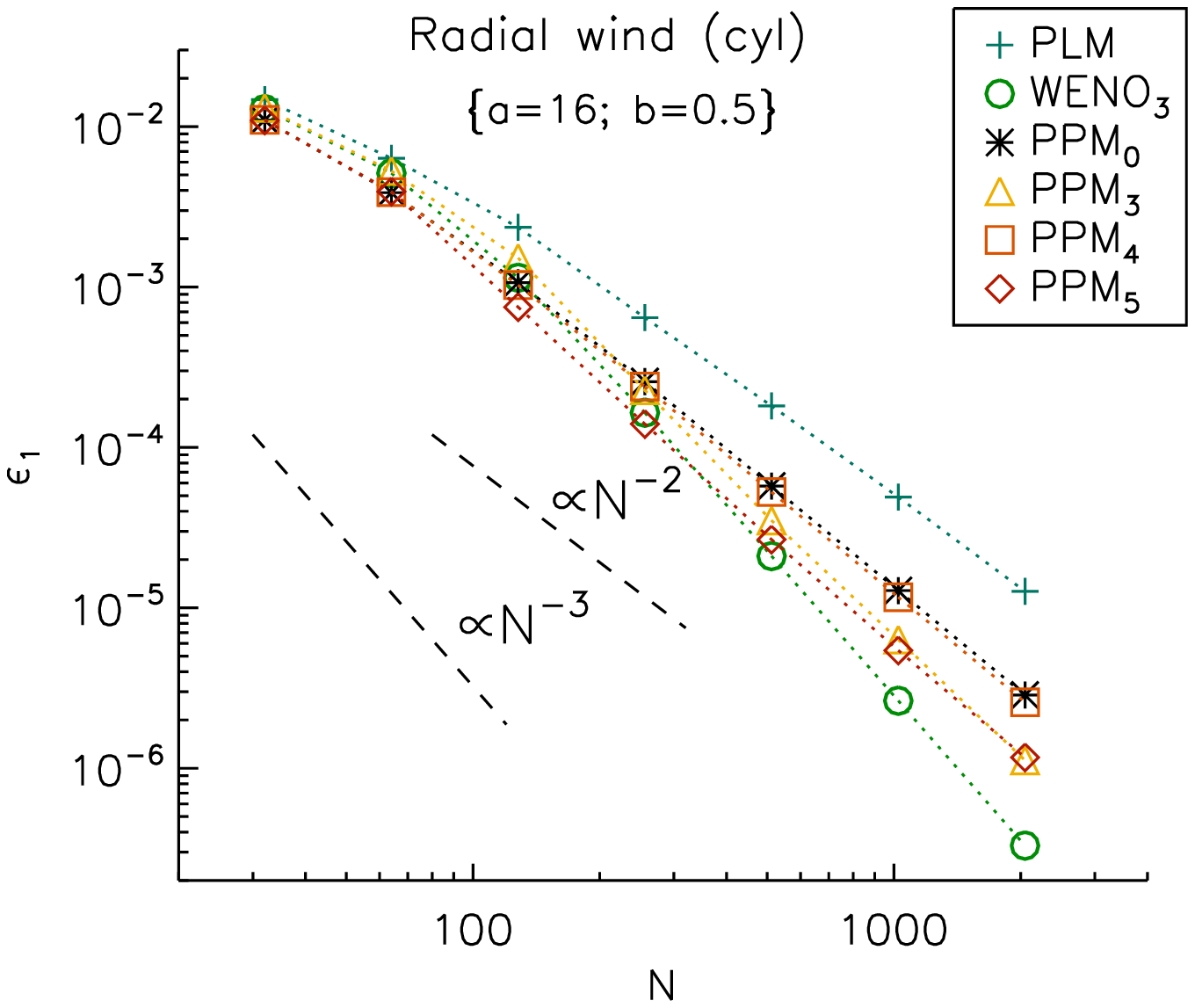}

 \vspace{12pt}

 \includegraphics[width=0.4\textwidth]{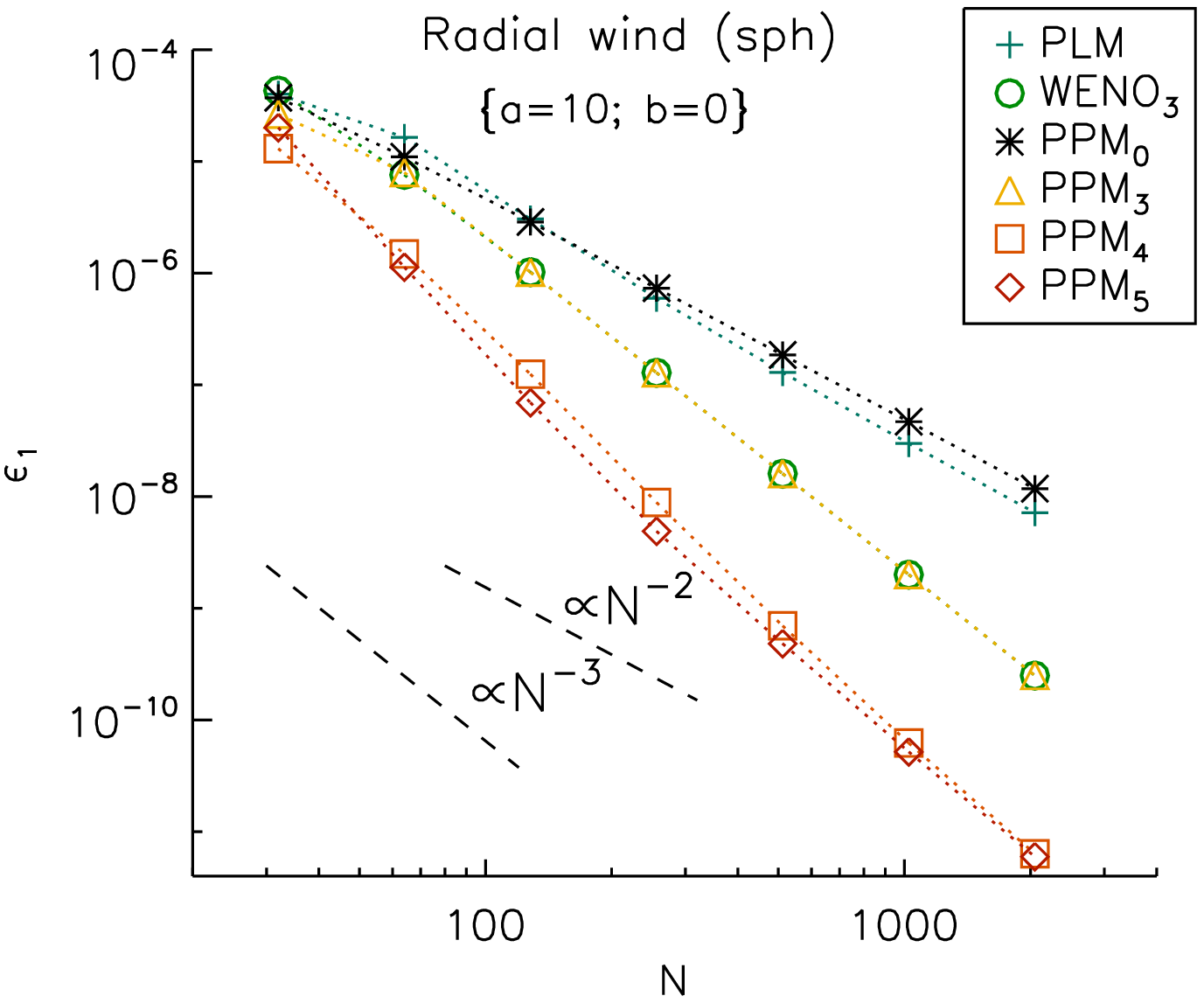}%
 \includegraphics[width=0.4\textwidth]{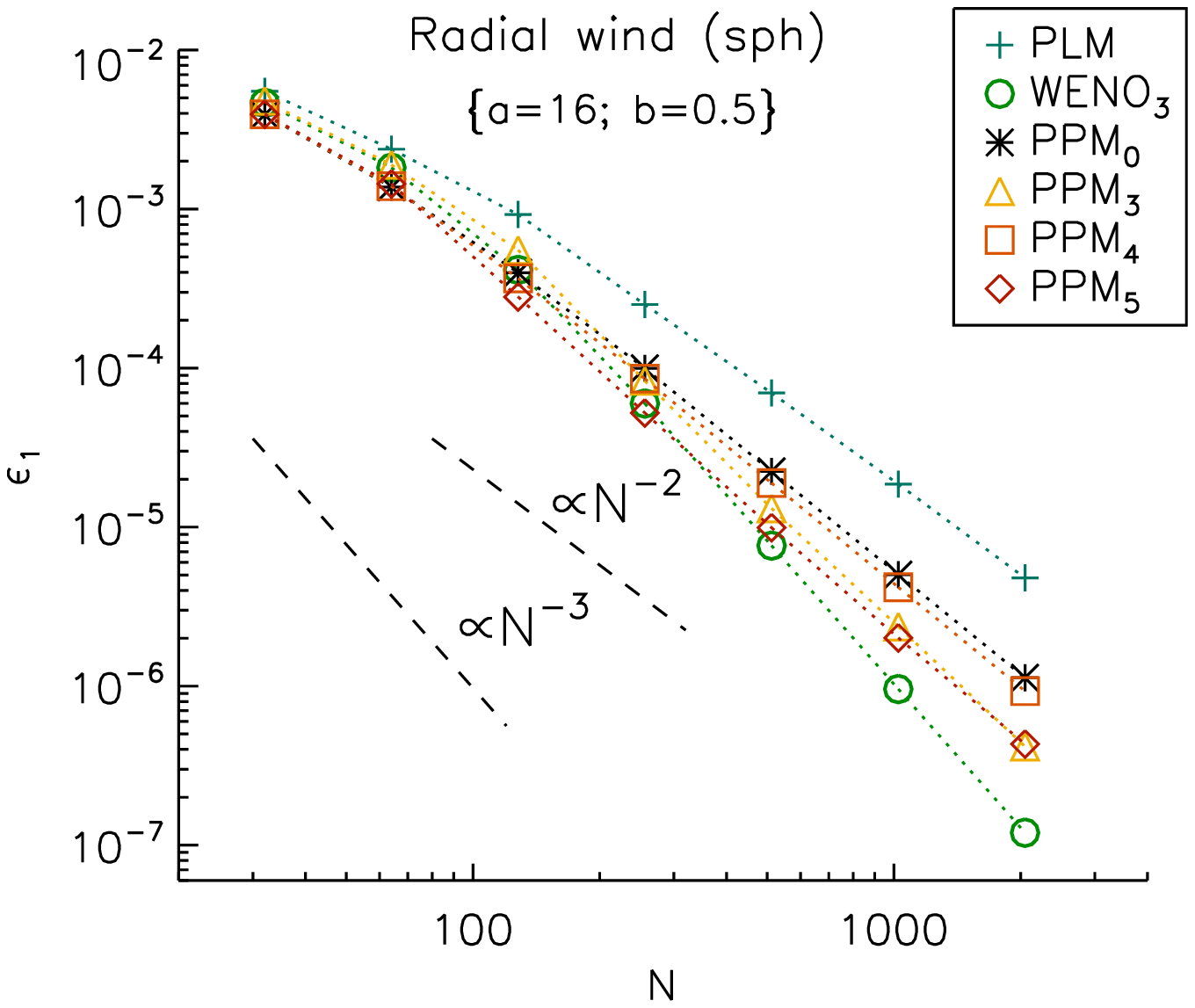}
 \caption{\footnotesize $L_1$ error measurements for the radial wind test problem in cylindrical coordinates (top panels) and spherical coordinates (bottom panels) as function of the resolution. 
 Left and right panels refer, respectively, to computations performed with $a=10$, $b=0$ (left) and $a=16$, $b=1/2$ (right).}
 \label{fig:rad_wind_err}
\end{figure}

As a second benchmark, the Euler equations are evolved with the adiabatic Equation of state (\ref{eq:Energy}) and initial condition given by $\alpha_0 = 5$ and $\rho_0(\xi) = 1 + \exp[-a^2(\xi - b)^2]$ where $a$ and $b$ are constants.
In analogy to Section \ref{sec:radial_advection}, computations are carried out using $\{a=10,\,b=0\}$ corresponding to a monotonically decreasing profile and $\{a=16,\, b=1/2\}$ corresponding to a Gaussian profile with a maximum at $r=b$.
The density profiles obtained with the different schemes in the two cases are plotted, together with the exact solution, in Fig. \ref{fig:rad_wind_pro} at $t=0.4$ on $64$ equidistant zones.
Fig. \ref{fig:rad_wind_err} shows the $L_1$ norm errors for density at different resolutions for the selected reconstruction schemes in cylindrical and spherical coordinates at $t=0.4$.
The corresponding numerical values and the order of convergence are listed in Table \ref{tab:rad_wind}.

In the monotonic profile case, PPM$_5$ and PPM$_4$ achieve the best accuracy in both cylindrical and spherical coordinates with $\epsilon_1\lesssim 10^{-10}$ (at the largest resolution) and order of convergence between $3$ and $4$, followed by PPM$_3$ and WENO$_3$ with $10^{-10}\lesssim\epsilon_1\lesssim 10^{-9}$ and ${\cal O}_{L_1}\sim 3$ and, lastly, by linear interpolation ($\epsilon\approx 6\times 10^{-8}$).
On the contrary, the uncorrected PPM$_0$ scheme performs very poorly for this particular configuration and converges linearly in analogy with the findings of Section \ref{sec:radial_advection}.

In the non-monotonic profile case, WENO$_3$ converges with the expected third-order accuracy yielding the smallest errors at the largest resolution, that is, $\epsilon_1\approx 3\times 10^{-7}$ and $\epsilon_1\approx 10^{-7}$ in cylindrical and spherical coordinates, respectively.
Third- and fifth-order PPM schemes have order of convergence between $2$ and $3$ with very similar errors, $\epsilon\approx 10^{-6}$ and $\epsilon_1\approx 4\times10^{-7}$ in the two coordinate systems.
Finally, the original PPM$_0$ and PPM$_4$ do not show strong difference for this configuration as already observed in section \ref{sec:radial_advection}.

These results lead to the conclusion that geometrical corrections in the reconstruction algorithm are particularly important in regions of large curvature and strengthen the behavior already settled in Section \ref{sec:radial_advection} for a scalar conservation law. 

It is worth mentioning that several numerical experiments (not shown here) have evidenced that straightforward application of the proposed reconstruction schemes to conservative variables (rather than primitive as done here) leads to considerable numerical oscillations for this particular test problem.


\begin{table}[!ht]
\caption{\footnotesize Errors and orders of convergence (in $L_1$ norm) for the radial wind problem in cylindrical (column 3-6) and spherical (columns 7-10) coordinates at $t=0.4$ for selected reconstruction schemes.
Errors are given for different sets of the constants $\{a,b\}$ used to define the initial density profile.}
\label{tab:rad_wind}
\centering
\footnotesize
\begin{tabular*}{\textwidth}{@{\extracolsep{\fill}} lr rrrr rrrr}\hline
        &   &  \multicolumn{4}{c}{Cylindrical} 
            &  \multicolumn{4}{c}{Spherical} \\
  \cline{3-6} \cline{7-10}
        &   &  \multicolumn{2}{c}{$\{a=10,b=0\}$} 
            &  \multicolumn{2}{c}{$\{a=16,b=1/2\}$} 
            &  \multicolumn{2}{c}{$\{a=10,b=0\}$} 
            &  \multicolumn{2}{c}{$\{a=16,b=1/2\}$} \\
  \cline{3-4} \cline{5-6} \cline{7-8} \cline{9-10}
Method & $N_r$  & $\epsilon_1\left(Q\right)$  & ${\cal O}_{L_1}$
                & $\epsilon_1\left(Q\right)$  & ${\cal O}_{L_1}$ 
                & $\epsilon_1\left(Q\right)$  & ${\cal O}_{L_1}$
                & $\epsilon_1\left(Q\right)$  & ${\cal O}_{L_1}$ \\
 \noalign{\smallskip} \hline\noalign{\smallskip}
\hline PLM
 & 32 & 4.73E-004 & - & 1.47E-002 & - & 4.01E-005 & - & 5.51E-003 & -\\ \noalign{\smallskip}
 & 64 & 1.56E-004 & 1.61 & 6.35E-003 & 1.21 & 1.65E-005 & 1.29 & 2.38E-003 & 1.21\\ \noalign{\smallskip}
 & 128 & 2.77E-005 & 2.49 & 2.36E-003 & 1.43 & 3.05E-006 & 2.43 & 9.25E-004 & 1.36\\ \noalign{\smallskip}
 & 256 & 5.31E-006 & 2.38 & 6.44E-004 & 1.87 & 5.96E-007 & 2.36 & 2.51E-004 & 1.88\\ \noalign{\smallskip}
 & 512 & 1.13E-006 & 2.23 & 1.82E-004 & 1.83 & 1.29E-007 & 2.21 & 6.97E-005 & 1.85\\ \noalign{\smallskip}
 & 1024 & 2.60E-007 & 2.12 & 4.90E-005 & 1.89 & 2.99E-008 & 2.11 & 1.86E-005 & 1.90\\ \noalign{\smallskip}
 & 2048 & 6.22E-008 & 2.06 & 1.27E-005 & 1.95 & 7.17E-009 & 2.06 & 4.80E-006 & 1.96\\ \noalign{\smallskip}
\hline WENO$_3$
 & 32 & 4.03E-004 & - & 1.29E-002 & - & 4.31E-005 & - & 4.66E-003 & -\\ \noalign{\smallskip}
 & 64 & 7.64E-005 & 2.40 & 5.12E-003 & 1.33 & 7.59E-006 & 2.50 & 1.82E-003 & 1.35\\ \noalign{\smallskip}
 & 128 & 1.06E-005 & 2.85 & 1.14E-003 & 2.16 & 1.02E-006 & 2.89 & 4.15E-004 & 2.14\\ \noalign{\smallskip}
 & 256 & 1.35E-006 & 2.98 & 1.65E-004 & 2.79 & 1.28E-007 & 2.99 & 5.99E-005 & 2.79\\ \noalign{\smallskip}
 & 512 & 1.69E-007 & 3.00 & 2.10E-005 & 2.97 & 1.60E-008 & 3.00 & 7.65E-006 & 2.97\\ \noalign{\smallskip}
 & 1024 & 2.10E-008 & 3.00 & 2.64E-006 & 3.00 & 2.00E-009 & 3.00 & 9.58E-007 & 3.00\\ \noalign{\smallskip}
 & 2048 & 2.63E-009 & 3.00 & 3.30E-007 & 3.00 & 2.50E-010 & 3.00 & 1.20E-007 & 3.00\\ \noalign{\smallskip}
\hline PPM$_3$
 & 32 & 3.73E-004 & - & 1.31E-002 & - & 2.65E-005 & - & 4.79E-003 & -\\ \noalign{\smallskip}
 & 64 & 8.39E-005 & 2.15 & 5.28E-003 & 1.31 & 7.96E-006 & 1.73 & 1.90E-003 & 1.33\\ \noalign{\smallskip}
 & 128 & 1.05E-005 & 3.00 & 1.52E-003 & 1.80 & 1.02E-006 & 2.97 & 5.52E-004 & 1.78\\ \noalign{\smallskip}
 & 256 & 1.40E-006 & 2.90 & 2.27E-004 & 2.74 & 1.28E-007 & 2.99 & 8.32E-005 & 2.73\\ \noalign{\smallskip}
 & 512 & 1.80E-007 & 2.95 & 3.50E-005 & 2.70 & 1.60E-008 & 3.00 & 1.31E-005 & 2.67\\ \noalign{\smallskip}
 & 1024 & 2.26E-008 & 2.99 & 6.33E-006 & 2.47 & 2.00E-009 & 3.00 & 2.37E-006 & 2.47\\ \noalign{\smallskip}
 & 2048 & 2.80E-009 & 3.01 & 1.12E-006 & 2.50 & 2.50E-010 & 3.00 & 4.17E-007 & 2.50\\ \noalign{\smallskip}
\hline PPM$_4$
 & 32 & 1.77E-004 & - & 1.10E-002 & - & 1.30E-005 & - & 3.94E-003 & -\\ \noalign{\smallskip}
 & 64 & 2.30E-005 & 2.94 & 3.90E-003 & 1.50 & 1.49E-006 & 3.13 & 1.39E-003 & 1.50\\ \noalign{\smallskip}
 & 128 & 2.28E-006 & 3.33 & 1.03E-003 & 1.93 & 1.25E-007 & 3.58 & 3.67E-004 & 1.92\\ \noalign{\smallskip}
 & 256 & 1.88E-007 & 3.60 & 2.40E-004 & 2.10 & 8.86E-009 & 3.81 & 8.57E-005 & 2.10\\ \noalign{\smallskip}
 & 512 & 1.58E-008 & 3.58 & 5.27E-005 & 2.18 & 6.96E-010 & 3.67 & 1.89E-005 & 2.18\\ \noalign{\smallskip}
 & 1024 & 1.29E-009 & 3.61 & 1.16E-005 & 2.18 & 6.21E-011 & 3.49 & 4.18E-006 & 2.18\\ \noalign{\smallskip}
 & 2048 & 1.01E-010 & 3.68 & 2.58E-006 & 2.17 & 6.33E-012 & 3.29 & 9.30E-007 & 2.17\\ \noalign{\smallskip}
\hline PPM$_5$
 & 32 & 3.15E-004 & - & 1.09E-002 & - & 2.01E-005 & - & 3.95E-003 & -\\ \noalign{\smallskip}
 & 64 & 2.50E-005 & 3.65 & 3.93E-003 & 1.47 & 1.13E-006 & 4.15 & 1.44E-003 & 1.45\\ \noalign{\smallskip}
 & 128 & 2.37E-006 & 3.40 & 7.50E-004 & 2.39 & 6.92E-008 & 4.03 & 2.80E-004 & 2.36\\ \noalign{\smallskip}
 & 256 & 2.17E-007 & 3.45 & 1.40E-004 & 2.42 & 4.89E-009 & 3.82 & 5.23E-005 & 2.42\\ \noalign{\smallskip}
 & 512 & 2.06E-008 & 3.39 & 2.66E-005 & 2.39 & 4.81E-010 & 3.35 & 9.92E-006 & 2.40\\ \noalign{\smallskip}
 & 1024 & 1.98E-009 & 3.38 & 5.43E-006 & 2.30 & 5.18E-011 & 3.22 & 2.01E-006 & 2.30\\ \noalign{\smallskip}
 & 2048 & 1.90E-010 & 3.38 & 1.17E-006 & 2.22 & 5.96E-012 & 3.12 & 4.32E-007 & 2.22\\ \noalign{\smallskip}
\hline
\end{tabular*}
\end{table}
\normalsize

\subsubsection{Magnetic confinement of a cylindrical plasma column}
%
%
%

\begin{figure}[!ht]
 \centering
 \includegraphics[width=0.4\textwidth]
                 {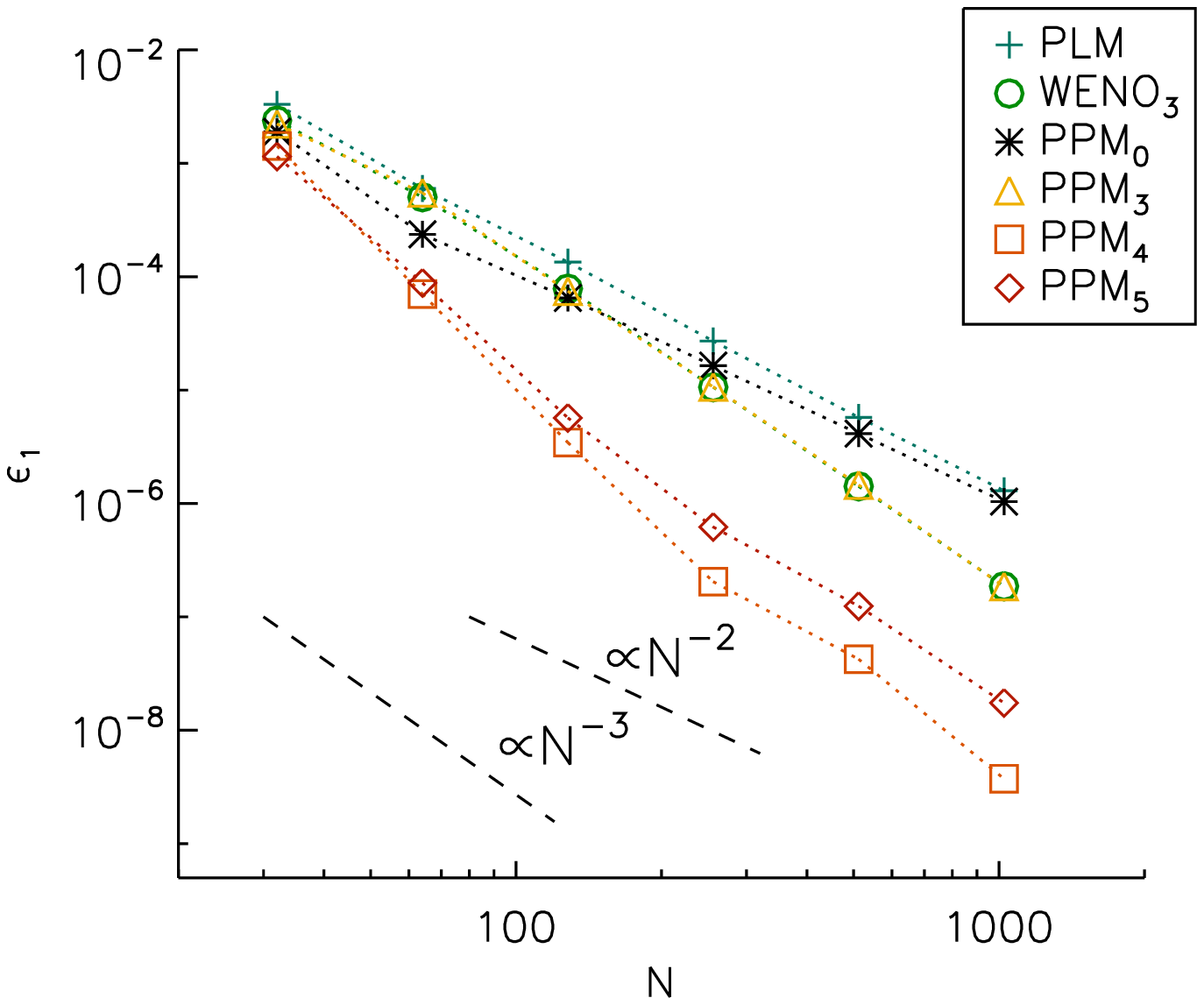}%
 \includegraphics[width=0.4\textwidth]
                 {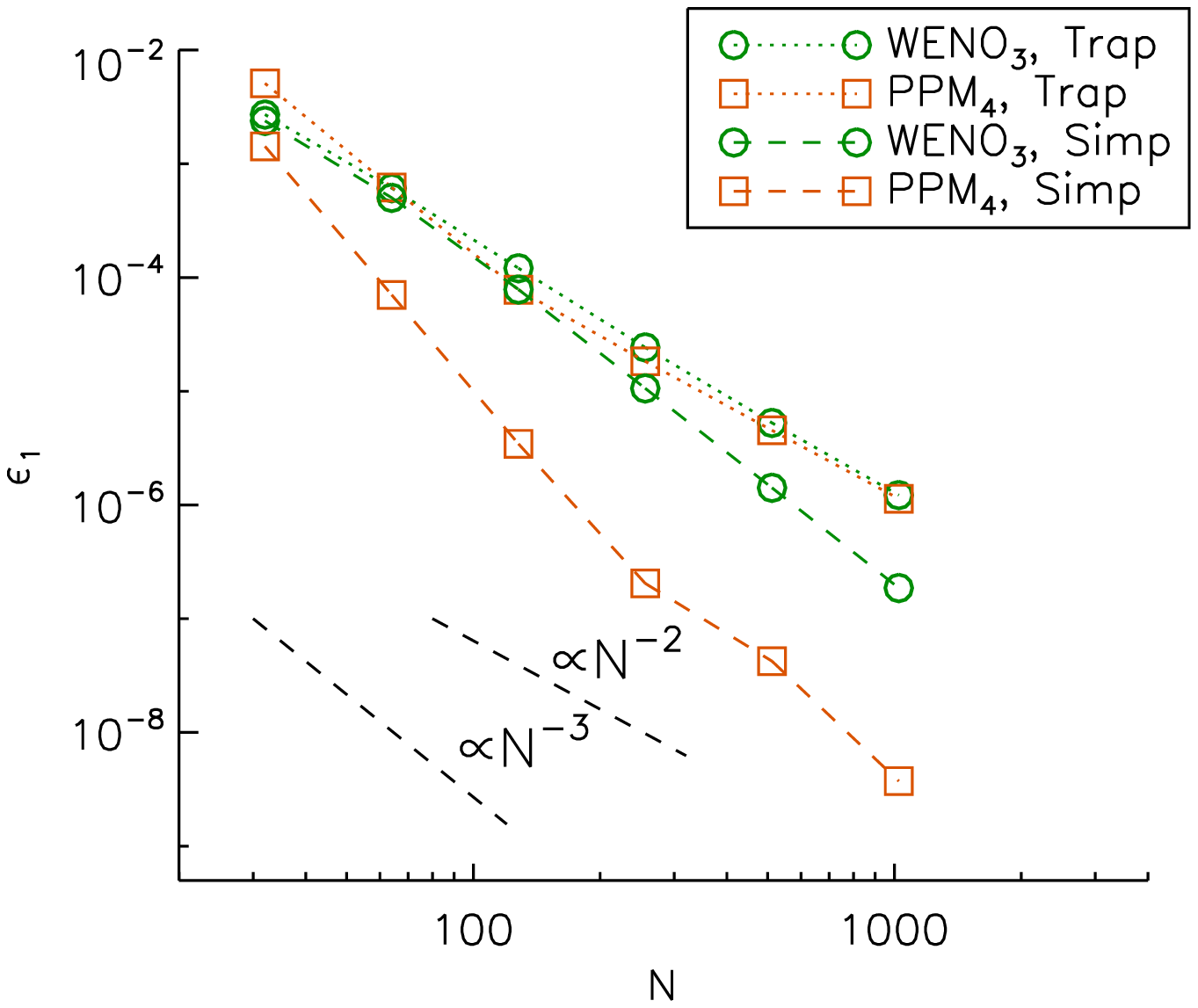}
 \caption{\footnotesize Errors in $L_1$ norm for the MHD plasma column test case as function of the resolutions.
Left panel: comparison between selected reconstruction schemes using Simpson quadrature rule to approximate the source term.
Right panel: errors obtained using the trapezoidal rule (dotted lines) and Simpson rule (dashed lines) for WENO$_3$ and PPM$_4$.
}
 \label{fig:mhd_equil}
\end{figure}

An equilibrium configuration describing a magnetically-confined hot plasma column is considered to evaluate the performance of the reconstruction methods in cylindrical coordinates.
The governing conservation laws are given by the ideal magnetohydrodynamics (MHD) equations in one dimension:
\begin{equation}\label{eq:MHD1D}
  \pd{}{t}\left(\begin{array}{l}
  \rho        \\ \noalign{\medskip}
  \rho v_R    \\ \noalign{\medskip}
  \rho v_\phi R \\ \noalign{\medskip}
  \rho v_z    \\ \noalign{\medskip}
  B_\phi    \\ \noalign{\medskip}
  B_z       \\ \noalign{\medskip}
  E
\end{array}\right)
 +
 \frac{1}{R} \pd{}{R}
\left(\begin{array}{c}
   \rho v_R R \\ \noalign{\medskip}
   (\rho v_R^2 - B_R^2 + p_t)R      \\ \noalign{\medskip}
   (\rho v_\phi R v_R - B_\phi B_R R)R   \\ \noalign{\medskip}
   (\rho v_z      v_R - B_zB_R)R   \\ \noalign{\medskip}
   (B_\phi v_R  - B_Rv_\phi)R   \\ \noalign{\medskip}
   (B_z    v_R  - B_Rv_z)R      \\ \noalign{\medskip}
   (E + p_t)v_R - (\vec{v}\cdot\vec{B})v_R
\end{array}\right)
  = 
\left(\begin{array}{c}
   0                                  \\ \noalign{\medskip}
   (p_t + \rho v_\phi^2 - B_\phi^2)/R \\ \noalign{\medskip}
   0                                  \\ \noalign{\medskip}
   0                                  \\ \noalign{\medskip}
   -(v_\phi B_R - v_RB_\phi)/R        \\ \noalign{\medskip}
   0                                  \\ \noalign{\medskip}
   0
\end{array}\right) \,,
\end{equation}
where, besides the usual gas-dynamical quantities, the magnetic field vector $\vec{B}=(B_R,\,B_\phi,\, B_z)$ has been introduced.
Owing to the solenoidal condition of $\vec{B}$, the radial component of magnetic field is constant $B_R=0$ and does not need to be evolved.
The total energy density $E$ and total pressure $p_t$ are defined, respectively, by
\begin{equation}\label{eq:total_energy}
 E   = \frac{p}{\Gamma - 1} + \frac{1}{2}\rho\vec{v}^2 + \frac{1}{2}\vec{B}^2
 \,;\qquad
 p_t = p + \frac{\vec{B}^2}{2} \,,
\end{equation}
where $\Gamma = 5/3$ is the specific heat ratio.

The equilibrium configuration consists of a static ($\vec{v}=0$), uniform density ($\rho=1$) hot plasma column with pressure and azimuthal magnetic field radial profiles given by
\begin{equation}
  p      = \frac{p_0}{(1 + R^2/R_0^2)^2} \,;  \qquad
  B_\phi = \frac{R}{R_0}\frac{\sqrt{2p_0}}{1 + R^2/R_0^2} \,,
\end{equation}
where $p_0=1$ is the thermal pressure at the axis and $R_0=1$ is a fiducial radius.
The equilibrium is thus determined by the mutual balance between gradient and source terms describing the combined action of pressure and Lorentz forces.

Eqs. (\ref{eq:MHD1D}) are solved on the computational domain $0\le R\le 10$ with axisymmetric boundary conditions at $R=0$ and fixed values in the ghost zones beyond $R=10$.
Fluid variables are initialized using a five-point Gaussian rule to obtain the correct volume averages over the cell.
The MHD equations are evolved using a Courant number $0.8$ until $t=10$.

The left panel in Fig (\ref{fig:mhd_equil}) shows the errors obtained with selected reconstruction schemes by doubling the resolution from $N=32$ up to $N=1024$.
The fourth- and fifth-order PPM schemes present the smallest errors while the uncorrected original scheme (PPM$_0$) yields, at large resolutions, errors which are more than two orders of magnitude larger than PPM$_4$ and a poor convergence rate.
WENO$_3$ and PPM$_3$ show very similar errors.
Here the source terms in Eq. (\ref{eq:MHD1D}) are integrated using the second of Simpson rules in Eq. (\ref{eq:S:simpson12}) with weights given by Eq. (\ref{eq:S:simpson_weights2}) with $m=1$.

For comparison, errors obtained by approximating the integral of the source term with the trapezoidal rule Eq. (\ref{eq:S:trapezoidal2} with $m = 1$) are plotted and compared to the previous ones in the right panel of Fig (\ref{fig:mhd_equil}) for WENO$_3$ and PPM$_4$.
The plot indicates that Simpson quadrature yields considerably smaller errors and third- or higher-order convergence rate when the underlying discretization scheme has the same order of accuracy.
For linear reconstruction (not shown), the two integrations yields substantially the same errors.

\subsubsection{Spherical wind in 2D cylindrical coordinates}
%

As a final test, the propagation of a spherically symmetric radial wind into a static ambient medium is investigated by solving the Euler equations of gas dynamics in 2D cylindrical coordinates $(R,z)$.
Denoting with $\rho$, $v_R$, $v_z$, $p$ and $E$ the density, radial velocity, vertical velocity, pressure and total energy of the flow, respectively, the gas-dynamical equations can be written as
\begin{equation}\label{eq:euler_2D}
  \pd{}{t}\left(\begin{array}{c}
  \rho \\ \noalign{\medskip}
  \rho v_R \\ \noalign{\medskip}
  \rho v_z \\ \noalign{\medskip}
   E     
 \end{array}\right)
 + 
 \frac{1}{R}\pd{}{R}\left(\begin{array}{c}
  \rho v_R R \\ \noalign{\medskip}
  (\rho v^2_R + p)R \\ \noalign{\medskip}
   \rho v_zv_RR \\ \noalign{\medskip}
   (E+p)v_RR
\end{array}\right)
  +
 \pd{}{z}\left(\begin{array}{c}
  \rho v_z \\ \noalign{\medskip}
  \rho v_Rv_z \\ \noalign{\medskip}
  \rho v_z^2 + p \\ \noalign{\medskip}
  (E+p)v_z
\end{array}\right)
  = 
 \left(\begin{array}{c}
  0 \\ \noalign{\medskip}
  p/R \\ \noalign{\medskip}
  0 \\ \noalign{\medskip}
  0 
 \end{array}\right) \,.
\end{equation}
where $E$ is the total energy of the fluid given by Eq. (\ref{eq:total_energy}) with $\vec{B}=0$.
The initial condition consists of a static ambient medium with constant density and pressure values,
\begin{equation}
  \rho = \frac{1}{4}   \,;\qquad
   v_R = v_z = 0  \,;\qquad
   p   = c_{s,a}^2\frac{\rho}{\Gamma} \,.
\end{equation}
where $c_{s,a} = 4\times 10^{-3}$ is the ambient sound speed.
Eqs. (\ref{eq:euler_2D}) are solved everywhere in the domain with the exception of the spherical region $r=\sqrt{R^2+z^2}\le 1$ where a radial spherically symmetric wind is prescribed by keeping fluid variables constant in time.
The wind is characterized by a constant mass outflow rate, radial velocity and pressure:
\begin{equation}\label{eq:stellar_wind}
  \rho v_r r^2  =  1                                 \,;\qquad
  \vec{v}       = \tanh\left(5r\right)\,\left(\frac{R}{r},\,\frac{z}{r}\right)   \,;\qquad
  p             = \frac{\rho^\Gamma c_{s,w}^2}{\Gamma} \,,
\end{equation}
where $c_{s,w} = 3\times 10^{-2}$ is the wind terminal sound speed.
In Eq. (\ref{eq:stellar_wind}), density and velocity are conveniently normalized to the physical properties of the wind in the sense that $\rho \sim 1$ and $v_r \sim 1$ at $r=1$.

The computational domain is defined by $0\le R\le 10$, $-10\le z\le 10$ with outflow boundary conditions applied on all sides except at the symmetry axis where density, pressure and vertical velocity have symmetric profiles while the radial velocity is anti-symmetric.
Computations are carried out using the HLL Riemann solver \cite{HLL.1983} on $256\times 512$ zones and a Courant number of $C_a=0.4$.
Reconstruction is applied directly to primitive variables.

\begin{figure}[!h]
 \centering
 \includegraphics[width=0.8\textwidth]{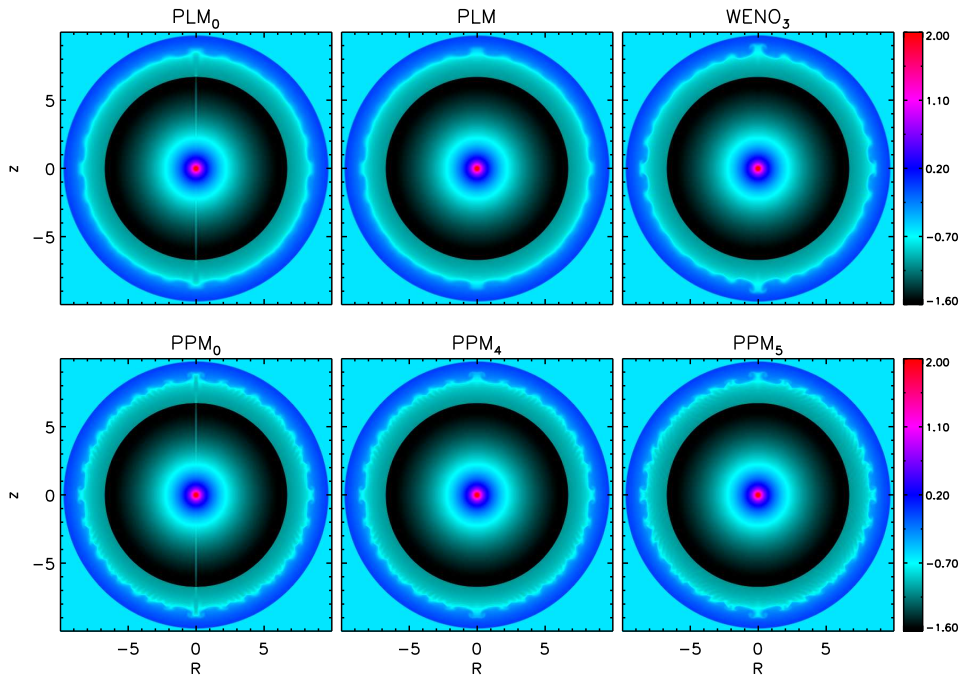}
 \caption{\footnotesize Log density maps for the stellar wind problem at $t=20$ using uncorrected (left panel, PLM$_0$ and PPM$_0$) and corrected (middle and right panels) schemes.
Notice the spurious numerical trails present at $R=0$ in the uncorrected schemes.}
 \label{fig:stellar_wind}
\end{figure}

Fig. \ref{fig:stellar_wind} compares the density maps obtained with selected reconstruction schemes with the uncorrected version of linear (PLM$_0$ using the standard van Leer, Eq. \ref{eq:VL_lim}) and PPM (PPM$_0$) at $t=20$.
The expansion pattern is made of an outermost shock wave enclosing a contact discontinuity and then an innermost shock front where the cold supersonic wind is suddenly heated and brought to subsonic speeds.
The initial spherical symmetry should be preserved throughout the evolution except at the contact wave where grid noise triggers Rayleigh-Taylor finger-like instabilities.
This behavior is correctly reproduced for our geometrically-corrected schemes while a spurious numerical trail appears on the symmetry axis when geometrical constraints are not considered in the underlying reconstruction algorithm (PLM$_0$ and PPM$_0$).
This feature persists at larger resolution and we found it to be a typical signature of an inconsistent formulation at the symmetry axis.
For increasing order of accuracy the amount of numerical diffusion reduces going from PLM to WENO$_3$ and then to PPM$_4$ and PPM$_5$.

\section{Summary}
%
%
%
%

In this paper, I have revised some among the most widespread reconstruction schemes for high-order finite volume discretizations of hyperbolic conservation laws in curvilinear coordinates.
Both scalar and nonlinear systems of conservation laws have been considered.
The problem of reconstruction from volume averages has been formulated in terms of a piecewise polynomial approximation to the solution values by demanding the method to be locally conservative in the neighborhood of a computational zone.
The interface states have been obtained as a linear combination of cell-averages with reconstruction weights that are no longer constant (as in the Cartesian case) but depends on the geometry, most appreciably close to the coordinate origin.
The reconstruction weights can be obtained by inverting a linear system of equations with a Vandermonde matrix whose coefficients depend on the moments of the Jacobian.
The formulation is general enough to be employed on uniform or irregular grids and the coefficients can be computed and stored only once at the beginning of the computation for efficiency purposes.
Explicit analytical expressions for $2^{\rm nd}$- up to $5^{\rm th}$-order accuracy have been derived for uniform radial grids in cylindrical and spherical coordinates and the reconstruction weights have been shown to depend only on the grid index but not on the physical location of the cell.

Although the reconstruction process should be rigorously carried out starting from the volume averages of the conservative variables it has been found, in the case of nonlinear systems, a more robust and equally accurate approach to perform the reconstruction on primitive variables (such density, velocity and pressure) insofar their volume averages are preliminarily computed to the sought order of accuracy.

In order to suppress unwanted oscillation, conventional limiting techniques such as linear Total Variation Diminishing (TVD), Weighted Essentially non-Oscillatory (WENO) and Piecewise Parabolic Method (PPM) have been consistently reforumlated to fulfill with geometrical constraints.


In addition, the numerical integration of curvilinear source terms has been addressed by introducing area-weighted trapezoidal and Simpson quadrature rules that make the computed integral exact for polynomials of degree $2$ and $3$.

Extensive numerical testing in one and two dimensions has demonstrated that geometrical corrections to the reconstruction algorithm are crucial in order to converge to the desired order of accuracy besides decreasing the truncation error.  
Neglecting such corrections can seriously degrade the accuracy of the method and generate spurious numerical artifacts which tend to become particularly pronounced close to the coordinate origin even for simple problems.

Finally, it is pointed out that the reconstruction methods presented in this work have been derived in a rather general way and, being one-dimensional, can be employed with little additional cost in other systems of orthogonal curvilinear coordinates as well.
The extension requires the knowledge of the scale factors and the computation of the moments of the (one-dimensional) Jacobian that can be carried out either analytically or numerically.
Multidimensional reconstruction and higher than second-order quadrature rules, on the other hand, will be investigated in forthcoming studies.

\vspace*{2ex}\par\noindent
{\bf Acknowledgements.}
The author wishes to thank J. Mackey and D. Meyer for fostering the initial discussion of the subjects developed in this work.

\clearpage
\appendix

\section{Derivation of the interpolation weights equations}
\label{app:linear_system_w}
%
%
%

In section \ref{sec:conservative_reconstruction} it has been shown (Eq. \ref{eq:interface_limit_1D}) that the left and right interface values can be computed as the limits from within the zone of the polynomial distribution (Eq. \ref{eq:polynomial}) approximating the cell averages,
\begin{equation}\label{eq:app:states}
  Q_i^{\pm} \equiv Q_i(\xi_{i\pm\HALF}) 
    = \sum_{n=0}^{p-1} a_{i,n} (\xi_{i\pm\HALF}-\xi^c_i)^n\,.
\end{equation}
The coefficients $\{a_{i,n}\}$ are the solution of the linear system (\ref{eq:linear_system}) which, in compact notations, we rewrite as
\begin{equation}\label{eq:app:system}
  \sum_{n=0}^{p-1}\tens{B}_{sn}a_{i,n} = \av{Q}_{i+s}\,,
\end{equation}
where $\tens{B}$ is the $p\times p$ square matrix appearing on the left hand side of Eq. (\ref{eq:linear_system}) with $s=-i_L,...,i_R$ spanning the rows and $n=0,..,p-1$ ranging across the columns.
The formal solution of Eq. (\ref{eq:app:system}) is 
\begin{equation}\label{eq:app:ain}
 a_{i,n} = \sum_{s=-i_L}^{i_R}\tens{C}_{ns}\av{Q}_{i+s} \,,
\end{equation}
where $\tens{C}=\tens{B}^{-1}$ is the inverse matrix of $\tens{B}$.

Written in this form Eq. (\ref{eq:app:states}) has the disadvantage of depending, through the definition of $a_{i,n}$, on a linear combination of the solution values and must therefore be recomputed at each time step and in each zone.
However, by inserting Eq. (\ref{eq:app:ain}) into (\ref{eq:app:states}) one obtains
\begin{equation}
  Q^{\pm}_i= \sum_{n=0}^{p-1}\left(\sum_{s=-i_L}^{i_R} 
   \tens{C}_{ns}\av{Q}_{i+s}\right)
  (\xi_{i\pm\HALF} - \xi^c_i)^n
  = \sum_{s=-i_L}^{i_R} \av{Q}_{i+s}
    \left(\sum_{n=0}^{p-1}
          \tens{C}_{ns}(\xi_{i\pm\HALF} - \xi^c_i)^n
    \right)
\end{equation}
showing that the interface values can indeed be written as a linear combination of the average values as in Eq. (\ref{eq:pm_states}) with coefficients
\begin{equation}\label{eq:app:w}
  w^{\pm}_{i,s} = \sum_n \tens{C}_{ns}(\xi_{i\pm\HALF} - \xi^c_i)^n \,.
\end{equation}
Furthermore, since $\tens{C}_{ns} = (\tens{C}^T)_{sn} = ((\tens{B}^T)^{-1})_{sn}$, Eq. (\ref{eq:app:w}) represents the solution of the linear system
\begin{equation}
  \sum_{s=-i_L}^{i_R}(\tens{B}^T)_{ns}w_{i,s}^{\pm} 
  = (\xi_{i\pm\HALF} - \xi^c_i)^n
\end{equation}
which is precisely Eq. (\ref{eq:linear_system_w}).

\section{Interpolation weights in the radial direction}
\label{app:interpolation_weights}
%
%

In the following, the reconstruction weights $w^\pm_{i,s}$ used to compute the leftmost and rightmost interface values defined by Eq. (\ref{eq:pm_states}) are given for a uniform grid.
Keeping the same notations as in \S\ref{sec:conservative_reconstruction}, the  stencil spans $i_L$ zones to the left of cell $i$ and $i_R$ zones to the right.
The order of accuracy is given by $p=i_L + i_R + 1$.
When the order of the reconstruction is even, left and right adjacent interface values are the same and only the coefficients $w^+_{i,s}$ are needed.

In the following the index $i$ will be used to label the computational zones and can be defined as
\begin{equation}
  i = \frac{\xi_{i+\HALF}}{\Delta \xi} \,.
\end{equation}
For completeness, interpolation weights are given not only for cylindrical and spherical coordinates but also for Cartesian geometry.

\subsection{Cartesian Coordinates}
\label{app:cartesian}
%
%

Cartesian reconstruction weights are simply obtained by solving Eq. (\ref{eq:linear_system_w}) using $\partial\vol /\partial\xi= 1$ in Eq. (\ref{eq:beta_coeff}).

\begin{itemize}

\item Reconstruction weights for $p=3\,, \ (i_L=1,\, i_R=1)$:

\begin{equation}
\begin{array}{lcl}
\DS (w^+_{i,-1},\, w^+_{i,0},\, w^+_{i,1}) &=& 
\DS \left(-\frac{1}{6},\,\frac{5}{6},\,\frac{1}{3}\right)
  \\ \noalign{\medskip}
\DS (w^-_{i,-1},\, w^-_{i,0},\, w^-_{i,1}) &=& 
\DS \left(\frac{1}{3},\,\frac{5}{6},\,-\frac{1}{6}\right)
\end{array}
\end{equation}

\item Reconstruction weights for $p=4\,, \ (i_L=1,\, i_R=2)$:

\begin{equation}
\left(w^+_{i,-1},\,w^+_{i,0},\,w^+_{i,1},\, w^+_{i,2}\right)  = 
  \DS \left( -\frac{1}{12},\, \frac{7}{12},\, \frac{7}{12},\,-\frac{1}{12}
\right)
\end{equation}
\vspace{0.5cm}

\item Reconstruction weights for $p=5\,, \ (i_L=2,\, i_R=2)$:

\begin{equation}
\begin{array}{lcl}
\DS \left(w^+_{i,-2},\, w^+_{i,-1},\, w^+_{i,0},\, w^+_{i,1},\, w^+_{i,2}\right) &=&
\DS \left(\frac{1}{30},\, -\frac{13}{60},\, \frac{47}{60},\,
          \frac{9}{20},\, -\frac{1}{20}\right)
  \\ \noalign{\medskip}
\DS \left(w^-_{i,-2},\, w^-_{i,-1},\, w^-_{i,0},\, w^-_{i,1},\, w^-_{i,2}\right) &=&
\DS \left( -\frac{1}{20},\, \frac{9}{20},\, \frac{47}{20}  
           -\frac{13}{60},\, \frac{1}{30}\right)
\end{array}
\end{equation}

\end{itemize}

\subsection{Cylindrical Coordinates}
%
%
%

Reconstruction weights for the cylindrical radial coordinate are obtained by inverting the linear system in Eq. (\ref{eq:linear_system_w}) with $\partial\vol /\partial\xi= \xi$ in Eq. (\ref{eq:beta_coeff}).
In the vanishing curvature limit ($i\to\infty$) the weights tend to the corresponding Cartesian expressions given in \ref{app:cartesian}.

\begin{itemize}

\item Reconstruction weights for $p=3\,, \ (i_L=1,\, i_R=1)$:

\begin{equation}\label{eq:iL1_iR1_cyl}\left\{\begin{array}{lcl}
w^+_{i,-1} & = & \DS \frac{- \left( 2\,i-3 \right)  \left( 2\,{i}^{2}-1 \right) 
}{\Delta} \\  \noalign{\bigskip}
w^+_{i,0} & = & \DS \frac{ \left( 10\,{i}^{2}-9\,i-11 \right)  \left( 2\,i-1 \right) 
}{\Delta} \\  \noalign{\bigskip}
w^+_{i,1} & = & \DS \frac{ \left( 2\,i+1 \right)  \left( 4\,{i}^{2}-9\,i+4 \right) 
}{\Delta}
\end{array}\right.
\qquad  \quad
\left\{\begin{array}{lcl}
w^-_{i,-1} & = & \DS \frac{ \left( 2\,i-3 \right)  \left( 4\,{i}^{2}+i-1 \right) 
}{\Delta} \\  \noalign{\bigskip}
w^-_{i,0} & = & \DS \frac{ \left( 10\,{i}^{2}-11\,i-10 \right)  \left( 2\,i-1 \right) 
}{\Delta} \\  \noalign{\bigskip}
w^-_{i,1} & = & \DS \frac{- \left( 2\,i+1 \right)  \left( 2\,{i}^{2}-4\,i+1 \right) 
}{\Delta}
\end{array}\right.\end{equation}
\vspace{0.5cm}
where $\Delta = 12\, \left( {i}^{2}-i-1 \right)  \left( 2\,i-1 \right) $.

\item Reconstruction weights for  $p=4\,, \ (i_L=1,\, i_R=2)$:

\begin{equation}\left\{\begin{array}{lcl}
w^+_{i,-1} & = & \DS \frac{- \left( 2\,i-3 \right)  \left( 5\,{i}^{3}+8\,{i}^{2}-3\,i-4 \right) }{\Delta} \\  \noalign{\bigskip}
w^+_{i,0} & = & \DS \frac{ \left( 2\,i-1 \right)  \left( 35\,{i}^{3}+24\,{i}^{2}-93\,i-60
 \right) 
}{\Delta} \\  \noalign{\bigskip}
w^+_{i,1} & = & \DS \frac{ \left( 2\,i+1 \right)  \left( 35\,{i}^{3}-24\,{i}^{2}-93\,i+60
 \right) 
}{\Delta} \\  \noalign{\bigskip}
w^+_{i,2} & = & \DS \frac{- \left( 2\,i+3 \right)  \left( 5\,{i}^{3}-8\,{i}^{2}-3\,i+4 \right) 
}{\Delta}
\end{array}\right.
\end{equation}
\vspace{0.5cm}
where $\Delta = 120\,{i}^{4}-360\,{i}^{2}+96$.

\item Reconstruction weights for $p=5\,, \ (i_L=2,\, i_R=2)$:

\begin{equation}\left\{\begin{array}{lcl}
w^+_{i,-2} & = & \DS \frac{4\, \left( 2\,i-5 \right)  \left( 3\,{i}^{4}-10\,{i}^{2}+4 \right) 
}{\Delta} \\  \noalign{\bigskip}
w^+_{i,-1} & = & \DS \frac{- \left( 2\,i-3 \right)  \left( 78\,{i}^{4}-75\,{i}^{3}-380\,{i}^{2}+
45\,i+164 \right) }{\Delta} \\  \noalign{\bigskip}
w^+_{i,0} & = & \DS \frac{ \left( 282\,{i}^{4}-525\,{i}^{3}-1300\,{i}^{2}+1395\,i+1276 \right) 
 \left( 2\,i-1 \right) 
}{\Delta} \\  \noalign{\bigskip}
w^+_{i,1} & = & \DS \frac{3\, \left( 2\,i+1 \right)  \left( 54\,{i}^{4}-175\,{i}^{3}-60\,{i}^{2}
+465\,i-228 \right) 
}{\Delta} \\  \noalign{\bigskip}
w^+_{i,2} & = & \DS \frac{-3\, \left( 2\,i+3 \right)  \left( 6\,{i}^{4}-25\,{i}^{3}+20\,{i}^{2}+
15\,i-12 \right) 
}{\Delta}
\end{array}\right.
\end{equation}
\vspace{0.5cm}
\begin{equation}\left\{\begin{array}{lcl}
w^-_{i,-2} & = & \DS \frac{-3\, \left( 2\,i-5 \right)  \left( 6\,{i}^{4}+{i}^{3}-19\,{i}^{2}-4\,i
+4 \right) 
}{\Delta} \\  \noalign{\bigskip}
w^-_{i,-1} & = & \DS \frac{3\, \left( 2\,i-3 \right)  \left( 54\,{i}^{4}-41\,{i}^{3}-261\,{i}^{2}
-36\,i+56 \right) 
}{\Delta} \\  \noalign{\bigskip}
w^-_{i,0} & = & \DS \frac{ \left( 282\,{i}^{4}-603\,{i}^{3}-1183\,{i}^{2}+1652\,i+1128 \right) 
 \left( 2\,i-1 \right) 
}{\Delta} \\  \noalign{\bigskip}
w^-_{i,1} & = & \DS \frac{- \left( 2\,i+1 \right)  \left( 78\,{i}^{4}-237\,{i}^{3}-137\,{i}^{2}+
628\,i-168 \right) 
}{\Delta} \\  \noalign{\bigskip}
w^-_{i,2} & = & \DS \frac{4\, \left( 2\,i+3 \right)  \left( 3\,{i}^{4}-12\,{i}^{3}+8\,{i}^{2}+8
\,i-3 \right) 
}{\Delta}
\end{array}\right.
\end{equation}
where $\Delta = 120\, \left( 2\,i-1 \right)  \left( 3\,{i}^{4}-6\,{i}^{3}-13\,{i}^{2}+
16\,i+12 \right)$.

\end{itemize}

\subsection{Spherical Coordinates}
%
%

Reconstruction weights for the spherical radial coordinate are computed by solving the linear system in Eq. (\ref{eq:linear_system_w}) with $\partial\vol /\partial\xi= \xi^2$ in Eq. (\ref{eq:beta_coeff}).
Note again that the vanishing curvature limit ($i\to\infty$) yields the corresponding Cartesian expressions given in \ref{app:cartesian}.

\begin{itemize}

\item Reconstruction weights for $p=3\,, \ (i_L=1,\, i_R=1)$:

\begin{equation}\label{eq:iL1_iR1_sph}\left\{\begin{array}{lcl}
w^+_{i,-1} & = & \DS \frac{- \left( 3\,{i}^{2}-9\,i+7 \right)  \left( 10\,{i}^{4}-9\,{i}^{2}+3
 \right) 
}{\Delta} \\  \noalign{\bigskip}
w^+_{i,0} & = & \DS \frac{ \left( 3\,{i}^{2}-3\,i+1 \right)  \left( 50\,{i}^{4}-90\,{i}^{3}-63\,
{i}^{2}+96\,i+69 \right) 
}{\Delta} \\  \noalign{\bigskip}
w^+_{i,1} & = & \DS \frac{2\, \left( 3\,{i}^{2}+3\,i+1 \right)  \left( 10\,{i}^{4}-45\,{i}^{3}+
72\,{i}^{2}-48\,i+12 \right) 
}{\Delta}
\end{array}\right.
\end{equation}
\vspace{0.5cm}
\begin{equation}\left\{\begin{array}{lcl}
w^-_{i,-1} & = & \DS \frac{2\, \left( 3\,{i}^{2}-9\,i+7 \right)  \left( 10\,{i}^{4}+5\,{i}^{3}-3
\,{i}^{2}-i+1 \right) 
}{\Delta} \\  \noalign{\bigskip}
w^-_{i,0} & = & \DS \frac{ \left( 3\,{i}^{2}-3\,i+1 \right)  \left( 50\,{i}^{4}-110\,{i}^{3}-33
\,{i}^{2}+100\,i+62 \right) 
}{\Delta} \\  \noalign{\bigskip}
w^-_{i,1} & = & \DS \frac{- \left( 3\,{i}^{2}+3\,i+1 \right)  \left( 10\,{i}^{4}-40\,{i}^{3}+51
\,{i}^{2}-22\,i+4 \right) 
}{\Delta}
\end{array}\right.
\end{equation}
where $\Delta = 18\left(
10\,{i}^{6}-30\,{i}^{5}+15\,{i}^{4}+20\,{i}^{3}-9\,{i}^{2}-6\,i+4\right)
$.

\item Reconstruction weights for $p=4\,, \ (i_L=1,\, i_R=2)$:

\begin{equation}\left\{\begin{array}{lcl}
w^+_{i,-1} & = & \DS \frac{- \left( 3\,{i}^{2}-9\,i+7 \right)  \left( 15\,{i}^{6}+48\,{i}^{5}+23
\,{i}^{4}-48\,{i}^{3}-30\,{i}^{2}+16\,i+12 \right) 
}{\Delta} \\  \noalign{\bigskip}
w^+_{i,0} & = & \DS \frac{ \left( 3\,{i}^{2}-3\,i+1 \right)  \left( 105\,{i}^{6}+144\,{i}^{5}-
487\,{i}^{4}-720\,{i}^{3}+510\,{i}^{2}+1008\,i+372 \right) 
}{\Delta} \\  \noalign{\bigskip}
w^+_{i,1} & = & \DS \frac{ \left( 3\,{i}^{2}+3\,i+1 \right)  \left( 105\,{i}^{6}-144\,{i}^{5}-
487\,{i}^{4}+720\,{i}^{3}+510\,{i}^{2}-1008\,i+372 \right) 
}{\Delta} \\  \noalign{\bigskip}
w^+_{i,2} & = & \DS \frac{- \left( 3\,{i}^{2}+9\,i+7 \right)  \left( 15\,{i}^{6}-48\,{i}^{5}+23
\,{i}^{4}+48\,{i}^{3}-30\,{i}^{2}-16\,i+12 \right) 
}{\Delta}
\end{array}\right.
\end{equation}
where $\Delta = 36\left(15\,{i}^{8}-85\,{i}^{6}+150\,{i}^{4}-60\,{i}^{2}+16\right)
$.

\item Reconstruction weights for $p=5\,, \ (i_L=2,\, i_R=2)$:

\begin{equation}\left\{\begin{array}{lcl}
w^+_{i,-2} & = &  \frac{2\, \left( 3\,{i}^{2}-15\,i+19 \right)  \left( 7\,{i}^{8}-45\,{i}^{6}+
94\,{i}^{4}-60\,{i}^{2}+16 \right) 
}{\Delta} \\  \noalign{\bigskip}
w^+_{i,-1} & = &  \frac{- \left( 3\,{i}^{2}-9\,i+7 \right)  \left( 91\,{i}^{8}-175\,{i}^{7}-
780\,{i}^{6}+930\,{i}^{5}+2417\,{i}^{4}-795\,{i}^{3}-1740\,{i}^{2}+240
\,i+508 \right) 
}{\Delta} \\  \noalign{\bigskip}
w^+_{i,0} & = & \frac{ \left( 3\,{i}^{2}-3\,i+1 \right)  \left( 329\,{i}^{8}-1225\,{i}^{7}-
1800\,{i}^{6}+8670\,{i}^{5}+3863\,{i}^{4}-20325\,{i}^{3}-5700\,{i}^{2}
+15120\,i+8132 \right) 
}{\Delta} \\  \noalign{\bigskip}
w^+_{i,1} & = & \frac{ \left( 3\,{i}^{2}+3\,i+1 \right)  \left( 189\,{i}^{8}-1225\,{i}^{7}+
1620\,{i}^{6}+4350\,{i}^{5}-11517\,{i}^{4}+1275\,{i}^{3}+16560\,{i}^{2
}-15120\,i+4212 \right) 
}{\Delta} \\  \noalign{\bigskip}
w^+_{i,2} & = & \frac{- \left( 3\,{i}^{2}+9\,i+7 \right)  \left( 21\,{i}^{8}-175\,{i}^{7}+
510\,{i}^{6}-510\,{i}^{5}-223\,{i}^{4}+645\,{i}^{3}-120\,{i}^{2}-240\,
i+108 \right) 
}{\Delta}
\end{array}\right.
\end{equation}

\vspace{0.5cm}

\begin{equation}\left\{\begin{array}{lcl}
w^-_{i,-2} & = & \frac{- \left( 3\,{i}^{2}-15\,i+19 \right)  \left( 21\,{i}^{8}+7\,{i}^{7}-
127\,{i}^{6}-51\,{i}^{5}+222\,{i}^{4}+96\,{i}^{3}-60\,{i}^{2}-16\,i+16
 \right) 
}{\Delta} \\  \noalign{\bigskip}
w^-_{i,-1} & = & \frac{ \left( 3\,{i}^{2}-9\,i+7 \right)  \left( 189\,{i}^{8}-287\,{i}^{7}-
1663\,{i}^{6}+1071\,{i}^{5}+4888\,{i}^{4}+1184\,{i}^{3}-1350\,{i}^{2}-
164\,i+344 \right) 
}{\Delta} \\  \noalign{\bigskip}
w^-_{i,0} & = &  \frac{ \left( 3\,{i}^{2}-3\,i+1 \right)  \left( 329\,{i}^{8}-1407\,{i}^{7}-
1163\,{i}^{6}+9431\,{i}^{5}+368\,{i}^{4}-21376\,{i}^{3}-310\,{i}^{2}+
15196\,i+7064 \right) 
}{\Delta} \\  \noalign{\bigskip}
w^-_{i,1} & = & \frac{- \left( 3\,{i}^{2}+3\,i+1 \right)  \left( 91\,{i}^{8}-553\,{i}^{7}+
543\,{i}^{6}+2329\,{i}^{5}-4388\,{i}^{4}-1544\,{i}^{3}+6850\,{i}^{2}-
3516\,i+696 \right) 
}{\Delta} \\  \noalign{\bigskip}
w^-_{i,2} & = & \frac{2\, \left( 3\,{i}^{2}+9\,i+7 \right)  \left( 7\,{i}^{8}-56\,{i}^{7}+
151\,{i}^{6}-122\,{i}^{5}-91\,{i}^{4}+132\,{i}^{3}+25\,{i}^{2}-42\,i+
12 \right) 
}{\Delta}
\end{array}\right.
\end{equation}
where $\Delta = 180\left(7\,{i}^{10}-35\,{i}^{9}+210\,{i}^{7}-161\,{i}^{6}-399\,{i}^{5}+390\,{i}^{4}+200\,{i}^{3}-164\,{i}^{2}-48\,i+48\right)$.

\end{itemize}

%
%
%




\bibliographystyle{elsarticle-num}
\bibliography{mignone}


%
%
\end{document}